\documentclass{article}

  \usepackage[main, final]{neurips_2026}

\usepackage[utf8]{inputenc} 
\usepackage[T1]{fontenc}    
\usepackage{hyperref}       
\usepackage{url}            
\usepackage{booktabs}       
\usepackage{amsfonts}       
\usepackage{nicefrac}       
\usepackage{microtype}      
\usepackage{xcolor}         

\usepackage{todonotes}
\usepackage{CJKutf8}

\title{\textsc{Felid}: A Flexible and Efficient Design for Transformer Fine-Tuning over Encrypted Data}

\author{
Linru Zhang \\ Nanyang Technological University 
\And 
Jun Jie Sim \\ Ant International, Ant Group
\And 
Xiangning Wang \\ Nanyang Technological University
\And 
Jiahao Zhong \\ Nanyang Technological University
\And 
Kaiyu Zhou \\ Nanyang Technological University
\And 
Huanyi Ye \\ Nanyang Technological University
\And 
Yongsen Zheng \\ Nanyang Technological University
\And 
Lushan Song \\ Ant International, Ant Group
\And 
Xiaojian Liang \\ Ant International, Ant Group
\And 
Yingting Liu \\ Ant International, Ant Group
\And
Yujing Sun \\ Nanyang Technological University
\And 
Huaxiong Wang \\ Nanyang Technological University
\And 
Pu Duan$^*$ \\ Ant International, Ant Group 
\And 
Kwok-Yan Lam\thanks{Corresponding author} \\ Nanyang Technological University
}

\usepackage[utf8]{inputenc} 
\usepackage[T1]{fontenc}    
\usepackage{hyperref}       
\usepackage{url}            
\usepackage{booktabs}       
\usepackage{amsfonts}       
\usepackage{nicefrac}       
\usepackage{microtype}      
\usepackage{xcolor}         
\usepackage{stmaryrd} 
\usepackage{float} 
\usepackage{csquotes} 
\usepackage{aliascnt} 
\usepackage{subfiles} 
\usepackage{lipsum} 
\usepackage{amsmath,amsfonts,amssymb,amsthm,bm} 
\usepackage{mathtools} 
\usepackage{enumitem} 
\usepackage{multirow} 
\usepackage{multicol} 
\usepackage{caption} 
\usepackage[dvipsnames]{xcolor} 
\usepackage{graphicx} 
\usepackage{tikz}
\usepackage[noabbrev,capitalize,nameinlink]{cleveref}
\usepackage{comment}
\usepackage{dcolumn}
\usepackage{threeparttable}
\usepackage{algorithm}
\usepackage{algorithmic}
\usepackage{makecell}
\usepackage{subfig}
\usepackage{subcaption}

\usepackage{xcolor}

\usepackage{pifont}

\newtheorem{definition}{Definition}[section]

\newcounter{note}[section]

\newcommand{\linru}[1]{\textcolor{PineGreen}{\bf (Linru: #1)}}

\newcommand{\dec}{\textrm{Dec}}
\newcommand{\ecd}{\textrm{Encode}}

\newcommand{\ckksrot}{\textrm{Rot}^{\textrm{HE}}}
\newcommand{\ckksrotint}{\textrm{Rot}^{\textrm{HE},\textrm{Internal}}}
\newcommand{\ckksrel}{\textrm{Relinearization}^{\textrm{HE}}}
\newcommand{\rot}{\textrm{Rot}}
\newcommand{\ckksks}{\textrm{KeySwitch}^{\textrm{HE}}}

\newcommand{\sk}{\textrm{sk}}

\newcommand{\grad}[1]{\frac{\partial L}{\partial #1}}

\newcommand{\R}{\mathbb{R}}

\newcommand{\C}{\mathbb{C}}

\renewcommand\vec{\mathbf}

\DeclareMathOperator{\layernorm}{LayerNorm}

\DeclareMathOperator*{\softmax}{Softmax}

\DeclareMathOperator*{\attention}{Attention}
\DeclareMathOperator*{\gelu}{GELU}
\DeclareMathOperator*{\add}{Add}
\DeclareMathOperator*{\mult}{Mult}
\DeclareMathOperator*{\conj}{Conj}
\DeclareMathOperator*{\plainmult}{PMult}

\DeclareMathOperator{\enc}{Enc}
\newcommand{\cmark}{\ding{51}}%
\newcommand{\xmark}{\ding{55}}%
\begin{document}

\maketitle

\begin{abstract} 

  Fully Homomorphic Encryption (FHE) enables non-interactive, privacy-preserving transformer fine-tuning, protecting highly sensitive data as inputs to large language models. Prior work has predominantly pursued efficiency gains through architectural refinements, yet overall performance remains bounded by the substantial cost inherent in encrypted computation.
  To address these issues, we target the bottleneck at the ciphertext operation level.
  Concretely, we propose a novel packing strategy that supports batching over a flexible number of inputs, making it well-suited for fine-tuning workloads.
  In addition, we design operation-specific matrix evaluation algorithms that significantly reduce (i) the number of ciphertext rotations and relinearizations, and (ii) the number of bootstrapping invocations.
  Additionally, we exploit functional bootstrapping to fold the evaluation of element-wise activation functions directly into the bootstrapping step, thereby reducing overall computational complexity.
  As a result, our framework achieves a $30\times$ reduction in ciphertext rotations and relinearizations across all matrix operations, as well as a $2.5\times$ reduction in FHE bootstrapping invocations over the entire computation flow, compared to the latest state-of-the-art work.
  It is worth noting that our system completes fine-tuning of a 2-layer BERT-style model with a batch size of 16 in 650 seconds, achieving a $8\times$ speedup over the state-of-the-art work. 
\end{abstract}


\section{Introduction and Related Works}
\label{sec:intro}

Given the growing privacy concerns over AI-as-a-Service, particularly for institutions with strict privacy requirements, such as financial institutions and law enforcement, Privacy-Preserving Machine Learning (PPML)~\cite{han2019logistic,lee2023hetal,lee2022low} has attracted significant attention.
However, despite continuous efforts to address this privacy problem via cryptography-based techniques, 
they mostly suffer from a critical limitation in efficiency, with Secure Multi-Party Computation (MPC)-based methods~\cite{yao1982protocols,chen2022the-x,lu2023bumblebee,pang2024bolt,limpcformer,zeng2025mpcache,dong2023puma} incurring substantial interactive communication overhead, while Fully Homomorphic Encryption (FHE)-based solutions face extremely high computational costs~\cite{zhang2024secure,cryptoeprint:2024/1881,park2024powerformer,cryptoeprint:2025/991,ARION}.

Given the practical limitations of interactive MPC-based methods, particularly the need for clients to remain continuously online and sustain stable, low-latency communication with the AI Cloud Service Provider (AI CSP) during a highly time-intensive process, non-interactive FHE-based approaches are draw growing interest.
Specifically, with the widespread adoption of Transformer architectures, particularly in Large Language Models (LLMs) that are commonly deployed as cloud services, make efforts are made in enabling privacy-preserving machine learning (PPML) for Transformers using non-interactive FHE. 
However, most work~\cite{zhang2024secure, cryptoeprint:2024/1881, park2024powerformer, cryptoeprint:2025/991, zimerman2024converting, cryptoeprint:2025/956} focuses on 
applying FHE to Transformers during the \textit{inference phase}, 
\textbf{leaving privacy-preserving fine-tuning during the \textit{training phase}} inadequately explored, which is crucial for task-specific adaptation and personalized responses.
Although ~\cite{lee2022privacy, lee2023hetal} have attempted to address this task, they are largely limited to fine-tuning the classification head while keeping the core Transformer layers untouched, failing to adequately address computational overhead.

Indeed, efficient privacy-preserving fine-tuning of Transformers with FHE is highly challenging, as FHE is mathematically secure but computationally expensive. To address this, two nontrivial aspects need to be considered:
\textsc{Q1). Model Design.} The efficiency of FHE varies across different operations. 
Replacing computationally intensive operations with more FHE-friendly alternatives can significantly improve overall performance.
\textsc{Q2). FHE Design.} More crucially, for each operation, the efficiency of FHE evaluation should be carefully optimized to reduce computational overhead.
Most recently, Rho et al.~\cite{rhoencryption} pioneered addressing FHE-friendly Model Design (\textsc{Q1}) via transforming standard Transformer models into a FHE computational friendly variant by adopting a LoRA-based fine-tuning strategy and replacing the softmax operation in attention with a Gaussian kernel (GK), but leave the more challenging problem of efficient FHE evaluation design (\textsc{Q2}) fully underexplored.

\subsection{Our Contributions}
To bridge this gap, based on the model architecture of~\cite{rhoencryption}, we propose \textbf{\textsc{FELID}}, a \textbf{\underline{F}}l\textbf{\underline{e}}xib\textbf{\underline{l}}e and Eff\textbf{\underline{i}}cient \textbf{\underline{D}}esign for transformer fine-tuning over encrypted data, focusing on a novel and efficient FHE evaluation design. 
We modify none of the model architecture or fine-tuning procedure, and thus our framework is fully compatible with existing model designs~\cite{rhoencryption}. 
Specifically, matrix evaluation and bootstrapping dominate the cost of FHE evaluation, accounting for approximately $70\% - 85\%$ of the runtime in~\cite{rhoencryption}.
We reduce the cost from all perspectives. 


\textbf{A Flexible Column-Wise Batch Packing Strategy} is first proposed for encrypting the input data. 
\textit{Batching multiple inputs} can improve the amortized latency. 
\footnote{For example in CPMM without batching, which computes linear combinations over the $d$ encrypted columns of one input in one ciphertext, aggregation of these columns with $O(\log(d))$ ciphertext rotations is required. 
Batching $b$ inputs with $b$ ciphertexts enables distributing the $d$ columns across $b$ ciphertexts, and the FHE rotations can be reduced from $O(\log(d))$ to $O(\log(d/b))$. 
When batching $d$ inputs with $d$ ciphertexts, 
the aggregation phase eliminates ciphertext rotations entirely, substantially lowering the per-input computational cost.}
\textit{Flexible} batch sizes make it well-suited for Transformer fine-tuning with variable batch size (typically 4-64), and adaptable to diverse tasks and training settings.\footnote{As shown in Table~\ref{tab:comp-packing}, existing works in FHE-based transformer evaluation either do not support batching or rely on a fixed batch size strategy, both of which are unsuitable for fine-tuning scenarios with variable batch sizes.} 
And \textit{column-wise} manner, compared to row-wise manner, further accelerates matrix multiplication. 

\textbf{Efficient Matrix Operations} are presented to \textbf{(1).}  Reduce the number of expensive FHE operations in matrix evaluations to decrease the computational complexity. \footnote{Ciphertext rotations and relinearizations are widely recognized as the most expensive operations in FHE-based matrix operations~\cite{cryptoeprint:2024/1881,park2024powerformer,cryptoeprint:2025/991,ARION}. }
For ciphertext-plaintext matrix multiplication in attention head and feedforward layers, we achieve \textbf{$213\times$} fewer ciphertext rotations and \textbf{$2$} fewer multiplicative levels, 
and for ciphertext-ciphertext matrix multiplications in LoRA, we achieve \textbf{$21\times$} fewer ciphertext rotations and relinearizations and \textbf{$1$} fewer multiplicative level,  compared to~\cite{rhoencryption} in a batch size of 16. 
\textbf{(2)}. Reduce the multiplicative depth consumption to decrease the number of bootstrappings in the pipeline, which is required to refresh cipher-texts once the remaining levels reach 0.  
For instance, in kernel attention, we propose ``column packing $\rightarrow$ diagonal packing'' evaluation pipeline that significantly lowers the multiplicative depth required to compute $\Vert Q_{i,:}-K_{j,:}\Vert_2^2$, reducing it from $7$ in~\cite{rhoencryption} to $3$. 
\footnote{$Q_{i,:}$ and $K_{j,:}$ are the $i$-th and the $j$-th row of $Q,K$. Our matrix evaluation algorithms differ from those in MOAI~\cite{cryptoeprint:2025/991} and ARION~\cite{ARION}. Although both use a ``column packing $\rightarrow$ diagonal packing '' strategy in attention block, they compute $QK^{\intercal}$, whereas our method targets $l_2$-norm of the rows from $Q$ and $K$ (kernel definition), so the underlying matrix evaluation algorithm is totally different. }


Our contributions are summarized as:
\begin{itemize}[leftmargin=*]
    \item We proposed a novel FHE Evaluation Design that significantly improved the privacy-preserving fine-tuning of transformers, via addressing the computational bottleneck in Matrix Evaluation and Bootstrapping.

    \item From computational complexity perspective, our framework achieves a $30\times$ reduction in ciphertext rotations and relinearizations across all matrix operations, as well as a $2.5\times$ reduction in FHE bootstrapping invocations over the entire computation flow, compared to SoTA~\cite{rhoencryption}. 

    \item From practical perspective, we complete fine-tuning of a 2-layer BERT-style model with a batch size of 16 in 650 seconds, achieving a $8\times$ speedup over SoTA~\cite{rhoencryption}. 
\end{itemize}

\begin{table}[tbp]
    \centering
    \caption{Comparison of packing methods in FHE-based non-interactive privacy-preserving transformer inference/fine-tuning schemes.}
    \label{tab:comp-packing}
    \resizebox{\textwidth}{!}{
    \begin{tabular}{c|cccccccc}
    \toprule
         Scheme 
         & NEXUS~\cite{zhang2024secure} 
         & THOR~\cite{cryptoeprint:2024/1881} 
         & Powerformer~\cite{park2024powerformer} 
         & MOAI~\cite{cryptoeprint:2025/991} 
         & Tricycle~\cite{cryptoeprint:2025/1200}
         & ARION~\cite{ARION} 
         & Rho et al.~\cite{rhoencryption}
         & Our work\\ \midrule
         Task & Inference & Inference &Inference &Inference &Inference &Inference &Fine-tuning & Fine-tuning \\
         Batch inputs & \cmark & \xmark & \xmark & \cmark & \xmark & \cmark & \xmark & \cmark
         \\
         Flexible size & \xmark & - & - & \xmark & - & \xmark & - & \cmark
         \\
    \bottomrule
    \end{tabular}
    }
\end{table}

\section{Preliminaries}
\textbf{Notations.}
For a positive integer $m$, let $[m]$ be the set of $\{0,1,2,...,m-1\}$. 
Given a modulus $p$ we use $a\%p$ for $a\ (\textrm{mod } p)$.
We use uppercase letters to denote a matrix $A$, and use $A[i][j]$ to represent the element at the $i$-th row and $j$-th column of $A$. 
$A[i][:]$ is used to denote the $i$-th row of $A$. 
We use bold lowercase letters to represent vectors: $\bm{v}$, and use $v[i]$ or $v_i$ to represent the $i$-th element of $\bm{v}$.
The Hadamard product, or the element-wise multiplication is denoted with $\odot$, and is applied between vectors or matrices of identical sizes $\bm{v_1}\odot\bm{v_2}$, $A\odot B$. 
For vector $\bm{v}$ we define its rotation by $j$ elements to the left as $\rot_j(\bm{v})=(v_j,v_{j+1},...v_0,...,v_{j-1})$. 

    

\subsection{Fully Homomorphic Encryption (FHE) }
FHE is a special cryptographic primitive that allows one to perform computations on encrypted data without decryption. 
This property makes FHE one of the most powerful and widely applied privacy-enhancing technologies. 
In this work, we will use the Cheon-Kim-Kim-Song (CKKS) FHE scheme \cite{cheon2017homomorphic,cheon2018bootstrapping} which supports arithmetic of complex numbers in $\C$ in single-instruction-multiple-data (SIMD) style for any given ciphertext.
More concretely, given $N$ the encryption parameter which is a power-of-$2$, 
the CKKS scheme can pack at most $N$ complex numbers. 
Let $ct_1:=\enc(\bm{m}_1)$, $ct_2:=\enc(\bm{m}_2)$, the CKKS scheme supports the following encrypted computations: \\
$\bullet\ \add(ct_1,ct_2)$. It outputs $\enc(\bm{m}_1+\bm{m}_2)$. We use $ct_1+ct_2$ for short. \\
$\bullet\ \mult(ct_1,ct_2)$. It outputs $\enc(\bm{m}_1\odot\bm{m}_2)$. We use $ct_1\odot ct_2$ for short. \\
$\bullet\ \plainmult(\bm{v},ct_1)$. It outputs $\enc(\bm{v}\odot\bm{m}_1)$. We use $\bm{v}\odot ct_1$ for short. \\
$\bullet\ \ckksrot_j(ct_1)$ outputs $\enc(\rot_j(\bm{m}_1))$; $\conj(ct_1)$ outputs $\enc(\bar{\bm{m}}_1)$, the conjugation of $\bm{m}_1$.  

\textbf{Bootstrapping and Functional Bootstrapping. }
In the CKKS FHE scheme, homomorphic multiplications inject noise into the underlying message of a ciphertext, and each ciphertext is assigned a level to track the noise growth.
A fresh ciphertext is considered to be at level $L$, and $\mult(\cdot,\cdot)$ and $\plainmult(\cdot,\cdot)$ consume one level each.
A ciphertext at level $0$ would yield a decryption error.
Bootstrapping~\cite{cheon2018bootstrapping} is a technique that \enquote{refreshes} a ciphertext by restoring levels to enable more computations. 
Functional bootstrapping~\cite{alexandru2025general,yang2025rboot} extends this idea by combining ciphertext refreshing with function evaluation in a single procedure, incurring little to no additional overhead.

\subsection{Privacy-Preserving Language Model Fine-Tuning}


\paragraph{Kernel Attention.}
The softmax operator in the standard attention mechanism (more details are in Appendix~\ref{app:softmax}) can be viewed as a row-wise normalized exponential kernel over query--key similarities, or more generally, a kernel-based weighting mechanism in the sense of classical kernel methods~\cite{berlinet2011reproducing}. 
In FHE, however, the softmax operator is difficult to evaluate as it involves several hard-to-approximate operations such as exponentiation, division, and row-wise maximum used for numerical stabilization \cite{cho2024fast,park2024powerformer,rhoencryption}. 
A natural encryption-friendly alternative is therefore to replace softmax with a FHE-friendly kernel function. 
\cite{rhoencryption} considered the \emph{Gaussian kernel attention}, defined for each head $h\in[H]$ as
\[
    \mathrm{GK\,Attn}(Q_h,K_h,V_h):=U_hV_h,\qquad
    U_h[i][j]:=
    \exp\!\left(
    -\frac{1}{2\sqrt{n_2/H}}
    \|Q_h[i][:]-K_h[j][:]\|_2^2
    \right),
\]
where $Q_h,K_h,V_h\in \R^{n_1\times (n_2/H)}$, $U_h\in \R^{n_1\times n_1} $. 
$n_1$ denotes token length, $n_2$ the hidden dimension, and $H$ the number of heads. 
Compared to softmax attention, this formulation eliminates the need to evaluate the division and max function over encrypted data. 
Moreover, the exponential function is only evaluated on non-positive inputs, where it is bounded and more amenable to polynomial approximation for FHE.

Recently, the Epanechnikov kernel has been studied in dense associative memory and compact kernel regression, where its bounded support naturally induces sparse attention scores~\cite{hoover2026dense,santos2026sparse}. 
In an unnormalized FHE-friendly form, \emph{Epanechnikov kernel attention} can be written as
\[
    \mathrm{EK\,Attn}(Q_h,K_h,V_h):=U_hV_h,\qquad
    U_h[i][j]:=
    \mathrm{ReLU}\!\left(
    1-\frac{1}{2\eta_h^2}
    \|Q_h[i][:]-K_h[j][:]\|_2^2
    \right),
\]
where $\eta_h>0$ is a learnable parameter for the $h$-th head.
In our implementation, $\eta_h$ is initialized as $\sqrt{n_2/H}$ and is fixed during fine-tuning.
Both the Gaussian and Epanechnikov kernel are viable alternatives to softmax attention. 


\paragraph{LoRA Fine-Tuning.}
Low-Rank Adaptation (LoRA)~\cite{hu2022lora} is a parameter-efficient fine-tuning method that keeps the set of pre-trained weights fixed and learns only a low-rank update, which is used to reduce the computational overhead in FHE-based privacy-preserving fine-tuning~\cite{rhoencryption}.
For a linear layer with weight matrix $W\in\mathbb{R}^{n_2\times n_2'}$, LoRA parameterizes the updatable matrices as
\[
    \Delta W = AB\in \R^{n_2\times n_2'},\qquad
    A\in\mathbb{R}^{n_2\times n_{\text{LoRA}}},\quad
    B\in\mathbb{R}^{n_{\text{LoRA}}\times n_2'},\quad
    n_{\text{LoRA}}\ll \min(n_2,n_2').
\]
Only the low-rank factors $A$ and $B$ are trained, while the backbone weights $W$ are frozen.

\section{Our Method}
To address the computational bottleneck in FHE-based transformer fine-tuning, we focus on designing efficient FHE evaluation methods tailored to fine-tuning workloads, which involves:
(1) a flexible batching strategy for encrypting input data, enabling support for variable batch sizes in transformer fine-tuning. 
(2) efficient matrix operations to reduce the computational complexity and decrease the number of bootstrapping operations in the pipeline, and
(3) flexible interleaved column packing for vector-wise non-linear functions to avoid ``internal rotation'' and functional bootstrapping for element-wise non-linear functions to reduce computational cost.

\textbf{Problem Statement and Threat Model. }
We consider a server–client computing model under a privacy-preserving threat setting.
The server hosts a language model while the client holds private data and the FHE secret key.
The embedding layer is assumed to be known to the client and is computed locally on the client side. 
During fine-tuning, the client first computes the token embeddings and encrypts them before sending the encrypted inputs to the server. 
The server then performs the forward pass and evaluates the loss function on encrypted data. Based on this process, encrypted weight updates are obtained and used to update the transformer parameters in the encrypted form.
During inference, the server executes the forward pass on encrypted inputs and returns encrypted predictions to the client, who decrypts them to obtain the final results.
We assume an \textit{honest-but-curious} server: it honestly follows the protocol but may attempt to infer information about the client’s private data from all the encrypted computations.


\subsection{A Flexible Column-Wise Batch Packing Strategy}
In transformer fine-tuning, inputs are processed in batches, where the batch size is a tunable hyperparameter and typically ranges from 4 to 64. 
So that a practically and effective packing solution to input data should support flexible batch packing, to better adapt to varying batch sizes and improve slot utilization within ciphertexts, enhancing overall performance.
For example, when the batch size is 16, flexible packing can significantly reduce slot underutilization compared to fixed-size schemes, where up to $50\%$ of slots in NEXUS and $93\%$ in MOAI and ARION may otherwise remain unused.

However, the only state-of-the-art work for privacy-preserving transformer finetuing by Rho et al.~\cite{rhoencryption} does not support true batching at all. Instead, it adopts a row-packing strategy in which each input sample is encoded independently into ciphertexts, with matrix elements arranged row-wise, and can only “simulate” batching by assigning one sample per GPU across eight GPUs. 
Thought works on privacy-preserving transformer inference, NEXUS~\cite{zhang2024secure}, MOAI~\cite{cryptoeprint:2025/991} and ARION~\cite{ARION} explored reducing amortized computational overhead via batching, 
their batch sizes are fixed (e.g., 32 in NEXUS and 256 in MOAI and ARION for BERT-base). 


Therefore, we propose a \textit{Flexible Column-Wise Batch Packing Strategy}, which supports flexible batch sizes within a unified FHE framework, maximizes batching efficiency to reduce amortized cost, and avoids slot under-utilization in ciphertexts. 
Our packing strategy packs columns across multiple inputs into ciphertexts, with elements from different inputs interleaved across slots.
More precisely, a batch of inputs are organized via \textit{Flexible Interleaved Column Packing} and encrypted. 

\begin{figure}
    \centering
    \includegraphics[width=0.9\linewidth]{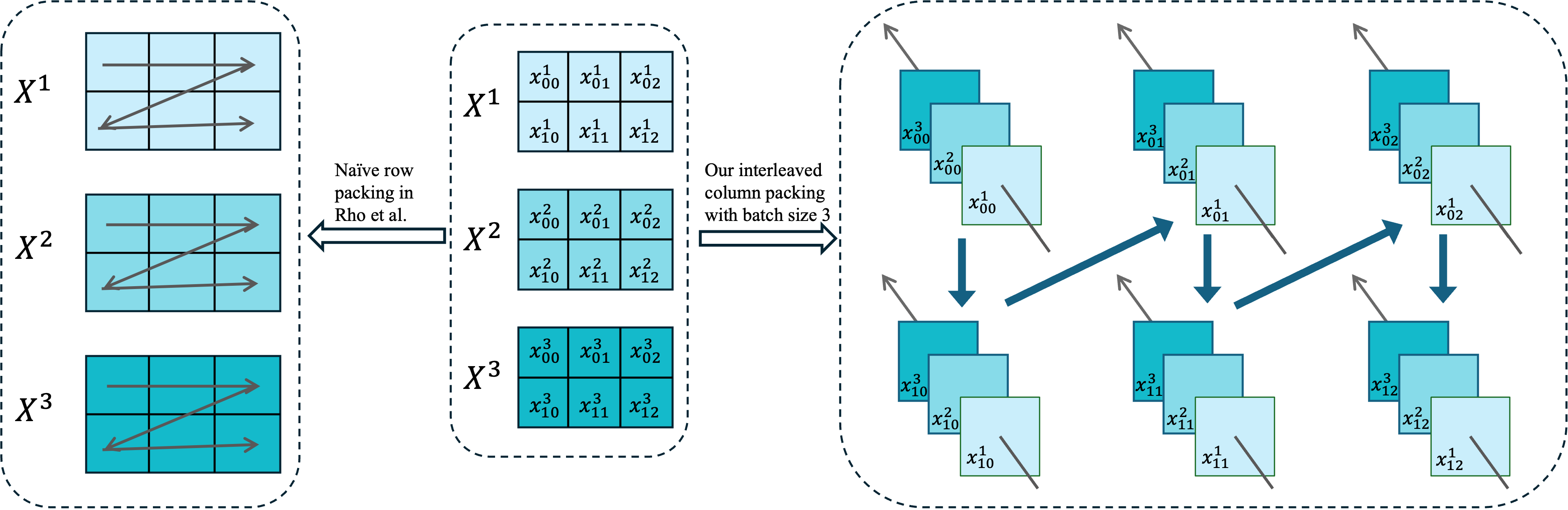}
    \caption{Toy example of comparison between row packing in Rho et al.~\cite{rhoencryption} and our flexible interleaved column packing. Here $n_1=2,n_2=3,b=3$. If $N=6$, then the $3$-flexible interleaved column packing divides $\{X^1,X^2,X^3\}$ into 3 ciphertexts encrypting: $(x^1_{00},x^2_{00},x^3_{00},x^1_{10},x^2_{10},x^3_{10}), 
     (x^1_{01},x^2_{01},x^3_{01},x^1_{11},x^2_{11},x^3_{11}),(x^1_{02},x^2_{02},x^3_{02},x^1_{12},x^2_{12},x^3_{12})$. }
    \label{fig:inter_col}
    \vspace{-1em}
\end{figure}

In Figure~\ref{fig:inter_col}, we compare the row-wise packing in~\cite{rhoencryption} with our flexible interleaved column-wise packing. 
Compared to row packing, our method avoids “internal rotations” within each row when computing layers in form of $\enc(X)W$. 
Furthermore, during the evaluation of transformer operations, to avoid expensive format conversion and minimize encrypted matrix transpositions, our framework further incorporates \textit{flexible interleaved diagonal packing}, enabling efficient evaluation across different types of computations.

The description of our \textit{Flexible Interleaved Packing} is given in Definition~\ref{def:inter-col-informal}, and a more detailed definition is in Definition~\ref{def:inter-col} and Definition~\ref{def:diag-packing} from Appendix~\ref{app:matrix}. 

\begin{definition}[(informal) Flexible interleaved column/diagonal packing]
\label{def:inter-col-informal}
Let $\{X^i\}_{i\in[b]}$ be a batch of matrices with 
$X^i \in \mathbb{R}^{n_1 \times n_2}$ for all $i \in [b]$. 
Given row index $r$ and column index $c$, we first pack all the $(r,c)$-th element from each $X^i$, denoted as $\{X^i[r][c]\}_{i\in[b]}$. 
We say we pack $\{X^i\}_{i\in[b]}$ in $b$-\textbf{interleaved column packing} if we concatenate sets $\{X^i[r][c]\}_{i\in[b]}$ in column-major:
\[
\{X^i[0][0]\}_{i\in[b]} \Big\| \{X^i[1][0]\}_{i\in[b]}\Big\|...\Big\|\{X^i[n_1-1][0]\}_{i\in[b]}\Big\|\{X^i[0][1]\}_{i\in[b]}\Big\|...
\]
We say we pack $\{X^i\}_{i\in[b]}$ in $b$-\textbf{interleaved (upper) diagonal packing} if $X^i$ is square and we concatenate sets $\{X^i[r][c]\}_{i\in[b]}$ in (upper) diagonal-major: 
\[
\{X^i[0][0]\}_{i\in[b]} \Big\| \{X^i[1][1]\}_{i\in[b]}\Big\|...\Big\|\{X^i[n_1-1][n_1-1]\}_{i\in[b]}\Big\|\{X^i[0][1]\}_{i\in[b]}\Big\|...
\]
\end{definition}

\subsection{Matrix Operations in Flexible Interleaved Packing}
\label{sec:mm-operations}
In this section, we describe our efficient matrix operations, mainly including 4 types: (1) Ciphertext-plaintext matrix multiplication, which is used to multiply the input with the plaintext weights like $W_Q, W_K, W_V$ and weight matrices in feedforward block; (2) Ciphertext-ciphertext matrix multiplication for LoRA; (3) Ciphertext-ciphertext matrix operation for kernel attention, i.e., the $l2$-norm of $Q$ and $K$, and $\times V$; (4) Ciphertext-ciphertext matrix multiplication for classification head. 
These matrix operations are also employed in the backward pass. 
Further details are provided in Appendix~\ref{app:backward}. 

\paragraph{Ciphertext-Plaintext Matrix Multiplication (CPMM).}
The $r$-th column in $X^iW$ is a linear combination of all the columns from $X^i$, with weights being the $r$-th column of $W$. 
Therefore, the CPMM, i.e., the linear combination of columns contains two phases: \\
(1) Multiplication: With properly chosen elements from plaintext matrix $W$, we can construct plaintext vectors of size $N$ for the weight multiplications. This step costs $1$ multiplicative level. \\
(2) Accumulation: 
The results from phase (1) are accumulated to form partial weighted sums of columns, and finally are combined using FHE rotations to produce ciphertexts encrypting columns of $X^iW$. This step does not costs multiplicative level.
The detailed algorithm is elaborated in Algorithm~\ref{alg:CPMM} and Algorithm~\ref{alg:CPMM-C} from Appendix~\ref{app:matrix}. 
Compare with the CPMM in \cite{rhoencryption}, ours with $b$-interleaved column packing requires about $213\times$ fewer FHE rotations for the CPMM in attention block, when evaluating $b=16$ inputs. 
In addition, our CPMM only costs $1$ multiplicative level, while  \cite{rhoencryption} costs 3. 

\paragraph{Ciphertext-Ciphertext Matrix Multiplication for LoRA.}
Compute $(XA)B$ instead of $X(AB)$ can reduce the FHE operation complexity from $O(n_2^2)$ to $O(n_2n_{\text{LoRA}})$ \cite{rhoencryption}. 
We use LoRA-friendly packing for $A\in\R^{n_2\times n_{\text{LoRA}}}$ and $B\in \R^{n_{\text{LoRA}}\times n_2}$, in Algorithm~\ref{alg:pack-A-LoRA} and~\ref{alg:pack-B-LoRA-C} respectively. 
Informally, let $\enc(X)$ contains $d:=\lceil \frac{bn_1n_2}{N}\rceil $ ciphertexts, we use $2d$ ciphertexts to pack $A$ or $B$, so that either $\enc(X)\enc(A)$ or $(\enc(X)\enc(A))\enc(B)$ are simplified to one round of ciphertext-ciphertext (ctct) multiplication and one round of accumulation. 
When evaluating $b=16$ inputs, ours requires $21\times$ fewer key-switching operations in the attention block compared to \cite{rhoencryption}. 
The detailed algorithm is elaborated in Algorithm~\ref{alg:CCMM-A-LoRA} and Algorithm~\ref{alg:CCMM-B-LoRA-C} from Appendix~\ref{app:matrix}. 
Ours only costs $2$ multiplicative level, while the one in \cite{rhoencryption} costs 3. 

\paragraph{Ciphertext-Ciphertext Matrix Operations for Kernel Attention.}
We compute matrix $S_h\in\R^{n_1\times n_1}$ s.t. $S_h[i][j]:=\Vert Q_h[i][:]-K_h[j][:] \Vert_2^2$. 
\cite{rhoencryption} computes $S_h[i][j]$ by $\Vert Q_h[i][:]\Vert_2^2+\Vert K_h[j][:] \Vert_2^2 - Q_h[i][:](K_h[j][:])^\intercal$. 
However, this step requires several matrix transpositions and consumes too many multiplicative levels. 
We propose a novel method to avoid matrix transpositions. 
In addition, our novel method generates $S_h$ in diagonal-major packing, making it possible to apply the optimization from \cite{ARION} to significantly reduce the number of FHE rotations. 
Our method is based on:
\\
(1) Compute $\{Q_h[i][:]-K_h[i][:]\}_{i\in[n_1]}$ by $Q_h-K_h$. 
Then the $l_2$-norm can be obtained by multiplication, FHE rotation and summations.
Observe that the result is the main diagonal of $S_h$. \\
(2) Rotating \textit{each column} of $K_h$ by $d$ steps, we can compute $\{Q_h[i][:]-K_h[(i+d)\%n_1][:]\}_{i\in[n_1]}$. Then we can obtain the $d$-th (upper) diagonal of $S_h$.\\
We integrate the optimizations from \cite{ARION} and elaborate the detailed algorithms for $S_h$ in Algorithm ~\ref{alg:CCMM-QK-doubleBSGS} and for the multiplication with $V_h$ in Algorithm~\ref{alg:CCMM-V-doubleBSGS}.
With all these optimizations, we use total 6 multiplicative levels to compute $S_h$ and $\sigma(S_h)V_h$, while in \cite{rhoencryption} it costs 12. 
Our algorithm has $1.40\times$ reduction in the number of key-switching compared to \cite{rhoencryption}. 

\paragraph{Ciphertext-Ciphertext Vector-Matrix Multiplication for Classification Head.}
To optimize the classification head within our framework, we have designed a pipeline that efficiently handles the transition from transformer layers to final predictions. 
We utilize special packing strategies for weight matrices in Algorithms~\ref{alg:pack-Wc1} and \ref{alg:pack-Wc2} respectively, and a ``repeated packing'' format for the token in Algorithm~\ref{alg:ext-first-row}. 
This approach ensures that the two multiplication steps cost minimal key-switching operations, resulting in a $1.4\times$ to $17.4\times$ reduction in the number of key-switching compared to \cite{rhoencryption}. 
Detailed formal descriptions of these packing algorithms and the corresponding homomorphic evaluation procedures are provided in Appendix~\ref{app-ccmm-class}.

\subsubsection{Analysis of Computation Overheads}
\label{sec:mm-analysis}

Key-switching, which is involved in HE rotations and relinearizations, is recognized as the most expensive operation in HE-based matrix operations such as matrix multiplications and kernel evaluations. 
Therefore, existing works~\cite{cryptoeprint:2024/1881,park2024powerformer,cryptoeprint:2025/991,ARION} usually use the number of key-switching as the benchmark in comparing different matrix operations. 

Table~\ref{tab:comp-MM} presents a comprehensive comparison between our algorithms and~\cite{rhoencryption} in terms of number of key-switching operations across all major matrix operations. 
Aggregated over all matrix operations in fine-tuning a 2-layer BERT-style model, our approach reduces key-switching operations by up to $30\times$ while reducing the total multiplicative depth by approximately half.

In kernel computation $\{S_h[u][v]:=|| Q_h[u][:]-K_h[v][:]||_2^2\}_{u,v\in[n_1]}$, our method incurs approximately $1.16\times$ more key switching operations. 
This trade-off is intentional for two reasons: 
(1) it significantly reduces the multiplicative depth from 7 to 3, which is critical for decreasing the frequency of bootstrapping; 
and (2) we jointly optimize the kernel computation and the subsequent $\times V_h$ operation using a double BSGS technique. While this slightly increases the number of rotations in the kernel computation, it leads to a larger reduction in rotations in the $\times V_h$ step.

Overall, the results demonstrate that our algorithms provide substantial efficiency gains across the entire transformer pipeline by reducing both the number of expensive homomorphic operations and required multiplicative levels.

\begin{table}[t]
    \caption{Comparisons with~\cite{rhoencryption}. Numbers in the table are based on evaluating all kinds of matrix operations (CPMM, CCMM for LoRA, CCMM for kernel, CCMM for $\times V$, and CCMM in classification head) in the forward pass of a single BERT layer with kernel attention on a batch of inputs (batch size 
b). The concrete parameters are as follows: $X^i \in \R^{128 \times 768}, i \in [16]$, $W_Q, W_K, W_V \in R^{768 \times 768}$, $A \in \R^{768 \times n_{\mathrm{LoRA}}}$, $B \in \R^{n_{\mathrm{LoRA}} \times 768}$, 
$W_{\mathrm{c2}} \in \R^{768 \times 2}$. We have: number of heads $H = 12$, batch size $b = 16$, and $n_{\mathrm{LoRA}}=2$, which are consistently used in both our experiments and those of~\cite{rhoencryption}. }
    \label{tab:comp-MM}
    \centering
    \resizebox{\textwidth}{!}{
    \begin{tabular}{cc|c|cccc}
    \toprule
    Transformer operations& Matrix operations & Schemes & $\#(\ckksrot)$ & $\#(\ckksrel)$ & \makecell[c]{Total \\ $\#(\ckksks)$} & \makecell[c]{Multiplicative\\ level used} \\ \midrule

    \multirow{2}{*}{\makecell[c]{$X^iW_*$,\\ $i \in [b],\ *=Q, K,$ and $V$}} & \multirow{2}{*}{CPMM}
    & Ours & 480 & 0 & 480 ($213\times$ fewer) & 1 (2 levels fewer)
    \\[.5ex]
    &  & \cite{rhoencryption} & 102480 & 0 & 102480 & 3
    \\ [.5ex] 
    \hline 
    
    \multirow{2}{*}{\makecell[c]{$X^iA_*B_*$,\\ $i \in [b],\ *=Q, K,$ and $V$}} & \multirow{2}{*}{CCMM-LoRA}
    & Ours & 24 & 78 &  102 ($21\times$ fewer)  & 2 (1 level fewer)
    \\ [.5ex]
    & & \cite{rhoencryption} & 2016 & 192  & 2208 & 3 
    \\ [.5ex] 
    \hline 

    \multirow{2}{*}{\makecell[c]{$S_h$, \\ $i \in [b], h \in   [H]$}} & \multirow{2}{*}{CCMM-kernel}
    & Ours & 21912 & 1536 & 23448 & 3 (4 levels fewer)
    \\ [.5ex]
    & & \cite{rhoencryption} & 19776 & 384 & 20160 & 7  
    \\ [.5ex] \hline

    \multirow{2}{*}{\makecell[c]{$\sigma(S^i_h)V^i_h$, \\  $i \in [b], h \in   [H]$}} &\multirow{2}{*}{ CCMM-V }
    & Ours & 7320 & 192 & 7512 ($3.1\times$ fewer) & 3 (2 levels fewer)
    \\ [.5ex]
    & & \cite{rhoencryption} & 23232 & 192 & 23424 & 5
    \\ [.5ex] \hline




    \multirow{2}{*}{\makecell[c]{$X^i_{[CLS]} W_{c1}$, $i \in [b]$ }} & \multirow{2}{*}{CCMM-classification-1}\protect\footnotemark
    & Ours & 361 & 6 & 367 ($1.4\times$ fewer) & 2 (1 level fewer)
    \\
    & & \cite{rhoencryption}  & 512 & 32 & 544 & 3 
    \\ \hline

    \multirow{2}{*}{\makecell[c]{$\tanh(\cdot)W_{c2}$, $i \in [b]$ }} & \multirow{2}{*}{CCMM-classification-2}
    & Ours & 10 & 1 & 11 ($17.4\times$ fewer) & 1 
    \\
    & & \cite{rhoencryption} & 176 & 16 & 192 & 1

    \\ 
    \bottomrule
    \end{tabular}
    }
\end{table}
\footnotetext{In~\cite{rhoencryption}, $W_{c1}$ is replaced with two dense block of LoRA-style shapes with rank=32, thus they can use CCMM for LoRA in this step. 
While in our scheme, we do not modify the model architecture and use $W_{c1} \in R^{768 \times 768}$ as the original model.}
\subsection{Non-Linear Function Evaluation}
\label{sec:non-linear}

We split non-linear functions into two broad categories: element-wise functions, like $\mathrm{exp}$, $\mathrm{ReLU}$, $\mathrm{GELU}$, and $\mathrm{Tanh}$; and vector-wise functions, like $\mathrm{LayerNorm}$ in transformer layers and $\mathrm{SoftMax}$ in the classification head.


For vector-wise functions, we organize inputs using a flexible interleaved column packing. Compared to the row-packing strategy in~\cite{rhoencryption}, (1) column-packing avoids “internal rotations” within rows when computing reductions such as $\sum_{i \in [n_1]}X[\cdot][i]$, thereby reducing the number of ciphertext rotations by half (a toy example is provided in Appendix~\ref{app-layernorm}), and (2) interleaved batching further reduce the amortized number of ciphertext rotations to $\log(N/bn_1)/b$, where $N$ is the slot count and $b$ is the batch size. 
Following~\cite{cryptoeprint:2024/1881, cryptoeprint:2025/991}, we approximate the inverse and inverse square root in $\mathrm{LayerNorm}$ and $\mathrm{SoftMax}$ using Goldschmidt division algorithm and Newton iteration. 
More details of the algorithms used are in Appendix~\ref{app-layernorm}.

Element-wise functions are evaluated during Functional Bootstrapping~\cite{yang2025rboot}, combining ciphertext refresh and activation function evaluation into a single operation with negligible additional cost.
Compared to the conventional “non-linear evaluation + bootstrapping” pipeline used in~\cite{rhoencryption, cryptoeprint:2025/991, cryptoeprint:2024/1881, park2024powerformer, zhang2024secure}, functional bootstrapping eliminates the need for a separate activation evaluation step, thereby improving the overall workflow. 
More details can be found in Appendix~\ref{app:fb}. 
In Table~\ref{tab:FT-time} from Section~\ref{sec:micro}, we show that our algorithm achieves $4.3\times$ speedup over~\cite{rhoencryption} in evaluating the activation functions in the feedforward block.

\section{Experimental Results}
\label{sec:exp}
\textbf{Model and Datasets. }
Experiments are conducted on 3 GLUE tasks~\cite{wang2018glue}, MRPC, RTE and STSB. 
Following~\cite{rhoencryption}, we use a compact BERT backbone with 2 transformer layers, hidden size is 768, feedforward size is 3072, and has 12 attention heads \footnote{\texttt{google/bert\_uncased\_L-2\_H-768\_A-12}~\cite{devlin2019bert}}. 
We refer the reader to Appendix~\ref{sec:app-exp-setup} for a more detailed justification behind our choices.
We use a batch size of $16$ for all tasks.

We compare softmax attention (SM), Gaussian kernel attention (GK), and Epanechnikov kernel attention (EK). The SM baseline is evaluated from the original pretrained checkpoint, and the GK baseline is taken directly from the published results of~\cite{rhoencryption}. 
More details are in Appendix~\ref{sec:app-exp-setup}. 
 
\textbf{FHE Parameters and Implementations. }
We set the degree of the polynomial ring as $2^{16}$, which supports $N=2^{15}$ plaintext slots in a ciphertext. 
The ciphertext modulus $Q$ is chosen to ensure a 128-bit security level~\cite{albrecht2021homomorphic}, and support $12$ multiplicative levels after bootstrapping.\footnote{ Our choice of ciphertext modulus and multiplicative depth is consistent with the baseline in~\cite{rhoencryption}. Specifically,~\cite{rhoencryption} reports 9 available levels before bootstrapping and a minimum requirement of 3 levels to perform bootstrapping, which matches the requirement for functional bootstrapping in our scheme. Their open-source implementation also confirms that 12 levels are restored after bootstrapping. }
Our scheme is initially implemented using the Microsoft SEAL library~\cite{sealcrypto} and subsequently migrated to the PhantomFHE library. \footnote{https://github.com/encryptorion-lab/phantom-fhe}
All of our experiments used a single NVIDIA H200 GPU. 
Our code is available. \footnote{https://github.com/LWEpsilon-FHE/FELID} 

\textbf{Baselines. }
The implementation in~\cite{rhoencryption} is based on the proprietary (non-open-sourced) HEaaN library and evaluated on an NVIDIA GeForce RTX 4090 GPU, whereas our implementation is built on PhantomFHE and runs on an NVIDIA H200 GPU. 
In Table~\ref{tab:micro-lib} in Appendix~\ref{app-lib-exp}, we benchmark the primitive operations of the CKKS scheme in both libraries under same parameters.
To ensure fair comparison, we reproduce their evaluation algorithms using the PhantomFHE library and measure runtime on the same hardware platform. 
\subsection{Microbenchmarks}
\label{sec:micro}
We first compare against Rho et al.~\cite{rhoencryption} to demonstrate the speedups achieved by our method on the following aspects. 
All experiments are conducted under the same parameter settings and hardware, using a single-threaded configuration.
 
\begin{figure}[tbp]
    \begin{minipage}[t]{0.25\linewidth}
        \centerline{\includegraphics[width=0.99\textwidth]{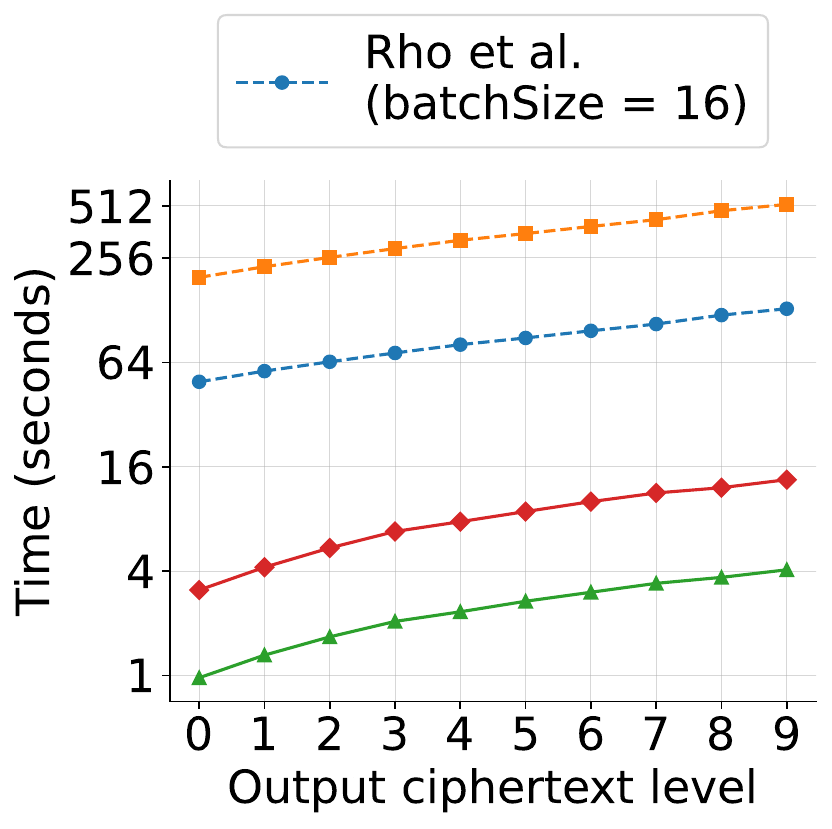}}
        \centerline{\makecell[c]{\footnotesize CPMM\\ \tiny $\R^{128 \times 768} \times \R^{768 \times 768}$}}
    \end{minipage}%
    \begin{minipage}[t]{0.25\linewidth}
        \centerline{\includegraphics[width=0.99\textwidth]{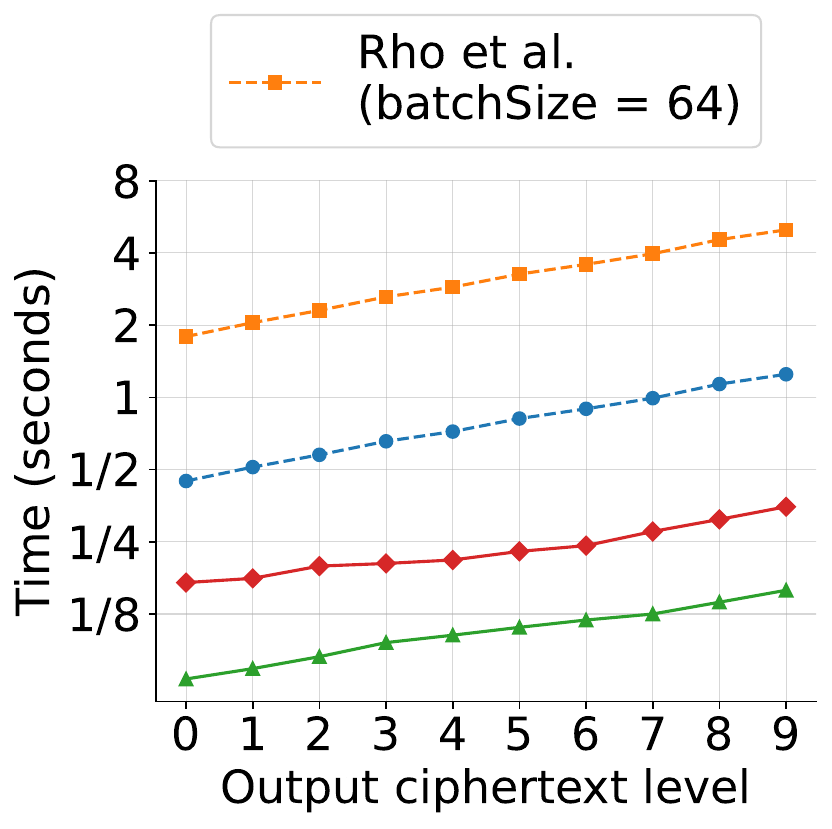}}
        \centerline{\makecell[c]{\footnotesize CCMM-LoRA\\ \tiny $\R^{128 \times 768} \times \R^{768 \times 2} \times \R^{2 \times 768}$}}
    \end{minipage}
    \begin{minipage}[t]{0.25\linewidth}
        \centerline{\includegraphics[width=0.99\textwidth]{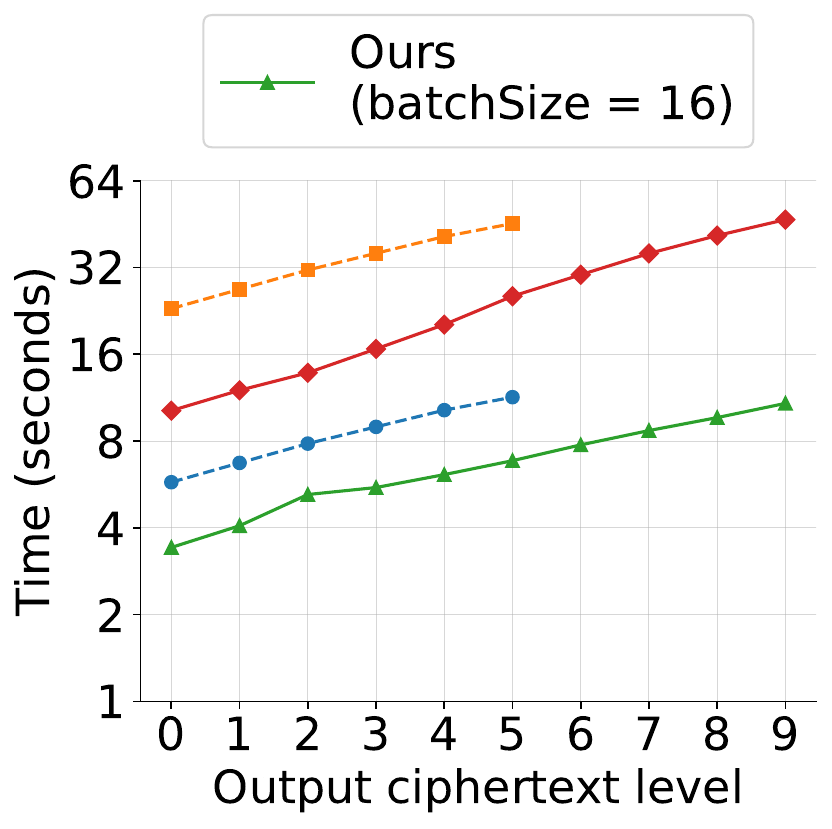}}
        \centerline{\makecell[c]{\footnotesize CCMM-kernel \\ \tiny $Q_h, K_h \in \R^{128 \times 128}$, $h \in [12]$}}
    \end{minipage}%
    \begin{minipage}[t]{0.25\linewidth}
        \centerline{\includegraphics[width=0.99\textwidth]{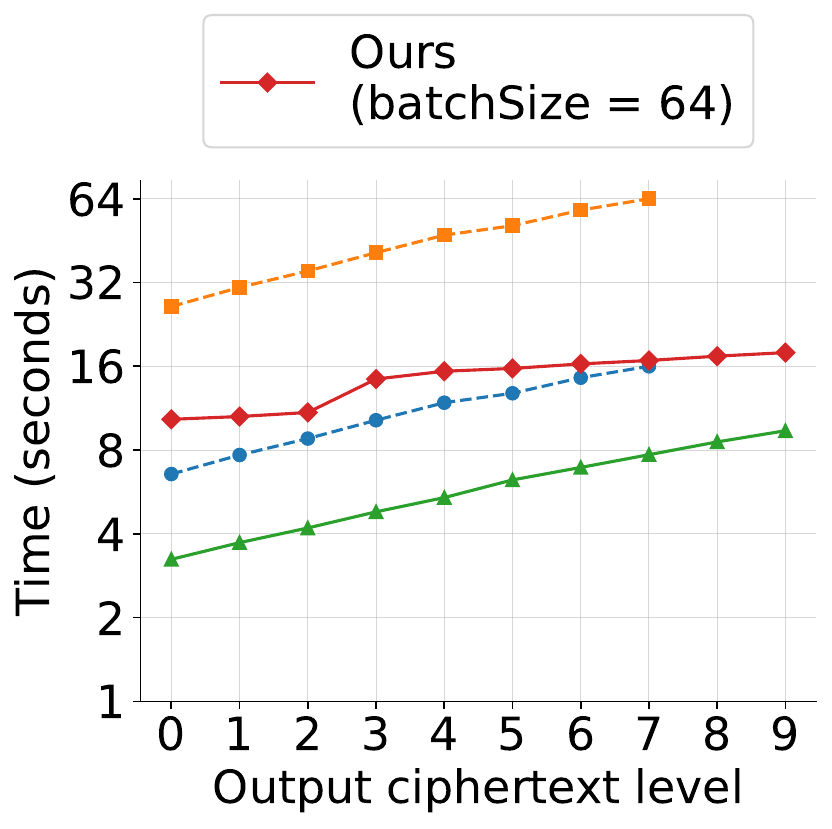}}
        \centerline{\makecell[c]{\footnotesize CCMM-V\\ \tiny $V_h \in \R^{128 \times 64}$, $h \in [12]$}}
    \end{minipage}%

    \caption{Runtime results of homomorphic matrix operations across different output ciphertext levels.}
    \label{fig:mmexp}
\vspace{-1.5em}
\end{figure}

\textbf{Matrix Operations.}
Figure~\ref{fig:mmexp} demonstrates the runtime performance of CPMM, CCMM for LoRA, CCMM for kernel, and CCMM for value matrix over a batch of inputs. 
Our scheme consistently outperforms Rho et al.~\cite{rhoencryption} across different output ciphertext levels in all four types of matrix operations. 
In particular, our scheme achieves $32\times$ to $63\times$ speedup in CPMM, $6.6\times$ to $14.5\times$ speedup in CCMM for LoRA, $1.5\times$ to $2.2\times$ speedup in CCMM for kernel, and $2\times$ to $3.8\times$ speedup in CCMM for value matrix. 
    
\textbf{Functional Bootstrapping.}
In Table~\ref{tab:funcBTS}, we compare the time cost of the functional bootstrapping~\cite{yang2025rboot} and the traditional ``evaluation + bootstrapping'' flow used in \cite{rhoencryption} in 
evaluating the activation function in the feedforward block from \cite{rhoencryption}, with input/output matrix in $\R^{128\times 1536}$. 
To demonstrate the improvement from functional bootstrapping, both experiments use the same row packing method as in \cite{rhoencryption}. 
As a result, our algorithm achieves $4.3\times$ speedup over~\cite{rhoencryption}. 
\begin{table}[tbp]
    \caption{The comparison between CKKS functional bootstrapping and the traditional ``evaluation + bootstrapping'' flow in \cite{rhoencryption}. The comparison is set to the activation in feedforward block. The input level and the output level are set to the same in the two settings. }
    \label{tab:funcBTS}
    \centering
    
    \begin{threeparttable}
        
    \resizebox{\textwidth}{!}{
    \begin{tabular}{c|c|c|c|c|c|c|c|c}
    \toprule
    & $\#$(input ct)& Input lv. & $\#$(BTS) & $\#$(FuncBTS) & $\#$(Output ct)& Output lv.& (Worst prec., Avg prec.)  & Time(s)\\
    \midrule
      \textsc{Felid} (ours)   &6  &4 &0  &6 &6 &12 & $(2^{-13.54},2^{-13.97})$\cite{yang2025rboot} & 4.02 $(4.33\times \text{faster})$\\
      Rho et al.~\cite{rhoencryption}\tnote{*}  &6  &4 &18 &0 &6 &12 & $(2^{-10},2^{-16.4})$\cite{rhoencryption} & 17.4\\
    \bottomrule
    \end{tabular}
    }
        \end{threeparttable}
        \begin{tablenotes}
            \scriptsize
            \item *For fairness, we reproduce their result under the same settings as ours using PhantomFHE. 
        \end{tablenotes}
\vspace{-1.2em}
\end{table}

\textbf{Runtime and Bootstrapping Counts.}
We count the number of bootstrapping of our scheme and~\cite{rhoencryption} in Table~\ref{tab:boot-time}. 
Our method reduces the number of (functional) bootstrapping operations by $2.5\times$, leading to a $2.69\times$ reduction in total bootstrapping runtime compared to Rho et al. \cite{rhoencryption}.

\begin{table}[htbp]
    \caption{Runtime (seconds) and total number of bootstrapping and functional bootstrapping operations in our scheme and in~\cite{rhoencryption} in LoRA fine-tuning, over a batch of input with batch size = 16. }
    \label{tab:boot-time}
    \centering
    \begin{threeparttable}
    \footnotesize
    \resizebox{\textwidth}{!}{
    \begin{tabular}{c|cc|cc|cc}
    \toprule
        \multirow{2}{*}{Scheme} & \multicolumn{2}{c|}{Forward pass} & \multicolumn{2}{c|}{Backward pass + Optimizer} & \multicolumn{2}{c}{Total}\\ \cmidrule(lr){2-3} \cmidrule(lr){4-5}  \cmidrule(lr){6-7} 
         & Number & Time (s) & Number & Time (s) & Number & Time (s)\\ \midrule
         \textsc{Felid} (ours) & 683 $(2.18\times \text{fewer})$  & 295.05  &  236 $(3.44\times \text{fewer})$& 111.95 & 919 $(2.5\times \text{fewer})$ & 407 $(2.69\times \text{faster})$ \\ \hline 
         Rho et al. \cite{rhoencryption} \tnote{*} & 1488 & 708 & 812 & 386 & 2300 & 1094 \\
         \bottomrule
    \end{tabular}
    }
    \end{threeparttable}
    \begin{tablenotes}
            \scriptsize
            \item *For fairness, we reproduce their result under the same settings as ours using PhantomFHE. 
        \end{tablenotes}
        \vspace{-1em}
\end{table}

\subsection{Fine-Tuning for 2-Layer BERT-Style Model}
\label{e2e_exp}


\paragraph{Fine-Tuning Time. }
In Table~\ref{tab:FT-time}, we report the execution time of each component, with a detailed breakdown of transformer operations provided in Appendix~\ref{app:FT-time-exp}. 
Benefiting from (1) optimized packing and more efficient matrix operations (Figure~\ref{fig:mmexp}); (2) eliminating separate point-wise activation evaluations via functional bootstrapping (Table~\ref{tab:funcBTS}); and (3) reducing the number of bootstrapping invocations (Table~\ref{tab:boot-time}), our method achieves consistent improvements across all components, reducing the fine-tuning time per batch from 5238 seconds to 650.6 seconds ($8\times$ speedup).

\begin{table}[htbp]
    \caption{Fine-tuning time (seconds) over a batch of input with batch size = 16.}
    \label{tab:FT-time}
    \centering
    \begin{threeparttable}
        
    \resizebox{\textwidth}{!}{
    \begin{tabular}{c|ccc|ccc|c|c|c}
    \toprule
    \multirow{2}{*}{Scheme} & \multicolumn{3}{c|}{Forward pass} & \multicolumn{3}{c|}{Backward pass} & \multirow{2}{*}{Optimizer} & \multirow{2}{*}{Total} & \multirow{2}{*}{Amortized}
    \\ \cmidrule(lr){2-4} \cmidrule(lr){5-7}
     & Matrix &  Non-linear & (Func)Boot & Matrix & Non-linear & (Func)Boot & & \\ \midrule 

    \textsc{Felid} (ours) & 67.79 & 9.56 & 295.05 & 164.04 & 1.91 & 92.45 & 19.9 & 650.6 & 40.66 $(8\times \text{faster})$  \\ \hline 

    Rho et al.~\cite{rhoencryption}\tnote{*} &  1429 & 291 & 708 & 2377 & 47 & 355 & 31 & 5238 & 327.38 \\ \bottomrule  
    
    \end{tabular}
    }
        \end{threeparttable}
        \begin{tablenotes}
            \scriptsize
            \item *For fairness, we reproduce their result under the same settings as ours using PhantomFHE.
        \end{tablenotes}
\end{table}

\textbf{Downstream Task Performance.}
We evaluate both full and LoRA fine-tuning in plaintext and LoRA fine-tuning in ciphertext. 
Under consistent plaintext settings, kernel-based attention preserves competitive accuracy compared to softmax, demonstrating that both Gaussian and Epanechnikov kernels are viable alternatives to softmax attention.
The ciphertext results in Table~\ref{tab:plaintext} indicate that the additional approximation error introduced by the FHE evaluation has a minimal impact on fine-tuning.

\begin{table}[htbp]
    \centering
    \caption{Plaintext and ciphertext GLUE results. Results marked with~\cite{rhoencryption} are taken from the published Gaussian-kernel transformer; ``ours'' denotes models trained using our continued-pretraining and distillation pipeline. To ensure a fair comparison, we fine-tune the models for $5$ epochs as per \cite{rhoencryption}. }
    \label{tab:plaintext}
    \resizebox{\textwidth}{!}{
    \begin{tabular}{l|cccc|ccc|ccc}
        \toprule
        & \multicolumn{4}{c|}{Plaintext Full fine-tuning} & \multicolumn{3}{c|}{Plaintext LoRA fine-tuning} & \multicolumn{3}{c}{Ciphertext LoRA fine-tuning} \\
        \cmidrule(lr){2-5} \cmidrule(lr){6-8} \cmidrule(lr){9-11}
        Task (Metric)
        & SM~\cite{rhoencryption} 
        & GK~\cite{rhoencryption} 
        & GK (ours) 
        & EK (ours) 
        & GK~\cite{rhoencryption} 
        & GK (ours)
        & EK (ours) 
        & GK~\cite{rhoencryption} 
        & GK (ours)
        & EK (ours) \\
        \midrule
        MRPC \footnotesize{(F1 $\uparrow$)}  & 0.8328 & 0.8431 & 0.8349 & 0.8381 
        & 0.8258 & 0.8298 & 0.8315 
        & 0.8147 & 0.8255 & 0.8323 \\
        RTE \footnotesize{(Accuracy $\uparrow$)}   & 0.5884 & 0.6101 & 0.5957 & 0.6173 
        & 0.5776 & 0.6029 & 0.6065 
        & 0.5993 & 0.5957 & 0.5921\\ 
        STS-B \footnotesize{(Pearson $\uparrow$)} & 0.8415 & 0.8215 & 0.8311 & 0.8230 
        & 0.8107 & 0.8184 & 0.8274 
        & 0.7997 & 0.8155 & 0.7824\\ 
        \midrule
        Average & 0.7542& 	0.7582& 	0.7539	& 0.7595& 	0.7380& 	0.7504	& 0.7551	& 0.7379	& 0.7456& 	0.7356\\
        \bottomrule
    \end{tabular}
    }
\end{table}

\section{Conclusions}
In this work, we propose \textsc{Felid}, a flexible and efficient fully homomorphic encryption framework for non-interactive transformer fine-tuning. 
\textsc{Felid} introduces (i) a novel packing strategy that enables flexible batching, (ii) operation-specific matrix evaluation algorithms tailored for kernel-based transformers, and (iii) the use of functional bootstrapping to eliminate separate evaluation of element-wise activations.
As a result, \textsc{Felid} reduces ciphertext rotations and relinearizations by $30\times$ and FHE bootstrapping invocations by $2.5\times$ across the entire computation flow compared to the state-of-the-art~\cite{rhoencryption}. Fine-tuning of a $2$-layer BERT-style model with a batch size of $16$ is completed in $650$ seconds, achieving an $8\times$ speedup over prior work.

\section*{Acknowledgments}
This work was conducted through the joint research laboratory of Ant International and Nanyang Technological University, with funding support from Ant International. The laboratory was established under a five-year partnership to advance digital trust and privacy-enhancing technologies (PETs). This research is supported by the National Research Foundation, Singapore and Infocomm Media Development Authority under its Trust Tech Funding Initiative. Any opinions, findings and conclusions or recommendations expressed in this material are those of the author(s) and do not reflect the views of National Research Foundation, Singapore and Infocomm Media Development Authority.

\bibliography{ref}
\bibliographystyle{plain}


\appendix
\section{Extended preliminaries}

\subsection{More detailed CKKS scheme}
The CKKS scheme \citep{cheon2017homomorphic} also includes the following building blocks:
\begin{itemize}
    \item $\textrm{Keygen}$: it generates the secret key $\sk$. 
    \item $\enc_{\sk}(\vec{z})$: for $\vec{z}\in\C^{N}$, it generates the ciphertext of complex vector $\vec{z}$ using $\sk$ and it can only be decrypted with $\sk$. 
    When $\sk$ is clear from the context, $\enc(\vec{z})$ is used. 
    \item $\dec_{\sk}(\enc_\sk(\vec{z}))$: it decrypts the ciphertext encrypting $\vec{z}$ with the secret key $\sk$. 
\end{itemize}

\paragraph{Encoding and decoding.}
The CKKS scheme does not directly encrypt vector in $\C^N$.  
It first encodes a complex vector into a polynomial with integer coefficients through a sequence of \textit{embedding} and \textit{scaling} operations. 
Let $\vec{z} \in \mathbb{C}^{N}$ denote the input vector. 
CKKS first applies an \textit{inverse canonical embedding} that maps $\mathbf{z}$ into a polynomial $z(X) \in \mathbb{C}[X]/(X^{2N} + 1)$, where $N$ is a power of two. 
The complex coefficients of $m(X)$ are scaled by a factor $\Delta > 0$ and rounded to obtain an integer-coefficient polynomial:
\[
\tilde{z}(X) = \lfloor \Delta \cdot m(X) \rceil \in \mathbb{Z}[X]/(X^{2N} + 1).
\]
This process is referred to as $\tilde{z}(X):=\ecd(\bm{z},\Delta)$, and it allows representation of complex numbers (of $\vec{z}$) with finite precision. 

Encryption is then performed over the ring $\mathbb{Z}_Q[X]/(X^{2N} + 1)$ for a large modulus $Q$. 
Given the secret key $\sk=s(X)$, a ciphertext is generated as a pair of polynomials:
\[
\enc_\sk(\tilde{z}(X)):=(c_0(X), c_1(X)) := \left( \tilde{z}(X) + e(X) - a(X) s(X),\; a(X) \right)
\in \big(\mathbb{Z}_Q[X]/(X^{2N} + 1)\big)^2,
\]
where $a(X)$ is sampled uniformly at random from $\mathbb{Z}_Q[X]/(X^{2N} + 1)$, and $e(X)$ are small noise polynomials sampled from a pre-defined discrete Gaussian distribution.
Therefore we can write $\enc_\sk(\ecd(\vec{z},\Delta))$ in to $\enc_\sk(\vec{z})$ for short as we do not modify and do not focus on encoding/decoding. 

Decryption recovers an approximation of $\tilde{m}(X)$ via:
\[
\tilde{m}(X) \approx c_0(X) + c_1(X) s(X) \mod Q,
\]
followed by \textit{decoding} (i.e., the inverse of $\ecd(\cdot, \Delta)$, scaling down by $\Delta$ and applying the canonical embedding) to obtain an approximate reconstruction of the original vector $\mathbf{z}$.

\subsection{Softmax attention transformer architecture}
\label{app:softmax}

Currently many large language models are based on transformer architecture. 
Such a model contains multiple layers and each layer has $H$ heads. 
Let $n_1$ be the token length of input, $n_2$ be the hidden dimension, $n_3=n_2/H$ be the dimension of each head. 
For simplicity bias vectors are omitted. 
In each layer, given input matrix $X$, the model first compute $Q_h=XW_{Q_h}, K_h=XW_{K_h}, V_h=XW_{V_h}$ for each head $h\in[H]$. 
For models with transformer architecture, like BERT\citep{devlin2019bert}, the layer will output
\[
X_h:=\attention(Q_h,K_h,V_h):=\softmax(\frac{Q_hK^\intercal_h}{\sqrt{n_3}})V_h,\ \forall h\in[H]~.
\]
\section{Matrix multiplication algorithms}
\label{app:matrix}





\begin{definition}[Flexible interleaved column packing]
\label{def:inter-col}
Suppose we have $b$ matrices $\{ X^i \in \R^{n_1 \times n_2}\}_{ i \in [b]}$ be:
\begin{equation*}
    X^i = 
    \left(
        \begin{array}{ccc}
           x^i[0][0]  & \cdots & X^i [0][n_2-1] \\
            \vdots & \ddots & \vdots \\
            x^i [n_1-1] [0] & \cdots & X^i [n_1-1] [n_2-1]
        \end{array}
    \right). 
\end{equation*}

Let $\{ \vec{m}^i \in \R^N \}_{i \in [r]}$, $r = \lceil n_2/ \lfloor N/(bn_1) \rfloor \rceil$ denote the plaintext vectors obtained by \textit{interleaved column packing} the matrices $X^0 ,..., X^{b-1}$. 
For simplicity, we assume $(bn_1) | N$ and $N | (bn_1 n_2)$. 
We define: 
\begin{equation*}
    m^i[j] = X^{j \% b} [(\lfloor \frac{j}{b} \rfloor )\%n_1] [i\frac{N}{bn_1}+\lfloor \frac{j}{bn_1} \rfloor ]
\end{equation*}
where $i \in [\frac{bn_1n_2}{N}]$, and $j \in [N]$.
We say that $\{\enc(\vec{m}^i)\}_{i \in [\frac{bn_1n_2}{N}]}$ is in $b$-\textbf{interleaved column packing}. 
\end{definition}


\subsection{Ciphertext-plaintext matrix multiplication (CPMM)}

Our CPMM takes an encrypted matrix in $b$-interleaved column packing with and a plaintext matrix, and output an encrypted matrix in $b$-interleaved column packing. 
For simplicity, we assume $(bn_1) | N$, $N | (bn_1 n_2)$, and $N | (bn_1 n_2')$.
The details of algorithm is in Algorithm~\ref{alg:CPMM}. 

\begin{algorithm}[htbp]
    \caption{Ciphertext-plaintext matrix multiplication. }
    \label{alg:CPMM}
    \begin{algorithmic}[1]  
        \REQUIRE Encrypted matrices $\{ X^i \in \R^{n_1 \times n_2}\}_{ i \in [b]}$ in $b$-interleaved column packing $\{\enc(\vec{m}^i) \}_{i \in [r_1]}$, $r_1 = \frac{bn_1n_2}{N}$, matrix $W \in \R^{n_2 \times n_2'}$ 
        \ENSURE Encrypted matrices $\{ X^iW \in \R^{n_1 \times n_2'}\}_{ i \in [b]}$ in $b$-interleaved column packing $\{ \enc(\vec{\hat{m}}^i) \}_{i \in [r_2]}$, $r_2= \frac{bn_1n_2'}{N}$. 

        \STATE Let $t_1 = \lceil \sqrt{\frac{N}{bn_1}} \rceil$, and $t_2 = \lceil \frac{N}{bn_1t_1} \rceil$. 
        
        \FOR{i from 0 to $r_1-1$}
            \FOR{k from 0 to $t_1-1$}
                \STATE Evaluate $ct_{it_1+k} = \ckksrot_{kbn_1}(\enc(\vec{m}^i))$. 
            \ENDFOR
        \ENDFOR

        \FOR{i from 0 to $r_2-1$}
            \FOR{$k_1$ from 0 to $t_2-1$}
                \STATE Let $\hat{ct} = \enc(\vec{0})$. 
                \COMMENT{Initialization of ciphertext}
                \FOR{j from 0 to $r_1-1$}
                    \FOR{$k_2$ from 0 to $t_1-1$}
                        \STATE Let $\vec{d} = \vec{0}$. 
                        \FOR{$l_1$ from 0 to $\frac{N}{bn_1}-1$}
                            \FOR{$l_2$ from 0 to $n_1-1$}
                                \FOR{$l_3$ from 0 to $b-1$}
                                    \STATE Let $\mathrm{Ind}_{row} = \frac{jN}{bn_1} +(k_1t_1+k_2+l_1)\% \frac{N}{bn_1}$, and $\mathrm{Ind}_{col} = \frac{iN}{bn_1}+l_1\% \frac{N}{bn_1}$. 
                                    \STATE Let $d[l_1bn_1+l_2b+l_3] = W[\mathrm{Ind}_{row}] [\mathrm{Ind}_{col}]$. 
                                    \COMMENT{Pack weights into vector $\vec{d}$}
                                \ENDFOR
                            \ENDFOR
                        \ENDFOR

                        \STATE Evaluate $\hat{ct}=\hat{ct}+\plainmult(\rot_{-k_1t_1bn_1}(\vec{d}),ct_{jt_1+k_2k}). $
                        \COMMENT{ct-pt multiplication}
                    \ENDFOR
                \ENDFOR
                \IF{$k_1 = 0$}
                    \STATE Evaluate $\enc(\vec{\hat{m}}^i) = \ckksrot_{k_1t_1bn_1}(\hat{ct})$. 
                \ELSE
                    \STATE Evaluate $\enc(\vec{\hat{m}}^i) = \enc(\vec{\hat{m}}^i) + \ckksrot_{k_1t_1bn_1}(\hat{ct})$. 
                \ENDIF
            \ENDFOR
        \ENDFOR
 
\end{algorithmic}
\end{algorithm}

\paragraph{Optimization with imaginary parts. }
Making use of imaginary parts of each CKKS slot will further save HE operations, especially in the ct-pt matrix multiplications. 
We modify Algorithm~\ref{alg:CPMM} so that each $X^iW$ in the encrypted output is compactly packed in the complex number field $\C$. 
$\forall W\in \R^{n_2\times n_2'}$.
Let $W_{even}\in\R^{n_2\times (n_2'/2)}$ include all the even-index columns of $W$ (starting from index $0$),
and let $W_{odd}\in\R^{n_2\times (n_2'/2)}$ include all the odd-index columns of $W$.
Define $W_{\C}:=W_{even}+\sqrt{-1}W_{odd}\in\C^{n_2\times (n_2'/2)}$. 
The details of algorithm is in Algorithm~\ref{alg:CPMM-C}.

\begin{algorithm}[htbp]
    \caption{Optimized Ciphertext-plaintext matrix multiplication. }
    \label{alg:CPMM-C}
    \begin{algorithmic}[1]  
        \REQUIRE Encrypted matrices $\{ X^i \in \R^{n_1 \times n_2}\}_{ i \in [b]}$ in $b$-interleaved column packing $\{\enc(\vec{m}^i\in\R^N) \}_{i \in [r_1]}$, $r_1 = \frac{bn_1n_2}{N}$, matrix $W \in \R^{n_2 \times n_2'}$ 
        \ENSURE Encrypted matrices $\{ X^{i}W_{\C} \in \C^{n_1 \times (n_2'/2)}\}_{ i \in [b]}$ in $b$-interleaved column packing $\{ \enc(\vec{\hat{m}}^i\in\C^N) \}_{i \in [r_2]}$, $r_2= \frac{bn_1n_2'}{2N}$. 

        \STATE Let $t_1 = \lceil \sqrt{\frac{N}{bn_1}} \rceil$, and $t_2 = \lceil \frac{N}{bn_1t_1} \rceil$. 

        \FOR{i from 0 to $r_1-1$}
            \FOR{k from 0 to $t_1-1$}
                \STATE Evaluate $ct_{it_1+k} = \ckksrot_{kbn_1}(\enc(\vec{m}^i))$. 
            \ENDFOR
        \ENDFOR
        \STATE Let $W_{even}\in\R^{n_2\times (n_2'/2)}$ include all the even-index columns of $W$. (starting from index $0$). 
        \STATE Let $W_{odd}\in\R^{n_2\times (n_2'/2)}$ include all the odd-index columns of $W$.
        \STATE Define $W_{\C}:=W_{even}+\sqrt{-1}W_{odd}\in \C^{n_2\times (n_2'/2)}$. 
        \COMMENT{Packing $W$ in $\C$}. 
        \FOR{i from 0 to $r_2-1$}
            \FOR{$k_1$ from 0 to $t_2-1$}
                \STATE Let $\hat{ct} = \enc(\vec{0})$. 
                \FOR{j from 0 to $r_1-1$}
                    \FOR{$k_2$ from 0 to $t_1-1$}
                        \STATE Let $\vec{d} = \vec{0}$. 
                        \FOR{$l_1$ from 0 to $\frac{N}{bn_1}-1$}
                            \FOR{$l_2$ from 0 to $n_1-1$}
                                \FOR{$l_3$ from 0 to $b-1$}
                                    \STATE Let $\mathrm{Ind}_{row} = \frac{jN}{bn_1} +(k_1t_1+k_2+l_1)\% \frac{N}{bn_1}$, and $\mathrm{Ind}_{col} = \frac{iN}{bn_1}+l_1\% \frac{N}{bn_1}$. 
                                    \STATE Let $d[l_1bn_1+l_2b+l_3] = W_{\C}[\mathrm{Ind}_{row}] [\mathrm{Ind}_{col}]$. 
                                \ENDFOR
                            \ENDFOR
                        \ENDFOR

                        \STATE Evaluate $\hat{ct}=\hat{ct}+\plainmult(\rot_{-k_1t_1bn_1}(\vec{d}),ct_{jt_1+k_2k}). $
                        \COMMENT{ct-pt multiplication}
                    \ENDFOR
                \ENDFOR
                \IF{$k_1 = 0$}
                    \STATE Evaluate $\enc(\vec{\hat{m}}^i) = \ckksrot_{k_1t_1bn_1}(\hat{ct})$. 
                \ELSE
                    \STATE Evaluate $\enc(\vec{\hat{m}}^i) = \enc(\vec{\hat{m}}^i) + \ckksrot_{k_1t_1bn_1}(\hat{ct})$. 
                \ENDIF
            \ENDFOR
        \ENDFOR
 
\end{algorithmic}
\end{algorithm}

\paragraph{Rotations in a batch. }
Number of HE rotations in Algorithm~\ref{alg:CPMM} is 
\[
    \#(\ckksrot) = r_1t_1 + r_2t_2 \approx \frac{bn_1n_2}{N} \cdot \frac{\sqrt{N}}{\sqrt{bn_1}} + \frac{bn_1n_2'}{N} \cdot \frac{\sqrt{N}}{\sqrt{bn_1}} = (n_2+n_2')\sqrt{\frac{bn_1}{N}}. 
\]

By using imaginary parts, we reduce the number of output ciphertexts by half, so the number of HE rotations in Algorithm~\ref{alg:CPMM-C} is 
\[
\#(\ckksrot) = r_1t_1 + \frac{r_2}{2}t_2 \approx (n_2+\frac{n_2'}{2})\sqrt{\frac{bn_1}{N}}.
\]

If we set $N = 32768$, $b = 16$, $X^i \in \R^{128 \times 768}$, and $W \in \R^{768 \times 768}$, then numbers of HE rotations in Algorithm~\ref{alg:CPMM} and Algorithm~\ref{alg:CPMM-C} are $384$ and $288$ respectively. 

\paragraph{Multiplicative level and ct-pt multiplications in a batch. }
CPMM consumes 1 multiplicative level with $r_2t_2r_1t_1$ ct-pt multiplications in Algorithm~\ref{alg:CPMM} and $\frac{r_2t_2r_1t_1}{2}$ ct-pt multiplications in Algorithm~\ref{alg:CPMM-C}.

If we set $N = 32768$, $b = 16$, $X^i \in \R^{128 \times 768}$, and $W \in \R^{768 \times 768}$, then numbers of HE ct-pt multiplications in Algorithm~\ref{alg:CPMM} and Algorithm~\ref{alg:CPMM-C} are $36864$ and $18432$ respectively. 

In summary, the required number of each HE operation for $\{\enc(X^i)W\}_{i \in [b]}$, where $X^i \in \R^{n_1 \times n_2}$, and $W \in \R^{n_2 \times n_2'}$ , is listed in Table~\ref{tab:CPMM-ours}. 

\paragraph{Remark.}
In the attention block, using LoRA fine-tuning we can set $W_*=W_{*,frozen}+AB$, where $*=Q,K,V$. 
$W_{*,frozen}$ are plaintext, so we can set 
$W_{attn,frozen}=(W_{Q,frozen}\Vert W_{K,frozen}\Vert W_{V,frozen})$. 
By Table~\ref{tab:CPMM-ours}, 
applying Algorithm~\ref{alg:CPMM-C} with $W_{attn,frozen}$ and $n_2'=3n_2$ is much better than applying Algorithm~\ref{alg:CPMM-C} 3 times, with $W_{Q,frozen}, W_{K,frozen}, W_{V,frozen}$ respectively. 

\begin{table}[htbp]
    \centering
    \caption{The number of each HE operation for $\{\enc(X^i)W\}_{i \in [b]}$ in our scheme}
    \label{tab:CPMM-ours}
    \begin{tabular}{cccccc}
    \toprule
         Operations & Rot & ct-ct Mult & (Relinearization) & ct-pt Mult & Level \\ \midrule
         CPMM-basic 
         & $(n_2+n_2')\sqrt{\frac{bn_1}{N}}$ 
         & 0 & 0
         &  $\frac{bn_1n_2n_2'}{N}$  
         & 1 
         \\ \midrule
         CPMM-optimized 
         & $(n_2+\frac{n_2'}{2})\sqrt{\frac{bn_1}{N}}$ 
         & 0 & 0 
         & $\frac{bn_1n_2n_2'}{2N}$  
         & 1 
         \\
    \bottomrule
    \end{tabular}
    
\end{table}

\subsection{Ciphertext-ciphertext matrix multiplication (CCMM) in LoRA}

LoRA fine-tuning updates low rank matrices $A,B$ in $W+AB$, where $W\in \R^{n_2\times n_2'}, A \in \R^{n_2 \times n_{\mathrm{LoRA}}}$, $B \in \R^{n_{\mathrm{LoRA}} \times n_2'}$, and $n_{\mathrm{LoRA}} \ll n_2'$. 
Notice that $A$ and $B$ are updating during fine-tuning, so they are encrypted matrices. 

Therefore, evaluating the linear layer $X(W+AB)=XW+(XA)B$ requires one CPMM (Algorithm~\ref{alg:CPMM-C}) to evaluate $XW$ and 2 CCMMs (Algorithm~\ref{alg:CCMM-A-LoRA} and~\ref{alg:CCMM-B-LoRA-C}) to evaluate $(XA)B$. 
For simplicity, we assume $(bn_1) | N$, $N | (bn_1 n_2)$, and $N | (bn_1 n_2')$. 
As shown in \cite{rhoencryption}, special packing and matrix multiplications for low rank matrices $A,B$ enables much better latency. 
The packing of $A$ and $B$ are elaborated in Algorithm~\ref{alg:pack-A-LoRA} and~\ref{alg:CCMM-B-LoRA}, respectively.

\begin{algorithm}[htbp]
    \caption{Repeated column packing for matrix A in LoRA. }
    \label{alg:pack-A-LoRA}
    \begin{algorithmic}[1]
        \REQUIRE Matrix $A \in \R^{n_2 \times n_{\mathrm{LoRA}}}$, $r_1 = \frac{bn_1n_2}{N}$.
        \ENSURE $\{\enc(\vec{a}^i) \}_{i \in [r_A]}$ where $r_A=r_1 n_{\mathrm{LoRA}}$. 

        \FOR{i from 0 to $n_{\mathrm{LoRA}}-1$}
            \FOR{j from 0 to $r_1-1$}
                \FOR{$k_1$ from 0 to $\frac{n_2}{r_1}-1$}
                    \FOR{$k_2$ from 0 to $bn_1-1$}
                        \STATE Let $\vec{a}^{ir_1+j}[k_1bn_1+k_2] = A[j\frac{n_2}{r_1} + k_1][i]$. 
                    \ENDFOR
                \ENDFOR
                \STATE Evaluate $\enc(\vec{a}^{ir_1+j})$. 
            \ENDFOR
        \ENDFOR
    \end{algorithmic}
\end{algorithm}

\begin{algorithm}[htbp]
    \caption{Repeated row packing for matrix B in LoRA. }
    \label{alg:pack-B-LoRA}
    \begin{algorithmic}[1]
        \REQUIRE Matrix $B \in \R^{n_{\mathrm{LoRA}} \times n_2'}$, $r_1 = \frac{bn_1n_2}{N}$. 
        \ENSURE $\{\enc(\vec{b}^i) \}_{i \in [r_B]}$ where $r_B=r_1 n_{\mathrm{LoRA}}$.

        \FOR{i from 0 to $n_{\mathrm{LoRA}}-1$}
            \FOR{j from 0 to $r_1-1$}
                \FOR{$k_1$ from 0 to $\frac{n_2'}{r_1}-1$}
                    \FOR{$k_2$ from 0 to $bn_1-1$}
                        \STATE Let $\vec{b}^{ir_1+j}[k_1bn_1+k_2] = B[i][j\frac{n_2'}{r_1}+k_1]$. 
                    \ENDFOR
                \ENDFOR
                \STATE Evaluate $\enc(\vec{b}^{ir_1+j})$.                 
            \ENDFOR
        \ENDFOR
    \end{algorithmic}
\end{algorithm}

\begin{algorithm}[htbp]
    \caption{Ciphertext-Ciphertext matrix multiplication for matrix A in LoRA.}
    \label{alg:CCMM-A-LoRA}
    \begin{algorithmic}[1]  
        \REQUIRE Encrypted matrices $\{ X^i \in \R^{n_1 \times n_2}\}_{ i \in [b]}$ in $b$-interleaved column packing $\{\enc(\vec{m}^i) \}_{i \in [r_1]}$, $r_1 = \frac{bn_1n_2}{N}$. Encrypted matrix $A \in \R^{n_2 \times n_{\mathrm{LoRA}}}$ in repeated column packing $\{\enc(\vec{a}^i) \}_{i \in [r_A]}$, $r_A=r_1 n_{\mathrm{LoRA}}$. 
        \ENSURE Encrypted matrices $\{ X^iA \in \R^{n_1 \times n_{\mathrm{LoRA}}}\}_{ i \in [b]}$: $\{ \enc(\vec{\hat{m}}^i) \}_{i \in [r_{\mathrm{XA}}]}$, $r_{\mathrm{XA}}= n_{\mathrm{LoRA}}$. 

        \FOR{i from 0 to $n_{\mathrm{LoRA}}-1$}
            \FOR{j from 0 to $r_1-1$}
                \STATE Evaluate $ct = \mult(\enc(\vec{m}^j), \enc(\vec{a}^{ir_1+j}))$.
                \STATE Let $\mathrm{Ind}_{rot} = N/2$.
                \IF{j = 0}
                    \STATE Let $\enc(\vec{\hat{m}}^i) = ct$.
                \ELSE
                    \STATE Evaluate $\enc(\vec{\hat{m}}^i) = \enc(\vec{\hat{m}}^i) + ct$.
                \ENDIF
            \ENDFOR
            \WHILE{$\mathrm{Ind}_{rot} \geq (bn_1)$}
                    \STATE Evaluate $\enc(\vec{\hat{m}}^i) = \enc(\vec{\hat{m}}^i) + \ckksrot_{\mathrm{Ind}_{rot}}(\enc(\vec{\hat{m}}^i))$.
                    \STATE Evaluate $\mathrm{Ind}_{rot} = \mathrm{Ind}_{rot}/2$.
                \ENDWHILE
        \ENDFOR

\end{algorithmic}
\end{algorithm}

\begin{algorithm}[htbp]
    \caption{Ciphertext-Ciphertext matrix multiplication for matrix B in LoRA.}
    \label{alg:CCMM-B-LoRA}
    \begin{algorithmic}[1]  
        \REQUIRE Encrypted matrices $\{ (XA)^i \in \R^{n_1 \times n_{\mathrm{LoRA}}}\}_{ i \in [b]}$ output by Algorithm \ref{alg:CCMM-A-LoRA}: $\{\enc(\vec{m}^i) \}_{i \in [r_{\mathrm{XA}}]}$, $r_{\mathrm{XA}} = n_{\mathrm{LoRA}}$. Encrypted matrix $B \in \R^{n_{\mathrm{LoRA}} \times n_2'}$ in repeated row packing: $\{\enc(\vec{b}^i) \}_{i \in [r_B]}$, $r_B=r_A$. 
        \ENSURE Encrypted matrices $\{ X^iAB \in \R^{n_1 \times n_2'}\}_{ i \in [b]}$ in $b$-interleaved column packing $\{ \enc(\vec{\hat{m}}^i) \}_{i \in [r_2]}$, $r_2= \frac{bn_1n_2'}{N}$. 

        \FOR{i from 0 to $r_2-1$}
            \FOR{j from 0 to $r_{\mathrm{XA}}-1$}
                \STATE Evaluate $ct = \mult(\enc(\vec{m}^j), \enc(\vec{b}^{jr_2+i}))$.
                \IF{j = 0}
                    \STATE Let $\enc(\vec{\hat{m}}^i) = ct$.
                \ELSE
                    \STATE Evaluate $\enc(\vec{\hat{m}}^i) = \enc(\vec{\hat{m}}^i) + ct$.
                \ENDIF
            \ENDFOR
        \ENDFOR

\end{algorithmic}
\end{algorithm}

\paragraph{Optimizations with imaginary part. }
According to Algorithm~\ref{alg:CPMM-C}, the encrypted $XW$ is packed in complex number field $\C$ instead of $\R$.   
Thus we hope encrypted $XAB$ has the same packing format as $XW$ so that we can add them directly. In addition, packing $XAB$ in $\C$ also saves many FHE operations. 
We apply this optimization by modifing the repeated row packing of $B$, as shown in Algorithm~\ref{alg:pack-B-LoRA-C}. 
The optimized matrix multiplication algorithm with $B$ is in Algorithm~\ref{alg:CCMM-B-LoRA-C}. 

Similarly we can make use of the imaginary part of the CKKS slots in the packing and multiplications of the encrypted matrix $B\in \R^{n_{LoRA}\times n_2'}$. 
By using imaginary part, we can reduce the number of output ciphertexts in both packing and multiplication by half. 
We present the packing and multiplication algorithms for $B$ in Algorithm~\ref{alg:pack-B-LoRA-C} and Algorithm~\ref{alg:CCMM-B-LoRA-C}, respectively. 

\begin{algorithm}[htbp]
    \caption{Repeated row packing in $\C$ for matrix B in LoRA. }
    \label{alg:pack-B-LoRA-C}
    \begin{algorithmic}[1]
        \REQUIRE Matrix $B \in \R^{n_{\mathrm{LoRA}} \times n_2'}$
        \ENSURE $\{\enc(\vec{b}^i\in \C^N) \}_{i \in [r_B]}$, $r_B=\frac{r_1 n_{\mathrm{LoRA}}}{2}$, $r_1 = \frac{bn_1n_2}{N}$.
        
        \FOR{i from 0 to $n_{\mathrm{LoRA}}-1$}
            \FOR{j from 0 to $\frac{r_1}{2}-1$}
                \FOR{$k_1$ from 0 to $\frac{n_2'}{r_1}-1$}
                    \FOR{$k_2$ from 0 to $bn_1-1$}
                        \STATE Let $\vec{b}^{\frac{ir_1}{2}+j}[k_1bn_1+k_2] = B[i][2j\frac{n_2'}{r_1}+k_1] + \sqrt{-1}B[i][(2j+1)\frac{n_2'}{r_1}+k_1]$. 
                    \ENDFOR
                \ENDFOR
                \STATE Evaluate $\enc(\vec{b}^{\frac{ir_1}{2}+j})$. 
            \ENDFOR
        \ENDFOR
    \end{algorithmic}
\end{algorithm}

\begin{algorithm}[htbp]
    \caption{Optimized Ciphertext-Ciphertext matrix multiplication for matrix B in LoRA.}
    \label{alg:CCMM-B-LoRA-C}
    \begin{algorithmic}[1]  
        \REQUIRE Encrypted matrices $\{ (XA)^i \in \R^{n_1 \times n_{\mathrm{LoRA}}}\}_{ i \in [b]}$ output by Algorithm \ref{alg:CCMM-A-LoRA}: $\{\enc(\vec{m}^i\in\R^N) \}_{i \in [r_{\mathrm{XA}}]}$, $r_{\mathrm{XA}} = n_{\mathrm{LoRA}}$. Encrypted matrix $B \in \R^{n_{\mathrm{LoRA}} \times n_2'}$ in optimized repeated row packing $\{\enc(\vec{b}^i\in\C^N) \}_{i \in [r_B]}$, $r_B=\frac{r_A}{2}$. 
        \ENSURE Encrypted matrices $\{ X^iAB \in \C^{n_1 \times (n_2'/2)}\}_{ i \in [b]}$ in $b$-interleaved column packing: $\{ \enc(\vec{\hat{m}}^i\in\C^N) \}_{i \in [r_2]}$, $r_2= \frac{bn_1n_2'}{2N}$. 

        \FOR{i from 0 to $r_2-1$}
            \FOR{j from 0 to $r_{\mathrm{XA}}-1$}
                \STATE Evaluate $ct = \mult(\enc(\vec{m}^j) , \enc(\vec{b}^{jr_2+i}))$.
                \IF{j = 0}
                    \STATE Let $\enc(\vec{\hat{m}}^i) = ct$.
                \ELSE
                    \STATE Evaluate $\enc(\vec{\hat{m}}^i) = \enc(\vec{\hat{m}}^i) + ct$.
                \ENDIF
            \ENDFOR
        \ENDFOR

\end{algorithmic}
\end{algorithm}

\paragraph{Rotations in a batch. }
Number of HE rotations in $(XA)B$ is 
\[
    \#(\ckksrot) = n_{\mathrm{LoRA}}  \cdot \log (N/(bn_1)). 
\]

If we set $N = 32768$, $b = 16$, $X^i \in \R^{128 \times 768}$, $A \in \R^{768 \times 2}$, and $B \in \R^{2 \times 768}$ then the number of HE rotations is $8$.

\paragraph{Relinearizations. }
Number of relinearizations in $(XA)B$ by using Algorithm~\ref{alg:CCMM-A-LoRA} and Algorithm~\ref{alg:CCMM-B-LoRA} (with lazy relinearization in Algorithm \ref{alg:CCMM-B-LoRA}) is 
\[
    \#(\ckksrel) = n_{\mathrm{LoRA}} + r_2 = n_{\mathrm{LoRA}}+\frac{bn_1n_2'}{N}.
\]

By using imaginary parts in $B$, we reduce the number of output ciphertexts by half, so the number of relinearizations in $(XA)B$ by using Algorithm~\ref{alg:CCMM-A-LoRA} and Algorithm~\ref{alg:CCMM-B-LoRA-C} (with lazy relinearization in Algorithm \ref{alg:CCMM-B-LoRA-C}) is 
\[
    \#(\ckksrel) = n_{\mathrm{LoRA}} + \frac{bn_1n_2'}{2N}. 
\]

If we set $N = 32768$, $b = 16$, $X^i \in \R^{128 \times 768}$, $A \in \R^{768 \times 2}$, and $B \in \R^{2 \times 768}$ then numbers of HE relinearzations are $50$ and $26$ respectively.

\paragraph{Multiplicative levels and ct-ct multiplications in a batch. }
Each CCMM consumes 1 multiplicative level. 
The two CCMMs in LoRA consume 2 multiplicative levels with $n_{\mathrm{LoRA}} (r_1 + r_2)$ ct-ct multiplications. 
$r_2$ is decided by the packing and multiplication algorithms of $B$. 
If we set $N = 32768$, $b = 16$, $X^i \in \R^{128 \times 768}$, $A \in \R^{768 \times 2}$, and $B \in \R^{2 \times 768}$ then numbers of HE ct-ct multiplications is $192$ and $144$ respectively.

In summary, the required number of each HE operation for obtaining $\{\enc(X^iAB)\}_{i \in [b]}$, where $X^i \in \R^{n_1 \times n_2}$, $A \in \R^{n_2 \times n_\mathrm{LoRA}}$ , and $B \in \R^{n_{\mathrm{LoRA}} \times n_2'}$,  is listed in Table~\ref{tab:LoRA-ours}. 

\begin{table}[htbp]
    \centering
    \caption{The number of each HE operation for $\{\enc(X^i)\enc(A)\enc(B)\}_{i \in [b]}$ in our scheme}
    \label{tab:LoRA-ours}
    \resizebox{\textwidth}{!}{
    \begin{tabular}{cccccc}
    \toprule
         Operations & Rot & ct-ct Mult & Relinearization & ct-pt Mult & Level \\ \midrule
         LoRA-CCMM-basic 
         & $n_{\mathrm{LoRA}}  \cdot \log (N/(bn_1))$ 
         & $\frac{bn_1n_{\mathrm{LoRA}}(n_2+n_2')}{N}$ 
         & $n_{\mathrm{LoRA}}+\frac{bn_1n_2'}{N}$ 
         &  0  
         & 2 
         \\ \midrule
         LoRA-CCMM-optimized 
         & $n_{\mathrm{LoRA}}  \cdot \log (N/(bn_1))$ 
         & $\frac{bn_1n_{\mathrm{LoRA}}(2n_2+n_2')}{2N}$ 
         & $n_{\mathrm{LoRA}} + \frac{bn_1n_2'}{2N}$
         & 0  
         & 2 
         \\
    \bottomrule
    \end{tabular}
    }
    
\end{table}

\subsection{Ciphertext matrix computations in kernel}
For the sake of simplicity we omit the index $i\in[b]$ in $X$ in the matrix computation flow, so thus $Q_h,K_h,V_h$ and $S_h$, unless such an index is necessary.  

The matrix computations in kernels, like Gaussian kernel and Epanechnikov kernel, consists of (1) computing $l_2$ norms of $Q_h[i][:] - K_h[j][:]$, where $i,j \in [n_1]$ and $h\in[H]$; (2) computing ciphertext-ciphertext matrix multiplication between kernel results and $V_h$. 

For head $h\in[H]$, the matrix that stores all $l_2$ norms for one head is defined by:
\begin{equation*}
    S_h
    = 
    \left(
        \begin{array}{ccc}
           \sum_{i=0}^{n_3-1}(Q_h[0][i]-K_h[0][i])^2  & \cdots & \sum_{i=0}^{n_3-1}(Q_h[0][i]-K_h[n_1-1][i])^2 \\
            \vdots & \ddots & \vdots \\
            \sum_{i=0}^{n_3-1}(Q_h[n_1-1][i]-K_h[0][i])^2 & \cdots & \sum_{i=0}^{n_3-1}(Q_h[n_1-1][i]-K_h[n_1-1][i])^2
        \end{array}
    \right),
\end{equation*}
where $S_h \in \R^{n_1 \times n_1}$, and $Q_h,K_h \in \R^{n_1 \times n_3}$, $n_3=n_2/H$. 

We propose an efficient matrix computation pipeline for this kind of kernels. 

\begin{definition}[Flexible interleaved diagonal packing]
\label{def:diag-packing}
Suppose we have $b$ matrices $\{ X^i \in \R^{n_1 \times n_1}\}_{ i \in [b]}$ be:
\begin{equation*}
    X^i = 
    \left(
        \begin{array}{ccc}
           X^i[0][0]  & \cdots & X^i [0][n_1-1] \\
            \vdots & \ddots & \vdots \\
            X^i [n_1-1] [0] & \cdots & X^i [n_1-1] [n_1-1]
        \end{array}
    \right). 
\end{equation*}

Let $\{ \vec{m}^i \in \R^N \}_{i \in [r]}$, $r = \lceil n_1/ \lfloor N/(bn_1) \rfloor \rceil$ denote the plaintext vectors obtained by \textit{interleaved column packing} the matrices $X^0 ,..., X^{b-1}$. 
For simplicity, we assume $(bn_1) | N$ and $N | (bn_1^2)$. 
We define: 
\begin{equation*}
    m^i[j] = X^{j \% b} \left[\lfloor \frac{j}{b} \rfloor \% n_1\right] \left[\left(\frac{iN}{bn_1}+\lfloor \frac{j}{bn_1} \rfloor + \lfloor \frac{j}{b} \rfloor \right)\% n_1 \right]
\end{equation*}
where $i \in [\frac{bn_1^2}{N}]$, and $j \in [N]$.
We say that $\{\enc(\vec{m}^i)\}_{i \in [\frac{bn_1^2}{N}]}$ is in $b$-\textbf{interleaved diagonal packing}. 
\end{definition}

Algorithm~\ref{alg:CCMM-QK} takes encrypted $Q_h,K_h$ in $b$-interleaved column packing as input and output an encrypted matrix $S_h$ in $b$-interleaved diagonal packing. 
\footnote{
Using this notation, the $b$-interleaved diagonal packing in \cite{cryptoeprint:2025/991,ARION} is $\frac{N}{n_1}$-interleaved diagonal packing with a pre-determined parameter $\frac{N}{n_1}$. Our packing parameter $b$ is flexible. 
}
Then Algorithm~\ref{alg:CCMM-V} takes encrypted $\sigma(S_h)$ in $b$-interleaved diagonal packing and encrypted $V_h$ in $b$-interleaved column packing as input, and outputs an encrypted matrix $\sigma(S_h)V_h$ in $b$-interleaved column packing. 

\begin{algorithm}
    \caption{Ciphertext-Ciphertext matrix computation for $S_h$.}
    \label{alg:CCMM-QK}
    \begin{algorithmic}[1]  
        \REQUIRE Encrypted matrices $Q_h$, and $K_h$ in $b$-interleaved column packing $\{\enc(\vec{q}^i) \}_{i \in [r'_2]}$, and $\{\enc(\vec{k}^i) \}_{i \in [r'_2]}$, $r'_2 = \frac{bn_1n_3}{N}$ where $n_3=n_2/H$. 
        \ENSURE Encrypted matrix $S_h$ in $b$-interleaved diagonal packing $\{ \enc(\vec{\hat{m}}^i) \}_{i \in [r_\mathrm{QK}]}$, $r_\mathrm{QK}= \frac{bn_1^2}{N}$. 

        \FOR{$\mathrm{ind}$ from 0 to $r_\mathrm{QK}-1$}
            \FOR{i from 0 to $\frac{n_1}{r_\mathrm{QK}}-1$}
                \STATE $\hat{ct} = \enc(\vec{0})$
                \FOR{j from 0 to $r'_2-1$}
                    \STATE Evaluate $ct' = \ckksrotint_{ \mathrm{ind}n_1/r_\mathrm{QK}+i}(\enc(\vec{k}^j))$. 
                    \STATE Evaluate $ct' = \mult(\enc(\vec{q}^j) - ct',\enc(\vec{q}^j) - ct')$. 
                    \STATE Let $\mathrm{Ind}_{rot} = N/2$.
                    \IF{j = 0}
                        \STATE $\hat{ct} = ct'$. 
                    \ELSE
                        \STATE Evaluate $\hat{ct} = \hat{ct} + ct'$
                    \ENDIF
                \ENDFOR
                \WHILE{$\mathrm{Ind}_{rot} \geq (bn_1)$}
                    \STATE Evaluate $\hat{ct} = \hat{ct} + \ckksrot_{\mathrm{Ind}_{rot}}(\hat{ct})$.
                    \STATE Evaluate $\mathrm{Ind}_{rot} = \mathrm{Ind}_{rot}/2$.
                \ENDWHILE
                \STATE Let $\vec{\mathbf{mask}} = \vec{0}$. 
                \FOR{j from $i\cdot bn_1$ to $(i+1)\cdot bn_1-1$}
                    \STATE Let $\vec{\mathbf{mask}}[j] = 1$. 
                \ENDFOR
                \STATE Evaluate $\hat{ct} = \plainmult(\hat{ct}, \mathbf{mask})$. 

                \IF{i = 0}
                    \STATE Let $\enc(\vec{\hat{m}}^{\mathrm{ind}}) = \hat{ct}$. 
                \ELSE 
                    \STATE Evaluate $\enc(\vec{\hat{m}}^{\mathrm{ind}}) = \enc(\vec{\hat{m}}^{\mathrm{ind}})+\hat{ct}$. 
                \ENDIF
            \ENDFOR
        \ENDFOR

\end{algorithmic}
\end{algorithm}

\begin{algorithm}
    \caption{Ciphertext-Ciphertext matrix computation for $\times V_h$. }
    \label{alg:CCMM-V}
    \begin{algorithmic}[1]  
        \REQUIRE Encrypted matrix $S_h$ in $b$-interleaved diagonal packing $\{ \enc(\vec{m}^i) \}_{i \in [r_\mathrm{QK}]}$, $r_\mathrm{QK}= \frac{bn_1^2}{N}$. Encrypted matrix $V_h$ in $b$-interleaved column packing $\{\enc(\vec{v}^i) \}_{i \in [r'_2]}$, $r'_2 = \frac{bn_1n_3}{N}$.  
        
        \ENSURE Encrypted matrix $\sigma(S_h)V_h$ in $b$-interleaved column packing $\{ \enc(\vec{\hat{m}}^i) \}_{i \in [r'_3]}$, $r'_3= \frac{bn_1n_3}{N}$.

        \FOR{i from 0 to $r_\mathrm{QK}-1$} 
                \FOR{j from 0 to $\frac{n_1}{r_{\mathrm{QK}}}-1$}
                    \STATE Let $\vec{\mathbf{mask}} = \vec{0}$. 
                    \FOR{k from $i\cdot bn_1$ to $(i+1)\cdot bn_1-1$}
                        \STATE Let $\vec{\mathbf{mask}}[k] = 1$. 
                    \ENDFOR
                    \STATE Evaluate $\mathrm{Ext}_j(\enc(\vec{m}^i)) = \plainmult(\enc(\vec{m}^i), {\mathbf{mask}})$. 
                    \COMMENT Extract $j$-th diagonal from $\enc(\vec{m}^i)$. 

                    \STATE Let $\mathrm{Ind}_{rot} = N/2$.
                    \WHILE{$\mathrm{Ind}_{rot} \geq (bn_1)$}
                        \STATE $\mathrm{Ext}_j(\enc(\vec{m}^i)) += \ckksrot_{\mathrm{Ind}_{rot}}(\mathrm{Ext}_j(\enc(\vec{m}^i)))$. 
                        \COMMENT Copy $j$-th diagonal to all slots. 
                        \STATE Evaluate $\mathrm{Ind}_{rot} = \mathrm{Ind}_{rot}/2$.
                    \ENDWHILE 
                \ENDFOR
            \ENDFOR

        \FOR{ $\mathrm{ind}$ from 0 to $r'_2-1$}
            \FOR{i from 0 to $r_\mathrm{QK}-1$}
                \FOR{j from 0 to $\frac{n_1}{r_{\mathrm{QK}}}-1$}



                    \STATE Evaluate $ct^* = \ckksrotint_{in_1/r_\mathrm{QK}+j}(\enc(\vec{v}^\mathrm{ind}))$. 
                    \STATE Evaluate $ct' = \mathrm{Ext}_j(\enc(\vec{m}^i)) \cdot ct^*$. 

                    \IF{i = 0 and j = 0}
                        \STATE Let $\enc(\vec{\hat{m}}^{\mathrm{ind}}) = ct'$. 
                    \ELSE
                        \STATE Evaluate $\enc(\vec{\hat{m}}^{\mathrm{ind}}) = \enc(\vec{\hat{m}}^{\mathrm{ind}}) + ct'$. 
                    \ENDIF
                \ENDFOR
            \ENDFOR
        \ENDFOR

    \end{algorithmic}
\end{algorithm}

\paragraph{Optimizations. }
Applying the baby-step-giant-step (BSGS) from \cite{halevi2014algorithms,lu2021pegasus} can reduce the required FHE rotations from $O(n_1)$ to $O(\sqrt{n_1})$ in the multiplication between plaintext matrix and ciphertext vector with dimension $n_1$. 
When the matrix is encrypted, \cite{cryptoeprint:2024/1851,ARION} proposed the double BSGS: the matrix is packed and encrypted in diagonal form, and all the diagonals are already rotated. 
By modifying Algorithm~\ref{alg:CCMM-QK}, the output diagonals are rotated as desired by the double BSGS. 
This modification can also reduce the number of required rotations in Algorithm~\ref{alg:CCMM-QK} from $O(n_1)$ to $O(\sqrt{n_1})$. 
The modified algorithm is shown in Algorithm~\ref{alg:CCMM-QK-doubleBSGS}. 

Further, if $V_h$ is packed in $\C$ instead of $\R$, then about half of the FHE operations can be saved because the number of ciphertexts to encrypt $V_h$ is halved.
Algorithm~\ref{alg:CCMM-QK-doubleBSGS} and Algorithm~\ref{alg:CCMM-V-doubleBSGS} show the optimized matrix computations for $S_h$ and $\sigma(S_h)\times V_h$, respectively. 


\begin{algorithm}
    \caption{Ciphertext-Ciphertext matrix computation for $S_h$ with double BSGS.}
    \label{alg:CCMM-QK-doubleBSGS}
    \begin{algorithmic}[1]  
        \REQUIRE Encrypted matrices $Q_h$, and $K_h$ in $b$-interleaved column packing $\{\enc(\vec{q}^i) \}_{i \in [r'_2]}$, and $\{\enc(\vec{k}^i) \}_{i \in [r'_2]}$, $r'_2 = \frac{bn_1n_3}{N}$. 
        \ENSURE Encrypted matrix $S_h$ in {rotated} $b$-interleaved diagonal packing $\{ \enc(\vec{\hat{m}}^i) \}_{i \in [r_\mathrm{QK}]}$, $r_\mathrm{QK}= \frac{bn_1^2}{N}$. 

        \FOR{$\mathrm{ind}$ from 0 to $r_\mathrm{QK}-1$}
            \FOR{j from 0 to $r'_2-1$}
                \STATE Let $ct^q_{\mathrm{ind}r'_2+j} = \ckksrotint_{-\mathrm{ind}n_1/r_\mathrm{QK}}(\enc(\vec{q}^j))$. 
            \ENDFOR
        \ENDFOR

        \FOR{i from 0 to $\frac{n_1}{r_\mathrm{QK}}-1$}
            \FOR{j from 0 to $r'_2-1$}
                \STATE Let $ct^k_{ir'_2+j} = \ckksrotint_{i}(\enc(\vec{k}^j))$. 
            \ENDFOR
        \ENDFOR

        \FOR{$\mathrm{ind}$ from 0 to $r_\mathrm{QK}-1$}
            \FOR{i from 0 to $\frac{n_1}{r_\mathrm{QK}}-1$}
                \STATE Initialize $\hat{ct} = \enc(\vec{0})$. 
                \FOR{j from 0 to $r'_2-1$}
                    \STATE Evaluate $ct' = \mult(ct^q_{\mathrm{ind}r'_2+j}-ct^k_{ir'_2+j},ct^q_{\mathrm{ind}r'_2+j}-ct^k_{ir'_2+j})$. 
                    \STATE Let $\mathrm{Ind}_{rot} = N/2$.
                    \IF{j = 0}
                        \STATE $\hat{ct} = ct'$. 
                    \ELSE
                        \STATE Evaluate $\hat{ct} = \hat{ct} + ct'$
                    \ENDIF
                \ENDFOR
                \WHILE{$\mathrm{Ind}_{rot} \geq (bn_1)$}
                    \STATE Evaluate $\hat{ct} = \hat{ct} + \ckksrot_{\mathrm{Ind}_{rot}}(\hat{ct})$.
                    \STATE Evaluate $\mathrm{Ind}_{rot} = \mathrm{Ind}_{rot}/2$.
                \ENDWHILE
                \STATE Let $\vec{\mathbf{mask}} = \vec{0}$. 
                \FOR{j from $i\cdot bn_1$ to $(i+1)\cdot bn_1-1$}
                    \STATE Let $\vec{\mathbf{mask}}[j] = 1$. 
                \ENDFOR
                \STATE Evaluate $\hat{ct} = \plainmult(\hat{ct}, pt_{\mathbf{mask}})$. 

                \IF{i = 0}
                    \STATE Let $\enc(\vec{\hat{m}}^{\mathrm{ind}}) = \hat{ct}$. 
                \ELSE 
                    \STATE Evaluate $\enc(\vec{\hat{m}}^{\mathrm{ind}}) = \enc(\vec{\hat{m}}^{\mathrm{ind}})+\hat{ct}$. 
                \ENDIF
            \ENDFOR
        \ENDFOR

\end{algorithmic}
\end{algorithm}

\begin{algorithm}
    \caption{Ciphertext-Ciphertext matrix computation for $\times V$ with double BSGS.}
    \label{alg:CCMM-V-doubleBSGS}
    \begin{algorithmic}[1]  
        \REQUIRE Encrypted matrix $\sigma(S_h)$ in rotated $b$-interleaved diagonal packing $\{ \enc(\rot_l(\vec{m}^i\in\R^N)) \}_{i \in [r_\mathrm{QK}]}$, $r_\mathrm{QK}= \frac{bn_1^2}{N}$. Encrypted matrix $V$ in $b$-interleaved column packing in complex number field: $\{\enc(\vec{v}^i\in \C^N) \}_{i \in [r^*_2]}$, $r^*_2 = r'_2/2 = \frac{bn_1n_3}{2N}$.  
        
        \ENSURE Encrypted matrix $\sigma(S_h)V_h$ in $b$-interleaved column packing $\{ \enc(\vec{\hat{m}}^i\in \C^N) \}_{i \in [r'_3]}$, $r'_3= \frac{bn_1n_3}{N}$. 

        \FOR{i from 0 to $r_\mathrm{QK}-1$} 
                \FOR{j from 0 to $\frac{n_1}{r_{\mathrm{QK}}}-1$}
                    \STATE Let $\vec{\mathbf{mask}} = \vec{0}$. 
                    \FOR{k from $i\cdot bn_1$ to $(i+1)\cdot bn_1-1$}
                        \STATE Let $\vec{\mathbf{mask}}[k] = 1$. 
                    \ENDFOR
                    \STATE $\mathrm{Ext}_j(\enc(\vec{m}^i))= \plainmult(\enc(\vec{m}^i), {\mathbf{mask}})$. 
                    \COMMENT Extract $j$-th diagonal from $\enc(\vec{m}^i)$. 

                    \STATE Let $\mathrm{Ind}_{rot} = N/2$.
                    \WHILE{$\mathrm{Ind}_{rot} \geq (bn_1)$}
                        \STATE $\mathrm{Ext}_j(\enc(\vec{m}^i)) += \ckksrot_{\mathrm{Ind}_{rot}}(\mathrm{Ext}_j(\enc(\vec{m}^i)))$. 
                        \COMMENT Copy $j$-th diagonal to all slots. 
                        \STATE Evaluate $\mathrm{Ind}_{rot} = \mathrm{Ind}_{rot}/2$.
                    \ENDWHILE 
                \ENDFOR
            \ENDFOR

        \FOR{ $\mathrm{ind}$ from 0 to $r^*_2-1$}
            \FOR{j from 0 to $\frac{n_1}{r_{\mathrm{QK}}}-1$}
                \STATE Let $ct^*_j = \ckksrotint_{j}(\enc(\vec{v}^\mathrm{ind}))$.
                \COMMENT Baby step. 
            \ENDFOR
            
            \FOR{i from 0 to $r_\mathrm{QK}-1$}
                \STATE Initialize $\hat{ct} = \enc(\vec{0})$. 
                \FOR{j from 0 to $\frac{n_1}{r_{\mathrm{QK}}}-1$}  


                    \STATE Evaluate $ct' = \mult(\mathrm{Ext}_j(\enc(\vec{m}^i)), ct^*_j)$. 
                    \IF{$j=0$}
                        \STATE Let $\hat{ct} = ct'$.
                    \ELSE
                        \STATE Evaluate $\hat{ct} = \hat{ct} + ct'$.
                    \ENDIF
                \ENDFOR

                \STATE Evaluate $\hat{ct} = \ckksrotint_{in_1/r_\mathrm{QK}}(\hat{ct})$. 
                \IF{$i=0$}
                    \STATE Let $\enc(\vec{\hat{m}}^{\mathrm{ind}}_\C) = \hat{ct}$.
                \ELSE 
                    \STATE Evaluate $\enc(\vec{\hat{m}}^{\mathrm{ind}}_\C) = \enc(\vec{\hat{m}}^{\mathrm{ind}}_\C)+ \hat{ct}$. 
                \ENDIF
                
            \ENDFOR
            \STATE \textbf{Extract:} 
            \COMMENT{Note that $r_3'=2r^*_2$}
            \STATE $\enc(\vec{\hat{m}}^{\mathrm{2ind}}):=\enc(\vec{\hat{m}}^{\mathrm{ind}}_\C)+\conj(\enc(\vec{\hat{m}}^{\mathrm{ind}}_\C))$.
            \STATE $\enc(\vec{\hat{m}}^{\mathrm{2ind+1}}):=\enc(\vec{\hat{m}}^{\mathrm{ind}}_\C)-\conj(\enc(\vec{\hat{m}}^{\mathrm{ind}}_\C))$.
        \ENDFOR

    \end{algorithmic}
\end{algorithm}

\paragraph{Rotations in a batch. }
Number of HE rotations in computing $S_h$ by using Algorithm~\ref{alg:CCMM-QK} is 
\begin{equation*}
    \#(\ckksrot) = n_1 \cdot (2\cdot r'_2  + \log (N/(bn_1)) )
    =\frac{2bn_1^2n_3}{N}+n_1\log (N/(bn_1)). 
\end{equation*}

Number of HE rotations in $\times V_h$ by using Algorithm~\ref{alg:CCMM-V} is also 
\begin{equation*}
    \#(\ckksrot) = n_1 \cdot (2\cdot r'_2  + \log (N/(bn_1)) )
    =\frac{2bn_1^2n_3}{N}+n_1\log (N/(bn_1)). 
\end{equation*}

Numbers of HE rotations in $S_h$ and $\times V_h$ are reduced in Algorithm~\ref{alg:CCMM-QK-doubleBSGS} and Algorithm~\ref{alg:CCMM-V-doubleBSGS} due to applying the double BSGS technique \cite{cryptoeprint:2024/1851,ARION}. 

Number of HE rotations in $S_h$ by using Algorithm~\ref{alg:CCMM-QK-doubleBSGS} is 
\begin{equation*}
    \#(\ckksrot) = 2 \cdot r_{\mathrm{QK}} \cdot r'_2 + 2 \cdot \frac{n_1}{r_{\mathrm{QK}}} \cdot r'_2 + n_1 \cdot \log (\frac{N}{bn_1}) )
    = \frac{bn_1n_3}{N}(\frac{2bn_1^2}{N}+\frac{2N}{bn_1})+n_1\log(\frac{N}{bn_1}). 
\end{equation*}

Number of HE rotations in $\times V_h$ by using Algorithm~\ref{alg:CCMM-V-doubleBSGS} is 
\begin{equation*}
    \#(\ckksrot) = 2 \cdot \frac{n_1}{r_{\mathrm{QK}}} \cdot r^*_2 + n_1 \cdot \log (\frac{N}{bn_1}) ) + 2 \cdot r_{\mathrm{QK}} \cdot r^*_2 +r^*_2
    = \frac{bn_1n_3}{2N}(\frac{2N}{bn_1} + \frac{2bn_1^2}{N} + 1)+n_1\log(\frac{N}{bn_1}). 
\end{equation*}

If we set $N = 32768$, $b = 16$, $H = 12$, $Q_h, K_h, V_h \in \R^{128 \times 64}$, $S_h\in \R^{128\times 128}$ then numbers of HE rotations in two versions of matrix operations of $\sigma(S_h)V_h$ are $3072$ and $1314$, respectively.

\paragraph{Relinearizations. }
Number of relinearizations in $S_h$ in either algorithm is 
\begin{equation*}
    \#(\ckksrel) = n_1. 
\end{equation*}

Numbers of relinearizations in $\times V_h$ by using Algorithm~\ref{alg:CCMM-V} (with lazy relinearization) are 
\begin{equation*}
    \#(\ckksrel) = r'_2 = \frac{bn_1n_3}{N}. 
\end{equation*}

Numbers of relinearizations in $\times V$ by using Algorithm~\ref{alg:CCMM-V-doubleBSGS} (with lazy relinearization) are 
\begin{equation*}
    \#(\ckksrel) = r^*_2r_{\mathrm{QK}} = \frac{b^2n_1^3n_3}{2N^2},
\end{equation*}

If we set $N = 32768$, $b = 16$, $H = 12$, $Q_h, K_h, V_h \in \R^{128 \times 64}$, $S_h\in \R^{128\times 128}$ then numbers of relinearizations in two versions of matrix operations of $\sigma(S_h)V_h$ are $ 132$ and $144$, respectively.

\paragraph{Multiplicative levels, ct-ct multiplications and ct-pt multiplications in a batch. }
In the basic version (Algorithm~\ref{alg:CCMM-QK} and Algorithm~\ref{alg:CCMM-V}), 
computing $S_h$ consumes 2 multiplicative levels for $\enc(Q_h)$ and 3 multiplicative levels for $\enc(K_h)$ because $\enc(K_h)$ requires internal rotations which consumes 1 level. 
The number of HE ct-ct multiplications is $n_1r'_2$, and the number of HE ct-pt multiplications is $n_1(r'_2+1)$. 
Computing $\times V_h$ consumes 2 multiplicative levels for both inputs with also $n_1r'_2$ HE ct-ct multiplications and $n_1r'_2+n_1$ HE ct-pt multiplications.

In the optimized version(Algorithm~\ref{alg:CCMM-QK-doubleBSGS} and Algorithm~\ref{alg:CCMM-V-doubleBSGS}), 
computing $S_h$ consumes 3 multiplicative levels for $\enc(Q_h)$ and $\enc(K_h)$ because both of them require internal rotations. 
The number of HE ct-ct multiplications is $n_1r'_2$, and the number of HE ct-pt multiplications is $r'_2(r_{\mathrm{QK}}+\frac{n_1}{r_{\mathrm{QK}}})+n_1$.
Computing $\times V_h$ consumes 3 multiplicative levels for both inputs with also $n_1r^*_2$ HE ct-ct multiplications and $r^*_2(\frac{n_1}{r_{\mathrm{QK}}}+r_{\mathrm{QK}})+n_1$ HE ct-pt multiplications.

If we set $N = 32768$, $b = 16$, $H = 12$, $Q_h, K_h, V_h \in \R^{128 \times 64}$, $S_h\in \R^{128\times 128}$, then numbers of HE ct-ct multiplications in two versions of matrix operations of $\sigma(S_h)V_h$ are $1024$ and $768$, respectively.  
Numbers of HE ct-pt multiplications in two versions of matrix operations of $\sigma(S_h)V_h$ is $1280$ and $400$, respectively.

In summary, the required number of each HE operation for $\{S_h^i\}_{i \in [b]}$ is listed in Table~\ref{tab:kernel-ours}. 
The required number of each HE operation for $\times V^i, i \in [b]$, where $V^i \in \R^{n_1 \times n_3}$, is listed in Table~\ref{tab:V-ours}. 

\begin{table}[htbp]
    \centering
    \caption{The number of each HE operation for $\{S_h^i\}_{i \in [b]}$ in one head in our scheme}
    \label{tab:kernel-ours}
    \resizebox{\textwidth}{!}{
    \begin{tabular}{cccccc}
    \toprule
         Operations & Rot & ct-ct Mult & Relinearization & ct-pt Mult & Level \\ \midrule
         $S_h$-basic 
         & $\frac{2bn_1^2n_3}{N}+n_1\log (N/(bn_1))$ 
         & $\frac{bn_1^2n_3}{N}$ 
         & $n_1$
         &  $\frac{bn_1^2n_3}{N}+n_1$  
         & 3 
         \\ \midrule
         $S_h$-optimized 
         & $\frac{bn_1n_3}{N}(\frac{2bn_1^2}{N}+\frac{2N}{bn_1})+n_1\log(\frac{N}{bn_1}) $
         & $\frac{bn_1^2n_3}{N}$ 
         & $n_1$
         & $\frac{b^2n_1^3n_3}{N^2}$  
         & 3 
         \\
    \bottomrule
    \end{tabular}
    }
    
\end{table}

\begin{table}[htbp]
    \centering
    \caption{The number of each HE operation for $\{\sigma(S^i_h) V^i_h\}, i \in [b]$ in our scheme}
    \label{tab:V-ours}
    \resizebox{\textwidth}{!}{
    \begin{tabular}{cccccc}
    \toprule
         Operations & Rot & ct-ct Mult & Relinearization & ct-pt Mult & Level \\ \midrule
         $\times V_h$-CCMM-basic 
         & $\frac{2bn_1^2n_3}{N}+n_1\log (N/(bn_1))$  
         & $\frac{bn_1^2n_3}{N}$ 
         & $\frac{bn_1n_3}{N}$
         &  $\frac{bn_1^2n_3}{N}+n_1$  
         & 2 
         \\ \midrule
         $\times V_h$-CCMM-optimized 
         & $\frac{bn_1n_3}{2N}(\frac{2N}{bn_1} + \frac{2bn_1^2}{N} + 1)+n_1\log(\frac{N}{bn_1})$ 
         & $\frac{bn_1^2n_3}{2N}$ 
         & $\frac{b^2n_1^3n_3}{2N^2}$
         & $\frac{bn_1n_3}{2N}(\frac{bn_1^2}{N}+\frac{N}{bn_1})+n_1$  
         & 3 
         \\
    \bottomrule
    \end{tabular}
    }
    
\end{table}

\subsection{Ciphertext-ciphertext matrix multiplication in classification head}
\label{app-ccmm-class}

The matrix computations in classification head consists of (1) Extracting the $[CLS]$ (usually the first row): $\vec{h} \in \R^{1 \times n_2}$ of the output of transformer layers; (2) Computing ciphertext-ciphertext matrix multiplication between $\vec{h}$ and $W_{c1}\in \R^{n_2 \times n_2}$; (3) Computing ciphertext-ciphertext matrix multiplication between $\vec{h}'=\mathrm{Tanh}(\vec{h}W_{c1}+\vec{b}_{c1})$ and $W_{c2} \in \R^{n_2 \times c}$. 

The packing of $W_{c1}$ and $W_{c2}$ are elaborated in Algorithm~\ref{alg:pack-Wc1} and Algorithm~\ref{alg:pack-Wc2}. 
The 3 computations are shown in Algorithm~\ref{alg:ext-first-row}, Algorithm~\ref{alg:ccmm-wc1}, and Algorithm~\ref{alg:CCMM-wc2}. 

\begin{algorithm}[htbp]
    \caption{Packing for matrix $W_{c1}$ in classification head. }
    \label{alg:pack-Wc1}
    \begin{algorithmic}[1]
        \REQUIRE Matrix $W_{c1} \in \R^{n_2 \times n_2}$
        \ENSURE $\{\enc(\vec{w}_{c1}^i\in \R^N) \}_{i \in [r_{c1}]}$, $r_{c1}=\frac{bn_2^2}{N}$.

        \FOR{i from 0 to $\frac{n_2}{n_1}-1$}
            \FOR{j from 0 to $\frac{bn_1n_2}{N}-1$}
                \STATE Let $\vec{d}=\vec{0}$.
                \FOR{$k_1$ from 0 to $\frac{N}{bn_1}-1$}
                    \FOR{$k_2$ from 0 to $n_1-1$}
                        \FOR{$k_3$ from 0 to $b-1$}
                            \STATE Let $d[k_1bn_1+k_2b+k_3]=W_{c1}[j\frac{N}{bn_1}+k_1][in_1+k_2]$.
                        \ENDFOR
                    \ENDFOR
                \ENDFOR
                \STATE Evaluate $\enc(\vec{w}_{c1}^{i\frac{bn_1n_2}{N}+j}) = \enc(\vec
                d)$. 
            \ENDFOR
        \ENDFOR
    \end{algorithmic}
\end{algorithm}

\begin{algorithm}
    \caption{Packing for matrix $W_{c2}$ in classification head}
    \label{alg:pack-Wc2}
    \begin{algorithmic}[1]
        \REQUIRE Matrix $W_{c2} \in \R^{n_2 \times c}$
        \ENSURE $\{ \enc(\vec{w}^i_{c2}\in \R^N) \} _{i \in [c/2] }$.

        \FOR{i from 0 to $\frac{c}{2}-1$}
            \FOR{j from 0 to $n_2-1$}
                \FOR{k from 0 to $b-1$}
                    \STATE Let $\vec{w}^i_{c2}[jb+k]=W_{c2}[j][2i]$. 
                    \STATE Let $\vec{w}^i_{c2}[N/2+jb+k]=W_{c2}[j][2i+1]$.
                \ENDFOR
            \ENDFOR
            \STATE Evaluate $\enc(\vec{w}^i_{c2})$. 
        \ENDFOR
    
    \end{algorithmic}
\end{algorithm}

\begin{algorithm}
    \caption{Extract the first row from output of transformer layers}
    \label{alg:ext-first-row}
    \begin{algorithmic}[1]
        \REQUIRE Encrypted matrix $\{ X^i \in \R^{n_1 \times n_2} \}_{i \in [b]}$ in $b$-interleaved column packing $\{ \enc(\vec{m}^i \in \R^N) \}_{i \in [r_1]}$, $r_1=\frac{bn_1n_2}{N}$. 
        \ENSURE Encrypted vectors $\{ \vec{h}^i \in \R^{n_2} \}_{i \in [b]}$ in repeated packing $\{ \enc(\vec{\hat{h}}^i) \}_{i \in [r_1]}$.

        \STATE Let $\vec{\mathbf{mask}} = \vec{0}$. 
        \FOR{i from 0 to $\frac{N}{bn_1}-1$}
            \FOR{j from 0 to $b-1$}
                \STATE Let $\vec{\mathbf{mask}}[ibn_1+j] = 1$.
            \ENDFOR
        \ENDFOR

        \FOR{i from 0 to $r_1-1$}
            \STATE Evaluate $ct'=\plainmult(\enc(\vec{m}^i), \vec{\mathrm{mask}})$. 
            \STATE Let $\mathrm{Ind}_{rot} = b$.
            \WHILE{$\mathrm{Ind}_{rot} < bn_1$}
                \STATE Evaluate $ct'=ct' + \ckksrot_{N-\mathrm{Ind}_{rot}}(ct')$. 
                \COMMENT{Fill all the slots by repeated packing. }
                \STATE $\mathrm{Ind}_{rot} = \mathrm{Ind}_{rot} \cdot 2$.
            \ENDWHILE
            \STATE Let $\enc(\vec{\hat{h}}^i)=ct'$.
        \ENDFOR

    \end{algorithmic}
\end{algorithm}

\begin{algorithm}
    \caption{Ciphertext-Ciphertext matrix multiplication for matrix $W_{c1}$ in classification head}
    \label{alg:ccmm-wc1}
    \begin{algorithmic}[1]
        \REQUIRE Encrypted vectors $\{ \vec{h}^i \in \R^{n_2} \}_{i \in [b]}$ in repeated packing $\{ \enc(\vec{\hat{h}}^i) \}_{i \in [r_1]}$, $r_1=\frac{bn_1n_2}{N}$, encrypted matrix $W_{c1} \in R^{n_2 \times n_2}$ output by Algorithm~\ref{alg:pack-Wc1}: $\{\enc(\vec{w}_{c1}^i\in \R^N) \}_{i \in [r_{c1}]}$, $r_{c1}=\frac{bn_2^2}{N}$

        \ENSURE Encrypted vectors $\{ \vec{h}^iW_{c1} \}_{i \in [b]}$: $\enc(\vec{h}')$

        \FOR{i from 0 to $\frac{r_{c1}}{r_1}-1$}
            \STATE Let $ct'_i = \enc(\vec{0})$.
            \FOR{j from 0 to $r_1-1$}
                \STATE Evaluate $ct'_i = ct'_i + \enc(\vec{\hat{h}}^j) \cdot \enc(\vec{w}_{c1}^{ir_1+j})$. 
            \ENDFOR
            \STATE Let $\mathrm{Ind}_{rot} = N/2$
            \WHILE{$\mathrm{Ind}_{rot} \geq bn_1$}
                \STATE Evaluate $ct'_i = ct'_i + \ckksrot_{\mathrm{Ind}_{rot}}(ct'_i)$.
                \STATE $\mathrm{Ind}_{rot} = \mathrm{Ind}_{rot}/2$. 
            \ENDWHILE

            \STATE Let $\vec{\mathrm{mask}} = \vec{0}$. 
            \FOR{j from $ibn_1$ to $(i+1)bn_1-1$}
                \STATE Let $\vec{\mathrm{mask}}[j] = 1$.
            \ENDFOR
            \STATE Evaluate $ct'_i=ct'_i \cdot \vec{\mathrm{mask}}$. 

            \IF{i = 0}
                \STATE Let $\enc(\vec{h}') = ct'_i$.
            \ELSE 
                \STATE Evaluate $\enc(\vec{h}') = \enc(\vec{h}') + ct'_i$.
            \ENDIF
            
        \ENDFOR

        \STATE Evaluate $\enc(\vec{h}') = \enc(\vec{h}') + \ckksrot_{N/2}(\enc(\vec{h}'))$. 
        
    \end{algorithmic}
\end{algorithm}

\begin{algorithm}
    \caption{Ciphertext-Ciphertext matrix multiplication for matrix $W_{c2}$ in classification head}
    \label{alg:CCMM-wc2}
    \begin{algorithmic}[1]
        \REQUIRE Encrypted vectors $\{ \vec{h}^i \}_{i \in [b]}$ output by Algorithm~\ref{alg:ccmm-wc1}: $\enc(\vec{h}')$, and encrypted matrix $W_{c2} \in \R^{n_2 \times c}$ output by Algorithm~\ref{alg:pack-Wc2}: $\{ \enc(\vec{w}^i_{c2}\in \R^N) \} _{i \in [c/2] }$. 

        \ENSURE Encrypted vectors $\{ \vec{h}^iW_{c2} \in R^c \}_{i \in [b]}$: $\{ \enc(\vec{\hat{h}}^i) \}_{i \in [c/2]}$. 

        \FOR{i from 0 to $c/2$}
            \STATE Evaluate $\enc(\vec{w}^i_{c2}) = \enc(\vec{w}^i_{c2}) \cdot \enc(\vec{h}')$. 
            \STATE Let $\mathrm{Ind}_{rot} = N/4$
            \WHILE{$\mathrm{Ind}_{rot} \geq b$}
                \STATE Evaluate $\enc(\vec{w}^i_{c2}) = \enc(\vec{w}^i_{c2}) + \ckksrot_{\mathrm{Ind}_{rot}}(\enc(\vec{w}^i_{c2}))$.
                \STATE $\mathrm{Ind}_{rot} = \mathrm{Ind}_{rot}/2$. 
            \ENDWHILE
        \ENDFOR
    \end{algorithmic}
\end{algorithm}

\paragraph{Rotations in a batch. }
Number of HE rotations in extract the first row is 
\[
    \#(\ckksrot) = r_1 \cdot \log(n_1) = \frac{bn_1n_2\log(n_1)}{N}. 
\]

Number of HE rotations in $(\vec{h}W_{c1})$ is 
\[
    \#(\ckksrot) = \frac{n_2}{n_1} \cdot \log (N/(bn_1))+1. 
\]

Number of HE rotations in $(\vec{h}W_{c2})$ is 
\[
    \#(\ckksrot) = \frac{c}{2} \cdot \log (N/(2b)). 
\]

If we set $N = 32768$, $b = 16$, $X^i \in \R^{128 \times 768}$, $W_{c1} \in \R^{768 \times 768}$, and $W_{c2} \in \R^{768 \times 2}$ then the number of HE rotations in each step is 336, 25, and 10, respectively.

\paragraph{Relinearizations. }
Number of relinearizations in $(\vec{h}W_{c1})$ is 
\[
    \#(\ckksrel) = \frac{r_{c1}}{r_1} = \frac{n_2}{n_1}. 
\]

Number of relinearizations in $(\vec{h}W_{c2})$ is 
\[
    \#(\ckksrel) = \frac{c}{2}. 
\]

If we set $N = 32768$, $b = 16$, $X^i \in \R^{128 \times 768}$, $W_{c1} \in \R^{768 \times 768}$, and $W_{c2} \in \R^{768 \times 2}$ then numbers of HE relinearzations are $6$ and $1$ respectively.

\paragraph{Multiplicative levels and ct-ct multiplications in a batch. }
Each CCMM consumes 1 multiplicative level, extraction also consumes 1 multiplicaitve level. 
The two CCMMs in classiciation heads consume 2 multiplicative levels with $\frac{bn_2^2}{N}$, and $\frac{c}{2}$ ct-ct multiplications. 
If we set $N = 32768$, $b = 16$, $X^i \in \R^{128 \times 768}$, $W_{c1} \in \R^{768 \times 768}$, and $W_{c2} \in \R^{768 \times 2}$ then numbers of HE ct-ct multiplications is $288$ and $1$ respectively.

In summary, the required number of each HE operation in classification head,  is listed in Table~\ref{tab:class-ours}. 

\begin{table}[htbp]
    \centering
    \caption{The number of each HE operation for matrix computations in classification head in our scheme}
    \label{tab:class-ours}
    \resizebox{\textwidth}{!}{
    \begin{tabular}{cccccc}
    \toprule
         Operations & Rot & ct-ct Mult & Relinearization & ct-pt Mult & Level \\ \midrule
         Extraction
         & $\frac{bn_1n_2\log(n_1)}{N}$ 
         & 0 
         & 0
         & $\frac{bn_1n_2}{N}$  
         & 1 
         \\ \midrule
         CCMM for $W_{c1}$ 
         & $\frac{n_2}{n_1} \cdot \log (N/(bn_1))+1$ 
         & $\frac{bn_2^2}{N}$ 
         & $\frac{n_2}{n_1}$
         & $\frac{n_2}{n_1}$  
         & 1 
         \\ \midrule
         CCMM for $W_{c2}$ 
         & $\frac{c}{2} \cdot \log (N/(2b))$ 
         & $\frac{c}{2}$ 
         & $\frac{c}{2}$
         & 0  
         & 1 
         \\
    \bottomrule
    \end{tabular}
    }
    
\end{table}

\section{Backward Propagation for 2-Layer BERT with Kernel Attention and LoRA}

\paragraph{Notations. }
We consider a 2-layer Transformer encoder with $H=12$ heads per layer. The standard softmax attention is replaced by a kernel function 
$
\phi(Q, K)\in \R^{n_1\times n_1}$ whose $(i,j)$-th entry is $\sigma\left(\|Q[i][:] - K[j][:]\|_2^2 \right)~.
$
For example, in the gaussian kernel: 
\[
\sigma(x)=\exp \left( -\frac{x}{2\sqrt{n_1}} \right),\quad
\phi(Q, K)[i][j] = \exp\left(-\frac{1}{2\sqrt{n_1}}\|Q[i][:] - K[j][:]\|_2^2 \right)~,
\]
where $n_1$ is the sequence length. 

Apply LoRA fine-tuning with rank $n_{LoRA}$ in attention blocks. 
We define:
\begin{align*}
    W_{Q}^{(\ell)}:=W_{Q,frozen}^{(\ell)} + A_{Q}^{(\ell)}B_{Q}^{(\ell)},\ 
    W_{K}^{(\ell)}:=W_{K,frozen}^{(\ell)} + A_{K}^{(\ell)}B_{K}^{(\ell)},\ 
    W_{V}^{(\ell)}:=W_{V,frozen}^{(\ell)} + A_{V}^{(\ell)}B_{V}^{(\ell)}~.
\end{align*}
In our scheme, $\{A_{*}^{(\ell)},B_{*}^{(\ell)}\}_{*=Q,K,V;\ell=0,1},W_{c1},W_{c2}$ are trainable parameters. 
Given vector $\vec{v}\in \R^l$, we have $\frac{\partial L}{\partial \vec{v}} \in \R^l$, i.e., the same shape of $\vec{v}$. 
Similarly, for matrix $M\in \R^{r\times l}$, we have $\frac{\partial L}{\partial M}\in \R^{r\times l} $ and thus $\frac{\partial L}{\partial M[i][j]}=\frac{\partial L}{\partial M}[i][j]$. 

In this section, we list how to compute the gradients using encrypted data of all the trainable parameters. 
All data are packed using the $b$-\textit{interleaved batching}, i.e., each batch contains $b$ inputs. 
Since the evaluations for different inputs are independent and are executed in a SIMD manner, they do not interfere with one another. 
Therefore, it suffices to focus on a single input within a batch to explain the FHE algorithm flow in this section. 

In Section~\ref{app:forward-plain}, we list the forward pass step-by-step. 
In Section~\ref{app:backward}, we list the backward pass, using plaintext formulas. 
The implementations under FHE is elaborated in the boxed paragraphs in Section~\ref{app:backward}. 
Finally, in Section~\ref{app:update} we include how to compactly compress all the gradients in the fewest ciphertexts so that the number of the expensive bootstrapping operations is minimized. After this compression and optimizer, we decompress all the gradients and perform the model weight update. 

\subsection{Forward Pass }
\label{app:forward-plain}
Let the initial input be $X^{(0)}\in \R^{n_1\times n_2}$, the first layer output be $X^{(1)}$, and the second layer output be $X^{(2)}$. 
Let $ \vec{y}_{\text{true}}\in\{0,1\}^c$ be the sample's label in one-hot encoding. 
\paragraph{Attention block:}
For the sake of simplicity, we omit all the bias vectors in the attention blocks. 
Define $W_Q^{(\ell)} = (W_{Q,0}^{(\ell)}\Vert W_{Q,1}^{(\ell)} \Vert ... \Vert W_{Q,11}^{(\ell)})\in \R^{n_2\times n_2}$. Similarly define $W_K^{(\ell)}$ and $W_V^{(\ell)}$. 
Let $U_h^{(\ell)}=\phi(Q_h^{(\ell)},K_h^{(\ell)})$. 
For layer $\ell \in \{0,1\}$ and attention head $h \in \{0,\dots,11\}$:
\begin{align*}
Q_h^{(\ell)} &= X^{(\ell)} W_{Q,h}^{(\ell)},\ K_h^{(\ell)} = X^{(\ell)} W_{K,h}^{(\ell)},\ V_h^{(\ell)} = X^{(\ell)} W_{V,h}^{(\ell)}\\
U_{h}^{(\ell)}[i][j] &= \sigma\left(\|Q_{i,h}^{(\ell)}[i][:] - K_{j,h}^{(\ell)}[j][:]\|_2^2 \right),\quad
X_h^{(\ell)} = U_h^{(\ell)} V_h^{(\ell)}
\end{align*}
Then the output of the attention block  can be obtained by:
\begin{align*}
X_O^{(\ell)} = (X_0^{(\ell)}\Vert \dots\Vert X_{11}^{(\ell)})W_O^{(\ell)}, \quad
X_{attn}^{(\ell)} = \layernorm(X^{(\ell)} + X_O^{(\ell)})
\end{align*}
\paragraph{Feedforward block:}
\begin{align*}
    X_{ffn}^{(\ell)} &= \gelu(X_{attn}^{(\ell)}W_{fc1}^{(\ell)}+\vec{b}_{fc1}^{(\ell)})W_{fc2}^{(\ell)}+\vec{b}_{fc2}^{(\ell)}\\
    X^{(\ell+1)} &= \layernorm(X_{ffn}^{(\ell)} + X_{attn}^{(\ell)})
\end{align*}

\paragraph{Pooling and classification head:}
\begin{align*}
\vec{h} = X^{(2)}_{[CLS]}\in \R^{n_2},\ \vec{y}= \tanh(\vec{h}W_{c1} + \vec{b}_{c1})W_{c2}+\vec{b}_{c2}, \quad \text{Loss function }L = \mathcal{L}(\softmax(\vec{y}), \vec{y}_{\text{true}})\in \R
\end{align*}


\subsection{Backward Propagation}
\label{app:backward}
To begin with, with the loss function being cross-entropy, we have $\grad{\vec{y}}=\softmax(\vec{y})-\vec{y}_{\text{true}}$, which is an encrypted row vector. 

\subsubsection{Pooling and classification head}
\begin{align*}
\frac{\partial L}{\partial W_{c2}} 
&= \tanh(\vec{h}W_{c1} + \vec{b}_{c1})^\intercal \, \frac{\partial L}{\partial\vec{y}},\quad
\frac{\partial L}{\partial W_{c1}} 
= \mathbf{h}^\intercal \, \left(\left(\frac{\partial L}{\partial\vec{y}}W_{c2}^\intercal\right)\odot \left(1 - \tanh^2(\vec{h}W_{c1} + \vec{b}_{c1})\right)\right)\\
\frac{\partial L}{\partial \vec{h}} 
&= \left( \left(\frac{\partial L}{\partial \vec{y}} \, W_{c2}^\intercal \right)
\odot \left(1 - \tanh^2(\vec{h} W_{c1} + \vec{b}_{c1})\right) \right) W_{c1}^\intercal.
\end{align*}

\paragraph{From $\frac{\partial L}{\partial \vec{h}}$ to $\frac{\partial L}{\partial X^{(2)}}$}
Recall that $\vec{h} = X^{(2)}_{[\mathrm{CLS}]} \in \R^{n_2}$ is the first row of $X^{(2)} \in \R^{n_1 \times n_2}$ corresponding to the $[\mathrm{CLS}]$ token. 
Then the gradient w.r.t. $X^{(2)}$ is zero everywhere except at the first row:
\begin{align*}
\frac{\partial L}{\partial X^{(2)}[i][:]} =
\begin{cases}
\frac{\partial L}{\partial \vec{h}}, & \text{if } i = 0,\\
\vec{0}, & \text{otherwise}.
\end{cases}
\end{align*}

\fbox{
\parbox{\textwidth}{%
\begin{enumerate}[leftmargin=*]
\item $\grad{\vec{y}}\in \R^{1\times c}$, $\tanh(\vec{h}W_{c1} + \vec{b}_{c1})\in \R^{1\times n_2}$ are pre-stored from forward pass. 
Using $\bm{1}\in\R^{1\times n_2}$ we first obtain the ciphertext of $\tanh(\vec{h}W_{c1} + \vec{b}_{c1})^\intercal(1,1)$ and $\bm{1}^\intercal \grad{\vec{y}}$ by $O(\log n_2)$ FHE rotations and additions. Then $\tanh(\vec{h}W_{c1} + \vec{b}_{c1})^\intercal\grad{\vec{y}}$ is obtained by $\bigg(\tanh(\vec{h}W_{c1} + \vec{b}_{c1})^\intercal(1,1)\bigg)\odot\bigg(\bm{1}^\intercal \grad{\vec{y}}\bigg) $. 
\\
\item Compute $\grad{\vec{y}}W_{c2}^\intercal$ can be done by linear combination of all columns of $W_{c2}$, using $O(\log c)$ FHE rotations and 1 ctct multiplication. 
Then multiply $(1-\tanh^2(\vec{h}W_{c1} + \vec{b}_{c1}))$ using Hadamard multiplication. 
The left multiplication with $\vec{h}^\intercal$ can be done similarly as the previous step. \\
\item 
Note that each sub-matrix of $W_{c1}$ is packed in \textit{row-major} packing, therefore each sub-matrix of $W_{c1}^\intercal$ is packed in \textit{column-packing}. 
$\grad{\vec{h}}$ is obtained by FHE multiplication between a row vector and a matrix in column-packing. 
This can be achieved by several inner-products - Hadamard multiplications followed by aggregations across slots. \\
\item $\grad{X^{(2)}}\in \R^{n_1\times n_2}$ is a matrix in which the first row is $\grad{\vec{h}}$ and all other entries are $0$. 
\end{enumerate}
}
}

\subsubsection{Feedforward block}
\paragraph{From $\frac{\partial L}{\partial X^{(\ell+1)}}$ to $\frac{\partial L}{\partial X_{attn}^{(\ell)}}$}
Let 
$X^{(\ell+1)} = \layernorm(Z), Z = X_{ffn}^{(\ell)} + X_{attn}^{(\ell)}$ 
and $G = \frac{\partial L}{\partial X^{(\ell+1)}}$. 
Given $i \in[n_1]$, $\layernorm$ parameters $\bm{\gamma},\bm{\beta}\in \R^{n_2}$, we have $\mu_i = \sum_{j\in[n_2]} Z[i][j]/n_2$ and:
\begin{align*}
\sigma_i^2 = \sum_{j\in[n_2]}(Z[i][j]-\mu_i)^2/n_2,\ 
\hat{Z}[i][j] = \frac{Z[i][j]-\mu_i}{\sqrt{\sigma_i^2+\epsilon}}, \ 
X^{(\ell+1)}[i][j] = \gamma_j \hat{Z}[i][j] + \beta_j
\end{align*}

We first find the backward formula of layer normalization for each $\grad{Z[i][j]}$: 
\begin{align*}
H[i][j] &= G[i][j]\,\gamma_j, \quad
m_i = \frac{1}{n_2} \sum_{j\in[n_2]} H[i][j], \quad
v_i = \frac{1}{n_2} \sum_{j\in[n_2]} H[i][j]\hat{Z}[i][j] \\
\frac{\partial L}{\partial Z[i][j]} &=
\frac{1}{\sqrt{\sigma_i^2+\epsilon}}
\left( H[i][j] - m_i - \hat{Z}[i][j] v_i \right)~.
\end{align*}
And then we can obtain its compact matrix form ($\vec{m},\vec{v},\bm{1}$ are row vectors):
\begin{align}
H &= G \odot (\bm{1}^\intercal\bm{\gamma}), \quad 
\vec{m}^\intercal = \frac{1}{n_2} H \bm{1}^\intercal, \quad
\vec{v}^\intercal = \frac{1}{n_2} (H \odot \hat{Z}) \bm{1}^\intercal \\
\label{eq:back_ln}
\frac{\partial L}{\partial Z} &=
\left(\left(\frac{1}{\sqrt{\sigma_0^2+\epsilon}},\frac{1}{\sqrt{\sigma_1^2+\epsilon}},...,\frac{1}{\sqrt{\sigma_{n_1}^2+\epsilon}}\right)\bm{1}\right) \odot \left( H - \vec{m}^\intercal\bm{1} - \hat{Z} \odot (\vec{v}^\intercal\bm{1}) \right)
\end{align}
We include Eq.(\ref{eq:back_ln}) in the backward process of the feedforward block, defined as $G_Z:=\frac{\partial L}{\partial Z}$. 
Then we have for residual split: 
\[
\frac{\partial L}{\partial X_{attn}^{(\ell)}}\Big|_{\text{res}} = G_Z, 
\quad 
\frac{\partial L}{\partial X_{ffn}^{(\ell)}} = G_Z
\]

\fbox{
\parbox{\textwidth}{%
\begin{enumerate}[leftmargin=*]
\item $H$ is obtained from $G=\grad{X^{(\ell+1)}}$ and plaintext matrix $\bm{1}^\intercal\bm{\gamma}$. 
$\vec{m}$ is obtained by FHE rotations and summations, computing the mean of each row in $H$. 
\item $\hat{Z}$ is pre-stored from the forward pass, so $\vec{v}$ is computing the mean of each row in $H\odot\hat{Z}$. 
\item $\left(\frac{1}{\sqrt{\sigma_0^2+\epsilon}},\frac{1}{\sqrt{\sigma_1^2+\epsilon}},...,\frac{1}{\sqrt{\sigma_{n_1}^2+\epsilon}}\right)$ is pre-stored from the forward pass. 
\end{enumerate}
}
}

The feedforward block also contains $X_{ffn}^{(\ell)} = \gelu(U) W_{fc2} + \vec{b}_{fc2}\text{ where } U = X_{attn}^{(\ell)} W_{fc1} + \vec{b}_{fc1}$.
Combining all parts together: 
\[
\frac{\partial L}{\partial X_{attn}^{(\ell)}} 
= \frac{\partial L}{\partial X_{attn}^{(\ell)}}\Big|_{\text{res}} + \frac{\partial L}{\partial X_{attn}^{(\ell)}}\Big|_{\text{ffn}}
= G_Z + \left( (G_Z W_{fc2}^T) \odot \text{GELU}'(U) \right) W_{fc1}^T
\]
\fbox{
\parbox{\textwidth}{%
\begin{enumerate}[leftmargin=*]
\item $W_{fc1}^\intercal, W_{fc2}^\intercal$ are plaintext matrices. 
The multiplications with the two matrices are ciphertext-plaintext matrix multiplications. 
\item $\text{GELU}(x)$ is approximated by:
$\text{GELU}(x)\approx \frac{1}{2} x \left(1 + \tanh\!\left(p(x)\right)\right)$
where $p(x):=\sqrt{\frac{2}{\pi}}\left(x + 0.044715\,x^3\right)$.
Then 
\begin{align*}
\text{GELU}'(x)
&= \frac{1}{2}\left(1 + \tanh(p(x))\right)
+ \frac{1}{2} x \left(1 - \tanh^2(p(x))\right)
\sqrt{\frac{2}{\pi}} \left(1 + 0.134145\,x^2\right)~.
\end{align*}
Therefore, given $x$ and $\tanh(p(x))$ pre-stored from the forward pass, 
we can obtain $\text{GELU}'(U)$ by FHE polynomial evaluations. 
\end{enumerate}
}
}

\subsubsection{Attention block}
The first step is $\layernorm$. 
The residual split:
$
\frac{\partial L}{\partial X^{(\ell)}}\Big|_{\text{res}}=
\frac{\partial L}{\partial X_O^{(\ell)}} = \frac{\partial L}{\partial (X^{(\ell)}+X^{(\ell)}_O)}
$
can be obtained by Eq.(\ref{eq:back_ln}). 
Let $X^{(\ell)}_{cat}:=(X_0^{(\ell)}\Vert \dots\Vert X_{11}^{(\ell)})$ and let $\quad G_h := \frac{\partial L}{\partial X_h^{(\ell)}}$: 
\[
\frac{\partial L}{\partial X^{(\ell)}_{cat}}=\frac{\partial L}{\partial X_O^{(\ell)}}(W_O^{(\ell)})^\intercal=
\left(\frac{\partial L}{\partial X_0^{(\ell)}}\Vert \frac{\partial L}{\partial X_1^{(\ell)}}\Vert ... \Vert \frac{\partial L}{\partial X_{11}^{(\ell)}}\right)=
(G_0\Vert G_1\Vert ...\Vert G_{11})~.
\]
Then we can obtain the gradients $G_h$ for each $h\in[12]$.
\begin{align*} 
&\textbf{1. }\text{From} X_h^{(\ell)} = U_h^{(\ell)} V_h^{(\ell)}: \quad
\frac{\partial L}{\partial U_h^{(\ell)}} = G_h (V_h^{(\ell)})^\intercal, 
\qquad
\frac{\partial L}{\partial V_h^{(\ell)}} = (U_h^{(\ell)})^\intercal G_h ~.
\\
\end{align*}
\fbox{
\parbox{\textwidth}{%
\begin{enumerate}[leftmargin=*]
\item For $\grad{U_h^{(\ell)}}$, We apply again the double BSGS algorithm from \cite{cryptoeprint:2024/1851,ARION} to compute two matrices in (interleaved) column packing, and the result is in (interleaved) diagonal packing and is pre-rotated (similar to Algorithm~\ref{alg:CCMM-V-doubleBSGS}). 
\item For $\grad{V_h^{(\ell)}}$, we first obtain $(U_h^{(\ell)})^\intercal$ by $n_1$ rotations. 
It is also in the (interleaved) diagonal packing and is pre-rotated for the next step. 
Next, we apply the BSGS matrix multiplication between $(U_h^{(\ell)})^\intercal$ and $G_h$. 
Note that matrices are already rotated. 
\end{enumerate}
}
}

\begin{align*}
&\textbf{2. Define: } S_{h}^{(\ell)}[i][j] := \|Q_{h}^{(\ell)}[i][:] - K_{h}^{(\ell)}[j][:]\|_2^2, 
\text{ and } D_{h}^{(\ell)}[i][j] := \frac{\partial L}{\partial S_{h}^{(\ell)}[i][j]} 
= \frac{\partial L}{\partial U_{h}^{(\ell)}[i][j]} \cdot \sigma'(S_{h}^{(\ell)}[i][j])~. \\
&\textbf{3. Gradients w.r.t. } Q_{h}^{(\ell)}[i][:], K_{h}^{(\ell)}[j][:]: \\
&\frac{\partial L}{\partial Q_{h}^{(\ell)}[i][:]} 
= \sum_j 2 D_{h}^{(\ell)}[i][j] \bigl(Q_{h}^{(\ell)}[i][:] - K_{h}^{(\ell)}[j][:]\bigr), 
\frac{\partial L}{\partial K_{h}^{(\ell)}[j][:]} 
= -\sum_i 2 D_{h}^{(\ell)}[i][j] \bigl(Q_{h}^{(\ell)}[i][:] - K_{h}^{(\ell)}[j][:]\bigr)~. \\
&\textbf{4. Matrix form:} \\
&\frac{\partial L}{\partial Q_h^{(\ell)}}
= 2\Bigl( \mathrm{diag}(D_h^{(\ell)} \bm{1})\, Q_h^{(\ell)} - D_h^{(\ell)} K_h^{(\ell)} \Bigr), 
\frac{\partial L}{\partial K_h^{(\ell)}}
= 2\Bigl( \mathrm{diag}((D_h^{(\ell)})^\intercal \bm{1})\, K_h^{(\ell)} - (D_h^{(\ell)})^\intercal Q_h^{(\ell)} \Bigr)~. \\
\end{align*}
\fbox{
\parbox{\textwidth}{%
\begin{enumerate}[leftmargin=*]
\item All $\sigma'(S_{h}^{(\ell)}[i][j])$ are pre-stored from the forward pass. 
Together with $\grad{U_h^{(\ell)}}$ from the previous step, we can obtain $D_h^{(\ell)},(D_h^{(\ell)})^\intercal$ and $\text{diag}(D_h^{(\ell)}\bm{1}),\text{diag}((D_h^{(\ell)})^\intercal\bm{1})$. 
\item $Q_h^{(\ell)}$ and $K_h^{(\ell)}$ are pre-stored from the forward pass. Combining together we can obtain $\grad{Q_h^{(\ell)}}$ and $\grad{K_h^{(\ell)}}$ for each $h\in[12]$. 
\end{enumerate}
}
}

\begin{align*}
&\textbf{5. Gradients w.r.t. head $h$:} 
\frac{\partial L}{\partial W_{Q,h}^{(\ell)}} 
= (X^{(\ell)})^\intercal \frac{\partial L}{\partial Q_h^{(\ell)}}, 
\frac{\partial L}{\partial W_{K,h}^{(\ell)}} 
= (X^{(\ell)})^\intercal \frac{\partial L}{\partial K_h^{(\ell)}}, 
\frac{\partial L}{\partial W_{V,h}^{(\ell)}} 
= (X^{(\ell)})^\intercal \frac{\partial L}{\partial V_h^{(\ell)}} ~.\\
&\textbf{6. Gradients concatenated all heads $h\in[12]$:}\\
&\frac{\partial L}{\partial W_{Q}^{(\ell)}} = \left(\frac{\partial L}{\partial W_{Q,0}^{(\ell)}}\Vert\frac{\partial L}{\partial W_{Q,1}^{(\ell)}} \Vert...\Vert \frac{\partial L}{\partial W_{Q,11}^{(\ell)}}  \right) 
= (X^{(\ell)})^\intercal \left( \frac{\partial L}{\partial Q_0^{(\ell)}}\Vert\frac{\partial L}{\partial Q_1^{(\ell)}} \Vert...\Vert \frac{\partial L}{\partial Q_{11}^{(\ell)}} \right)
= (X^{(\ell)})^\intercal \grad{Q^{(\ell)}}~,\\
&\frac{\partial L}{\partial W_{K}^{(\ell)}} = \left(\frac{\partial L}{\partial W_{K,0}^{(\ell)}}\Vert\frac{\partial L}{\partial W_{K,1}^{(\ell)}} \Vert...\Vert \frac{\partial L}{\partial W_{K,11}^{(\ell)}}  \right) 
= (X^{(\ell)})^\intercal \left( \frac{\partial L}{\partial K_0^{(\ell)}}\Vert\frac{\partial L}{\partial K_1^{(\ell)}} \Vert...\Vert \frac{\partial L}{\partial K_{11}^{(\ell)}} \right)
= (X^{(\ell)})^\intercal \grad{K^{(\ell)}}~,\\
&\frac{\partial L}{\partial W_{V}^{(\ell)}} = \left(\frac{\partial L}{\partial W_{V,0}^{(\ell)}}\Vert\frac{\partial L}{\partial W_{V,1}^{(\ell)}} \Vert...\Vert \frac{\partial L}{\partial W_{V,11}^{(\ell)}}  \right) 
\ =\ (X^{(\ell)})^\intercal \left( \frac{\partial L}{\partial V_0^{(\ell)}}\Vert\frac{\partial L}{\partial V_1^{(\ell)}} \Vert...\Vert \frac{\partial L}{\partial V_{11}^{(\ell)}} \right)
= (X^{(\ell)})^\intercal \grad{V^{(\ell)}}~.\\
&\textbf{7. Gradient w.r.t. input } X^{(\ell)}: \\
&\frac{\partial L}{\partial X^{(\ell)}}
= \frac{\partial L}{\partial X^{(\ell)}}\Big|_{\text{res}}
+ \frac{\partial L}{\partial Q^{(\ell)}} (W_{Q,frozen}^{(\ell)} + A_{Q}^{(\ell)}B_{Q}^{(\ell)}) ^\intercal
+ \frac{\partial L}{\partial K^{(\ell)}} (W_{K,frozen}^{(\ell)} + A_{K}^{(\ell)}B_{K}^{(\ell)}) ^\intercal\\
&\qquad \qquad + \frac{\partial L}{\partial V^{(\ell)}} (W_{V,frozen}^{(\ell)} + A_{V}^{(\ell)}B_{V}^{(\ell)}) ^\intercal~. 
\end{align*}
\fbox{
\parbox{\textwidth}{%
\begin{itemize}[leftmargin=*]
\item Note that (1) each of $\grad{Q^{(\ell)}},\grad{K^{(\ell)}},\grad{V^{(\ell)}}$ has the same size and the same packing method as $\grad{X^{(\ell)}}$; and (2) $(W_{*,frozen}^{(\ell)}+A_{*}B_{*})^\intercal=(W_{*,frozen}^{(\ell)})^\intercal+(B_{*})^\intercal(A_{*})^\intercal$, where $*=Q,K,V$. 
We can use the same matrix multiplication code block from the forward pass for the computation of $\frac{\partial L}{\partial *^{(\ell)}} (W_{*,frozen}^{(\ell)} + A_{*}^{(\ell)}B_{*}^{(\ell)}) ^\intercal$, where $*=Q,K,V$. 
\end{itemize}
}
}
\begin{align*}
&\textbf{8. Gradient w.r.t. LoRA matrices}: 
\frac{\partial L}{\partial A_*^{(\ell)}} 
=\frac{\partial L}{\partial W_*^{(\ell)}}(B_*^{(\ell)})^\intercal, \ 
\frac{\partial L}{\partial B_*^{(\ell)}} 
= (A_Q^{(\ell)})^\intercal \frac{\partial L}{\partial W_*^{(\ell)}},\ 
* = Q,K,V~.
\end{align*}
\fbox{
\parbox{\textwidth}{%
\begin{enumerate}[leftmargin=*]
\item Note that $\grad{W_*^{(\ell)}}=(X^{(\ell)})^\intercal\grad{*^{(\ell)}}$, we have
$\grad{B_*^{(\ell)}}=(X^{(\ell)}A_*^{(\ell)})^\intercal\grad{W_*^{(\ell)}}$. 
$X^{(\ell)}A_*^{(\ell)}$ is pre-stored from the forward pass. 
\item In $\grad{A_*^{(\ell)}}=(X^{(\ell)})^\intercal\grad{*^{(\ell)}}(B_*^{(\ell)})^\intercal$, we first compute the latter multiplication $\grad{*^{(\ell)}}(B_*^{(\ell)})^\intercal$ to reduce the number of ctct multiplications. 
\end{enumerate}
}
}

\subsection{Update model weights}
\label{app:update}
In this section, we describe how to compactly pack the gradients of all updatable parameters from Section~\ref{app:backward} into as few ciphertexts as possible. 
This is motivated by the fact that the number of bootstrapping operations required in the optimizer and during model weight updates is $\Theta(L)$, where L denotes the number of ciphertexts used to store these gradients.
\paragraph{Average. }
For each updatable matrix, the gradients have to be averaged across a batch of size $b$.
These can be obtained by $O(\log b)$ FHE rotations and additions. 
\paragraph{Packing compactly. }
To use the minimal number of ciphertexts, we can pack all the updatable weights compactly. 
In summary, all the updatable weights are matrices:
\begin{itemize}
    \item $A_Q^{(\ell)},A_K^{(\ell)},A_V^{(\ell)}\in \R^{n_2\times n_{\text{LoRA}}}$, $\ell = 0 , 1$. 
    \item $B_Q^{(\ell)},B_K^{(\ell)},B_V^{(\ell)}\in \R^{n_2\times n_{\text{LoRA}}}$, $\ell = 0 , 1$. 
    \item $W_{c1}\in \R^{n_2\times n_2}, W_{c2}\in \R^{n_2\times c}$. 
\end{itemize}

For example, given $n_2=768,n_{\text{LoRA}}=2$ and $c=2$, all the LoRA matrices and $W_{c2}$ have totally $6\times 768 \times 2 + 768\times 2= 10752<N$. So we can pack all these matrices compactly in one CKKS ciphertext, by masking, rotation and addition. $9$ ciphertexts are required to pack $W_{c1}$ compactly. 

In \textbf{Average} step and \textbf{Packing} step, it will cost totally 1 multiplicative level to pack all these matrices compactly. 
Therefore, we keep $1$ multiplicative level in Figure~\ref{fig:fullflow}. 

\paragraph{Optimizer. }
We apply the AdamW \citep{loshchilov2017decoupled} optimizer for better performance as suggested in \cite{zhang2024transformers}, and use \cite{rhoencryption}'s FHE-friendly modification ($\sqrt{\hat{v}_t}+\epsilon\rightarrow \sqrt{\hat{v}_t+\epsilon})$. 
We include it in Algorithm~\ref{alg:AdamW-HE} for completeness. 
\begin{algorithm}[htbp]
    \caption{AdamW-HE \citep{rhoencryption}}
    \label{alg:AdamW-HE}
    \begin{algorithmic}[1]  
    \STATE \textbf{Parameters:} $\gamma, \beta_1, \beta_2, \epsilon>0, \lambda$. 
    \STATE \textbf{Initialize:} $m_0=0$, $v_0=0$, $t=1$.  
    \STATE In round $t$:
        \STATE \hspace{\algorithmicindent} Compute gradients of all updatable matrices and pack them compactly $g_t$ in $\alpha$ ciphertexts.  
        \STATE \hspace{\algorithmicindent} Let $\theta_t$ be all the updatable matrices (in compact form). 
        \STATE \hspace{\algorithmicindent} Apply bootstrapping on $g_t$. 
        \COMMENT{to the highest multiplicative level}
        \STATE \hspace{\algorithmicindent} $\theta_t:=(1-\gamma\lambda)\theta_{t-1}$. 
        \STATE \hspace{\algorithmicindent} $m_t:=\beta_1m_{t-1}+(1-\beta_1)g_t$, $v_t:=\beta v_{t-1}+(1-\beta_2)(g_t\odot g_t)$. 
        \STATE \hspace{\algorithmicindent} Apply bootstrapping on $m_t,v_t$ if the remaining multiplicative level is not enough. 
        \STATE \hspace{\algorithmicindent} Store $m_t,v_t$ for the next iteration. 
        \STATE \hspace{\algorithmicindent} $\hat{m}_t:=m_t/(1-\beta_1^t)$, $\hat{v}_t:=v_t/(1-\beta_2^t)$. 
        \STATE \hspace{\algorithmicindent} Update matrices: $\theta_t:=\theta_t-\gamma\hat{m}_t/(\sqrt{\hat{v}_t+\epsilon})$. 
        \STATE \hspace{\algorithmicindent} Apply bootstrapping on $\theta_t$.

    \end{algorithmic}
\end{algorithm}

\paragraph{Repacking and updating. }
After homomorphically evaluate AdamW-HE and obtain $\theta_t$, we repack all the updatable matrices from $\theta_t$ and derive all the updated matrices in our required form (e.g., the repeated row packing for LoRA matrices). 

\cite{rhoencryption} uses 1 ciphertext for each updatable matrix except $W_{c1}$. 
By compactly packing the updatable matrices, the number of required ciphertexts in our scheme is reduced from 22 (if we follow \cite{rhoencryption}) to 10, which dramatically reduce the cost of bootstrapping. 

\section{FHE Matrix Multiplications from \cite{rhoencryption}}
\subsection{CPMM and CCMM in \cite{rhoencryption}}
\label{app:MM_iclr25}
In this section, we summarize the computational complexity of CPMM and CCMM in \cite{rhoencryption}.
In \cite{rhoencryption}, CPMM and CCMM involve (1) split the matrices as several block matrices which can be packed in a single ciphertext; (2) call JKLS algorithm~\cite{jiang2018secure}. 
The number of each HE operation in JKLS algorithm is listed in Table~\ref{tab:JKLS}. \footnote{The number of each HE operation of CCMM is taken from Table 2 in JKLS~\cite{jiang2018secure}. The number of each HE operation of CPMM is derived by considering $B$ as a plaintext matrix and adjusting the number of each HE operation in step 1-2, step 2 and step 3 in Table 2. }

\begin{table}[htbp]
    \centering
    \caption{The number of each HE operation for one $d \times d$ homomorphic matrix multiplication required by JKLS algorithm~\cite{jiang2018secure}}
    \label{tab:JKLS}
    \begin{tabular}{ccccc}
    \toprule
         Operations & Rot & ct-ct Mult & ct-pt Mult & Level \\ \midrule
         CPMM in \cite{jiang2018secure} & $2d+3\sqrt{d}$ & 0 & $4d$ & 3 \\ \hline 
         CCMM in \cite{jiang2018secure} & $3d+5\sqrt{d}$ & $d$ & $4d$ &  3 \\ 
    \bottomrule
    \end{tabular}
\end{table}

For example, as mentioned in Section D.1 of~\cite{rhoencryption}, when performing homomorphic matrix multiplication of size $128 \times 768$ by $768 \times 768$, it will first split the matrices as three and eighteen block matrices, respectively, then go through each block matrix multiplication to get a corresponding result. 
As the same left matrices are used several times to multiply by the right block matrices, it can compute and save the left matrices, so the total number of JKLS algorithm calls in this example is 21, instead of 36. 

In Table~\ref{tab:CPMM-iclr25}, we summarize the required number of each HE operation for $
\enc(X)W$ in \cite{rhoencryption}. 
$X$ is split into three block matrices and 
$W$ is split into eighteen block matrices, 
with block shapes $128 \times 256$ and $256 \times 128$, respectively. 
With pre-computation and reuse for the matrix $X$, the number of rectangular JKLS calls is reduced from $36$ to $21$ (see Section D.1 in \cite{rhoencryption}). 
Each call has input matrices with shape $l \times d$ by $d \times l$, where $l=128$ and $d=256$.

\begin{table}[htbp]
    \centering
    \caption{The number of each HE operation for CPMM ($\enc(X)W$) for a single input in~\cite{rhoencryption}.}
    \label{tab:CPMM-iclr25}
    \resizebox{\textwidth}{!}{
    \begin{tabular}{cccccc}
    \toprule
         Operations & Rot & ct-ct Mult &Relinearization & ct-pt Mult & Level \\ 
    \midrule
         CPMM in \cite{rhoencryption}
         & $21\left(2l+3\sqrt{d}+\log_2(d/l)\right)$
         & $0$ 
         & $0$
         & $21(3d+2l)$
         & $3$ \\ 
    \bottomrule
    \end{tabular}
    }
\end{table}

In Table~\ref{tab:V-CCMM-iclr25}, we summarize the required number of each HE operation for $\{\enc(\sigma(S_h)) \enc(V_h) \}_{h \in [H]}$ in \cite{rhoencryption}, where $S_h \in \R^{n_1 \times n_1}$, and $V_h \in \R^{n_1 \times (n_3)}$. 
Using \cite{rhoencryption}'s notation, let $m:=n_1$ and $g:=H/2$. 
In the BERT-style setting, $m=128$ and $g=6$. The value multiplication
uses $g$ packed groups. Each packed group consists of complex packing, one
packed matrix multiplication, and repacking; the whole value path additionally
uses $g/2$ masking and $g/2$ left/right collection
additions.

\begin{table}[htbp]
    \centering
    \caption{The number of each HE operation for value CCMM
    ($\{\enc(\sigma(S_h))\enc(V_h)\}_{h\in[H]}$) for a single input
    in~\cite{rhoencryption}, following the GitHub implementation.}
    \label{tab:V-CCMM-iclr25}
    \resizebox{\textwidth}{!}{
    \begin{tabular}{cccccc}
    \toprule
         Operations & Rot & ct-ct Mult & Relinearization & ct-pt Mult & Level \\
    \midrule
         Value CCMM in \cite{rhoencryption}
         & $g\left(\frac{3m}{2}+4\lceil\sqrt{m}\rceil+2\right)$
         & $\frac{mg}{2}$ 
         & $2g$
         & $g(3m+1)$
         & $3$ \\
    \bottomrule
    \end{tabular}
    }
\end{table}

Furthermore, \cite{rhoencryption} proposed an optimized LoRA CCMMs (Algorithm 5 in \cite{rhoencryption}) to reduce the required number of HE operations and multiplicative levels. 
The required number of each HE operation for $(\enc(X)\enc(A))\enc(B)$, where $X \in \R^{128 \times 768}$, $A \in \R^{768 \times n_{\mathrm{LoRA}}}$, and $B \in \R^{n_{\mathrm{LoRA}} \times 768}$ ($n_{\mathrm{LoRA}}$ is the LoRA rank), is listed in Table~\ref{tab:lora-iclr25}. \footnote{Numbers in the table is taken from Table 9 in \cite{rhoencryption}. }

\begin{table}[htbp]
    \centering
    \caption{The number of each HE operation for LoRA-style CCMMs for a single input in~\cite{rhoencryption}.}
    \label{tab:lora-iclr25}
    \begin{tabular}{cccccc}
    \toprule
         Operations & Rot & ct-ct Mult & Relinearization & ct-pt Mult & Level \\ \midrule
         LoRA-style CCMMs & $40n_{\mathrm{LoRA}}-1$ & $6n_{\mathrm{LoRA}}$ & $6n_{\mathrm{LoRA}}$ & $4n_{\mathrm{LoRA}}$ & 3 \\ 
    \bottomrule
    \end{tabular}
    
\end{table}

\subsection{FHE squared $l_2$-norm computation in \cite{rhoencryption}}
\label{app:kernel_iclr25}
In this section we include the FHE algorithms in \cite{rhoencryption} for computing the square matrix $S_h$ given input as encrypted $Q_h,K_h$ in head $h$. 
More concretely, in each head $h$ they compute
\[
U_h[i][j]:= \exp \left(-\frac{1}{2\sqrt{n}} \Vert Q_h[i][:]-K_h[j][:]\Vert_2^2 \right)~,
\]
where $Q,K\in\R^{n_1\times n_3}$, $Q_h[i][:],K_h[j][:]\in\R^{n_3}$ represent the $i$-th row of $Q$ and the $j$-th row of $K$ respectively. 
\cite{rhoencryption} did not provide the algorithm of computing $\exp \left(-\frac{1}{2\sqrt{n}} \Vert Q_h[i][:]-K_h[j][:]\Vert_2^2 \right)$ in their paper. 
This section is following their GitHub implementation. 

\cite{rhoencryption} packs multiple heads together in computing the Gaussian kernel. 
Let the query and key matrices be $\bar{Q},\bar{K}$ which have shape $n_1\times n_2$. 
Therefore, packing $\bar{Q}$ or $\bar{K}$ will require $s:=\lceil \frac{n_1n_2}{N} \rceil$ ciphertexts.
More concretely, with the assumption $n_1|N$, \cite{rhoencryption} first divide $\bar{Q}$ into $s:=\lceil \frac{n_1n_2}{N} \rceil$ sub-matrices, each of which has size $n_1\times \frac{N}{n_1}$. 
Then we can write $\bar{Q}:=(\bar{Q}_0\Vert \bar{Q}_1\Vert...\Vert \bar{Q}_{s-1})$, where $\bar{Q}_i$ has size $n_1\times \frac{N}{n_1}$ and $i\in[s]$. 
The notations are similar for $\bar{K}:=(\bar{K}_0\Vert \bar{K}_1\Vert...\Vert \bar{K}_{s-1})$. 
\cite{rhoencryption} also assume that $2n_3|\frac{N}{n_1}$ so that each head will be packed and can be computed within the same ciphertext. 

\paragraph{The matrix packing of \cite{rhoencryption}.}
\cite{rhoencryption} follows the \textit{row-wise packing} from \cite{jiang2018secure}. 
In this packing, each row of the matrix is \textit{sequentially} packed into a single long vector with size $N/2$.
When the number of entries of the matrix exceeds the number of ciphertext slots, multiple ciphertexts are used. 
To pack $\bar{Q}$, \cite{rhoencryption} uses row-wise packing for each $\bar{Q}_i,\bar{K}_i,i\in[s]$. 
Each head has $n_3$ columns and the number of heads is $H=\frac{n_2}{n_3}$. 
Let $t:=\frac{N}{2n_1n_3}$. Each matrix $\bar{Q}_i$ or $\bar{K}_i$ will have $t$ heads packed. 
For any $i\in[s]$, $Q_i,K_i$ contain $2t$ heads each:
    \[
    \bar{Q}_i=(Q_{2it}\Vert Q_{2it+1}\Vert...\Vert Q_{2it+2t-1})
    \]
    \[
    \bar{K}_i=(K_{2it}\Vert K_{2it+1}\Vert...\Vert K_{2it+2t-1})
    \]
Please refer to Algorithm~\ref{alg:kernel-iclr25} for the detailed algorithm of \cite{rhoencryption}. 
All the matrices are in the encrypted form. 
For the sake of simplicity, we avoid using the notation $\enc(\cdot)$ in Algorithm~\ref{alg:kernel-iclr25}. 
Their computations are based on the decomposition:
\[
\Vert Q_h[i][:]-K_h[j][:]\Vert_2^2=\Vert Q_h[i][:]\Vert_2^2+\Vert K_h[j][:]\Vert_2^2-2Q_h[i][:]K^\intercal_h[j][:]~.
\]

\begin{algorithm}
    \caption{Kernel computation ($\{S_h\}_{h\in[H]}$) in \cite{rhoencryption}.}
    \label{alg:kernel-iclr25}
    \begin{algorithmic}[1]  
        \REQUIRE Encrypted row-packed matrices $\{\bar{Q}_i,\bar{K}_i\}_{i\in[s]}$. 
        \ENSURE Encrypted row-packed matrices $S$. 
    
    \STATE For any $i\in[s]$, split $Q_i$ and $K_i$ into two parts: 
    \COMMENT{Cost $1$ level}
    \[
    \bar{Q}_i={Q}_{i,L}+{Q}_{i,R}:=
    (Q_{2it}\Vert \bm{O} \Vert...\Vert Q_{2it+2t-2} \Vert \bm{O} )+
    (\bm{O} \Vert Q_{2it+1}\Vert...\Vert \bm{O}\Vert Q_{2it+2t-1})
    \]
    \[
    \bar{K}_i={K}_{i,L}+{K}_{i,R}:=
    (K_{2it}\Vert \bm{O} \Vert...\Vert K_{2it+2t-2} \Vert \bm{O} )+
    (\bm{O} \Vert K_{2it+1}\Vert...\Vert \bm{O}\Vert K_{2it+2t-1})
    \]
    \STATE \textbf{Step I}: compute the square $l_2$-norm for each row in $Q_{i,L}, Q_{i,R}, K_{i,L},K_{i,R},i\in[s]$: 
    \STATE Each head has shape $n_1\times n_3$. 
    \STATE $Q_{i,L,sum}:=Q_{i,L}\odot Q_{i,L}$, $Q_{i,R,sum}:=Q_{i,R}\odot Q_{i,R}$. 
    \COMMENT{Rotate and Sum. }
    \FOR{$col\in\{1,2,4,...,n_3/2\}$}
    \STATE $tmp:=$ Left rotate the ciphertext of $Q_{i,L,sum}$ for $col$ slots. 
    \STATE $Q_{i,L,sum}+=tmp$. 
    \STATE $tmp:=$ Right rotate the ciphertext of $Q_{i,R,sum}$ for $col$ slots. 
    \STATE $Q_{i,R,sum}+=tmp$. 
    \ENDFOR
    \STATE Masking:
    \STATE \hspace{\algorithmicindent} Set $Q_{i,L,sum}$ to $0$ except the $2n_3l$-th column, $l\in[t]$. 
    \COMMENT{Cost 1 level}
    \STATE \hspace{\algorithmicindent} Set $Q_{i,R,sum}$ to $0$ except the $(2n_3(l+1)-1)$-th column, $l\in[t]$.
    \COMMENT{Cost 1 level}
    
    \FOR{$col\in\{1,2,4,...,n_3\}$}
    \STATE $tmp:=$ Right rotate the ciphertext of $Q_{i,L,sum}$ for $col$ slots. 
    \STATE $Q_{i,L,sum}+=tmp$. 
    \STATE $tmp:=$ Left rotate the ciphertext of $Q_{i,R,sum}$ for $col$ slots. 
    \STATE $Q_{i,R,sum}+=tmp$. 
    \ENDFOR
    \STATE Compute $K_{i,L,sum},K_{i,R,sum}$ using the same steps. 
    \STATE Transpose $K_{i,L,sum},K_{i,R,sum}$ by \cite{jiang2018secure}.
    \STATE \textbf{Step II}: Compute $Q_{i,L}K_{i,L}^T$: 
    \STATE Repack the columns in $Q_{i,L}$ in complex number field $\C$:
    \STATE \hspace{\algorithmicindent} $Q_{i,L,\C}:=Q_{i,L}$. 
    \FOR{$j\in[\frac{N}{n_1}]$}
    \STATE \hspace{\algorithmicindent} The $2j$-th column of $Q_{i,L,\C}:= \textrm{the } 2j\textrm{-th column of } Q_{i,L} + \big(\textrm{the } (2j+1)\textrm{-th column of } Q_{i,L}\big)\sqrt{-1}$.
    \COMMENT{Cost 1 level}
    \STATE \hspace{\algorithmicindent} The $2j+1$-th column of $Q_{i,L,\C}:=$ the $2j$-th column of $Q_{i,L,\C}$. 
    \ENDFOR
    \STATE Repack the columns in $K_{i,L}$ in complex number field $\C$:
    \STATE \hspace{\algorithmicindent} $K_{i,L,\C}:=K_{i,L}$. 
    \FOR{$j\in[\frac{N}{n_1}]$}
    \STATE \hspace{\algorithmicindent} The $2j$-th column of $K_{i,L,\C}:= \textrm{the } 2j\textrm{-th column of } K_{i,L} - \big(\textrm{the } (2j+1)\textrm{-th column of } K_{i,L}\big)\sqrt{-1}$.
    \STATE \hspace{\algorithmicindent} The $2j+1$-th column of $K_{i,L,\C}:=$ the $2j$-th column of $K_{i,L,\C}$. 
    \ENDFOR
    \STATE Transpose $K_{i,L,\C}$ by \cite{jiang2018secure}. 
    \STATE A modified matrix multiplication from \cite{jiang2018secure} to compute $Q_{i,L,\C}K_{i,L,\C}^T$. 
    \STATE $Q_{i,L}K_{i,L}^T:=\textrm{Re}(Q_{i,L,\C}K_{i,L,\C}^T)$. 
    \COMMENT{Cost 1 level}
    \STATE Similarly, compute $Q_{i,R,\C}K_{i,R,\C}^T$. 
    \STATE Return $2s$ ciphertexts: 
    $$Q_{i,L,sum}+K_{i,L,sum}-2\times Q_{i,L}K_{i,L}^T, Q_{i,R,sum}+K_{i,R,sum}-2\times Q_{i,R}K_{i,R}^T, \quad \forall i\in[s]~.$$ 
\end{algorithmic}
\end{algorithm}

\paragraph{Analysis. }
For each $i\in[s]$: 
\begin{itemize}
    \item Generating $\{Q_{i,L,sum},Q_{i,R,sum},K_{i,L,sum},K_{i,R,sum} \}$ costs $4(\log n_3+\log(n_3/2))$ rotations, 4 ct-ct multiplications and 8 ct-pt multiplications. 
    \item Generating $\{Q_{i,L,\C},Q_{i,R,\C},K_{i,L,\C},K_{i,R,\C} \}$ costs 8 rotations, 0 ct-ct multiplications and 12 ct-pt multiplications. 
    \item 4 matrix transposition are required, each costs $2\sqrt{n_1+n_3}-1$ rotations and $n_1+n_3-1$ ct-pt multiplications. 
    \item 2 matrix multiplications to generate $Q_{i,L,\C}K_{i,L,\C}^T,Q_{i,R,\C}K_{i,R,\C}^T$, each costs $\frac{3n_3}{2}-3+4\sqrt{n_3}$ rotations,
    $n_3/2$ ct-ct multiplications (with lazy relinearization) and 
    $3n_3-5$ ct-pt multiplications. 
    \item Extracting $Q_{i,L}K_{i,L}^T,Q_{i,R}K_{i,R}^T$ costs 2 conjugations, which has similar cost as HE rotations.
\end{itemize}

In summary, the required number of each HE operation for kernel computation is listed in Table~\ref{tab:kernel-iclr25}. 

\begin{table}[htbp]
    \centering
    \caption{The number of each HE operation for Kernel CCMM for a single input
    in~\cite{rhoencryption}.}
    \label{tab:kernel-iclr25}
    \resizebox{\textwidth}{!}{
    \begin{tabular}{cccccc}
    \toprule
         Operations & Rot & ct-ct Mult & Relinearization & ct-pt Mult & Level \\
    \midrule
         Kernel CCMM in \cite{rhoencryption}
         &
         \begin{tabular}{@{}c@{}}
         $s\big[
         4\log_2(n_3^2/2)+8\lceil\sqrt{n_1+n_3}\rceil$ \\
         $+{3n_3}
         +8\sqrt{n_3}
         \big]$
         \end{tabular}
         & $s\left(4+{n_3}\right)$ 
         & $8s$
         &
         $s(4n_1+10n_3+6)$
         &
         $3$ \\
    \bottomrule
    \end{tabular}
    }
\end{table}

\section{Details of non-linear evaluations}
\label{app-nonlinear}

\subsection{Layernorm evaluations}
\label{app-layernorm}

\begin{figure}
    \centering
    \includegraphics[width=0.9\linewidth]{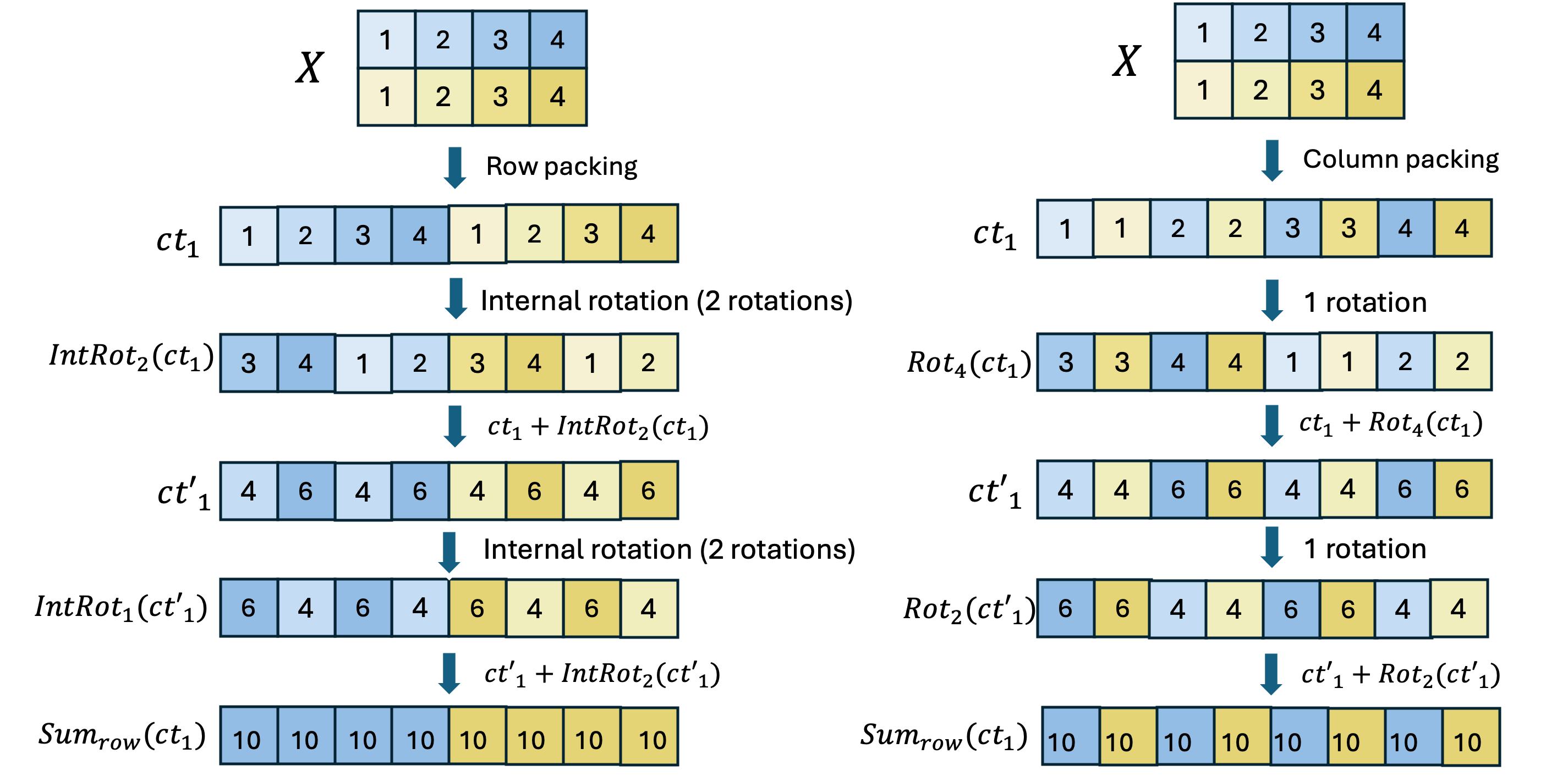}
    \caption{Toy example of summing elements within each row under row packing and column packing }
    \label{fig:introt_sum}
\end{figure}

Our Layernorm evaluation algorithm takes a set of ciphertexts in flexible interleaved column packing that encrypts $\{X^i\}_{i \in [b]}$, where $b$ is the batch size as input, and outputs a set of ciphertexts with the same packing format. 
Figure~\ref{fig:introt_sum} illustrates a toy example of summing elements within each row under row packing~\citep{rhoencryption} and our packing scheme. \footnote{For simplicity, the example considers a single input and omits interleaved batching. As discussed in our matrix operations, interleaved batching can further reduce the number of required rotations.}
In~\cite{rhoencryption}, a generic row-packing strategy is applied across transformer operations, which requires ``internal rotations'' within each row for such evaluations.
Each internal rotation is implemented via two ciphertext rotations, leading to a $2\times$ rotation cost compared to our column-based packing, which avoids such internal rotations. 

For higher precision of the inverse square root approximation, we perform the following transformation to $\mathrm{LayerNorm}$ for vector $\Vec{x}$:

\begin{align*}
    \mathrm{LayerNorm}(\Vec{x})[i] &= \gamma_i \frac{x_i-\mathrm{mean}(\Vec{x})}{\sqrt{\mathrm{var}(\Vec{x})}}+\beta_i \\
    &= \frac{\gamma_i}{\sqrt{c}} \frac{x_i-\mathrm{mean}(\Vec{x})}{\sqrt{\mathrm{var}(\Vec{x})/c}} + \beta_i
\end{align*}

where $c$ (as in \cite{zhang2024secure,rhoencryption} is a constant to ensure the input to the inverse square root is in an appropriate range. 

As discussed in~\cite{cryptoeprint:2025/991}, achieving sufficiently accurate Layernorm typically requires around 20 multiplicative levels, which exceeds our setting where only 12 levels remain after bootstrapping.
To address this, we introduce two additional bootstrapping steps: one applied to $\mathrm{var}(\Vec{x})/c$ before evaluating the inverse square root, and another within the inverse square root computation itself, ensuring sufficient multiplicative depth for accurate evaluation.
Notably, both bootstrapping operations are performed on a single aggregated ciphertext (after summation across inputs), rather than on all input ciphertexts. As a result, each Layernorm evaluation incurs a runtime of less than 2 seconds.

\subsection{Functional bootstrapping for element-wise functions}
\label{app:fb}
RBOOT framework \citep{yang2025rboot} accelerates secure neural network inference by fusing the evaluation of non-linear activation functions like ReLU directly into the CKKS bootstrapping process. 
The core innovation lies in exploiting the trigonometric structure of the modular reduction circuit.  
Normal CKKS bootstrapping usually uses $\sin(\cdot)$ and $\arcsin(\cdot)$ to approximate mod function $x (\text{mod }q_0)$. 

For example in ReLU function, 
the evaluation from RBOOT is integrated into the bootstrapping process using the following relationship for $x \in [-1/4, 1/4] + \mathbb{Z}$:
\begin{equation}
\label{eq:relu}
\text{ReLU}(x (\text{mod }1)) = \frac{1}{4\pi} \arcsin(\sin(2\pi x)) + \frac{1}{8\pi} \arcsin(-\cos(4\pi x)) + \frac{1}{16}
\end{equation}
As $\sin(\cdot)$ and $\cos(\cdot)$ can be obtained in bootstrapping process, 
and $\arcsin(\cdot)$ is approximated by polynomial already, 
we can obtain the result of $\text{ReLU}$ using Eq. (\ref{eq:relu}). 
Reported by \cite{yang2025rboot}, this reduces the multiplicative level consumption of ReLU evaluation in a targeted precision from 14 to 20 levels (typically required by high-degree polynomial approximations) down to approximately 8 levels. 
In general if we want to evaluate function $f(\cdot)$, we can replace $\arcsin(\cdot)$'s polynomial approximation with specialized minimax polynomials that approximate the composition of $f(cdot)$ and $\arcsin(\cdot)$.

\section{Full flow of \textsc{Felid}}
\label{sec:app-fullflow}

\begin{figure}
    \centering
    \includegraphics[width=\linewidth]{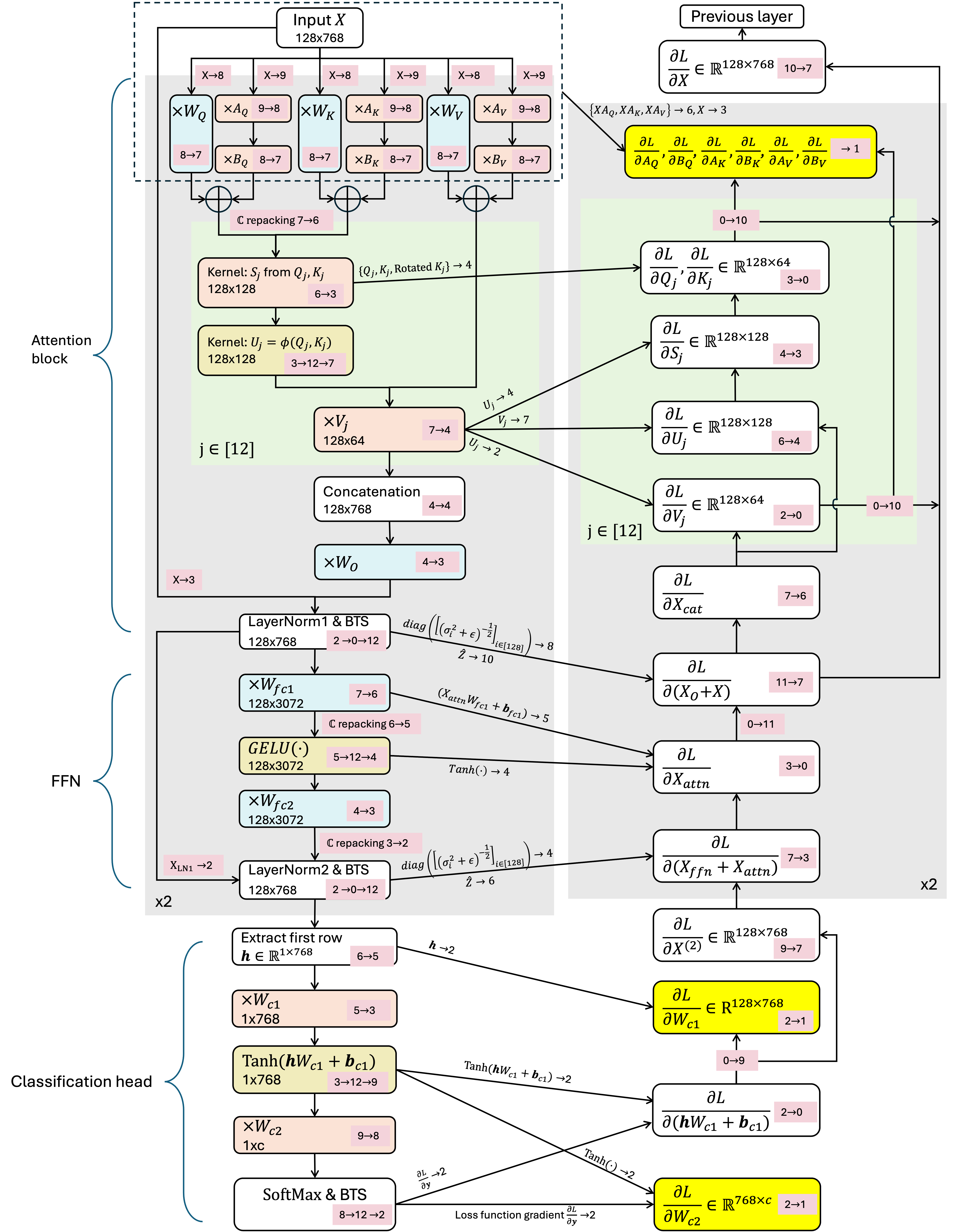}
    \caption{Full flow of our framework. }
    \label{fig:fullflow}
\end{figure}

Figure~\ref{fig:fullflow} illustrates the complete diagram of our evaluation flow to 
fine-tuning a 2-layer BERT-style model. 
The blue boxes are CPMM, while the green boxes are CKKS functional bootstrapping using \cite{yang2025rboot}. 
The changes of multiplicative levels are indicated in each node. 
Finally, all the yellow boxes are the gradients of updatable weights. 
They will be sent to compact packing process and the AdamW-HE optimizer~\citep{rhoencryption}. 

\section{Experimental details}
\label{sec:app-exp}

\subsection{Extended Details about Experimental Setup}
\label{sec:app-exp-setup}

\subsubsection{Continued-pretraining. }
For our kernel-attention models, we initialize from the same pretrained BERT checkpoint and replace softmax attention with either GK or EK. 
Instead of training a kernelized backbone from scratch, we start from the pretrained softmax BERT checkpoint and adapt only the components directly affected by the attention replacement. 
Specifically, we update the Q/K/V projections and kernel-specific parameters, while keeping the embeddings, feed-forward layers, and layer-normalization parameters frozen.
The continued-pretraining objective combines masked language modeling with distillation from the original softmax BERT teacher, using hidden-state mean-squared error and logits KL divergence.
Both our GK and EK models are trained for 500k steps. In our implementation, this continued-pretraining stage takes approximately 30--60 minutes, avoiding the need to train a kernelized backbone from scratch. 

\subsubsection{Justifications}
\label{app:limit}
\paragraph{Scope justification.}
We limit our primary evaluation to a 2-layer BERT-style architecture, matching the model size used in the baseline~\citep{rhoencryption}. 
We do not modify the feedforward blocks or the classification head.\footnote{In contrast,~\cite{rhoencryption} introduces two architectural modifications that we do not adopt: (1) splitting the output of the first feedforward layer into two parts and applying activation to only one part; and (2) replacing the first matrix in the classification head with two dense blocks of LoRA-style shapes with rank 32.}

This scope is architecturally sufficient because a 2-layer model encapsulates all fundamental interactions within transformer architecture, including the all the interactions between multi-head attention block, feedforward block, and sequential layer normalization blocks \citep{cryptoeprint:2025/1200,cryptoeprint:2026/198,rhoencryption}. 
Consequently, our improvements demonstrated at this scale are operationally fundamental and reflect the effectiveness of our operation-level and algorithmic optimizations. 
Furthermore, because transformer layers are evaluated sequentially and refreshed via bootstrapping to a fixed level at the start of each block, the total latency is strictly linear in the number of layers. 
As the SOTA FHE fine-tuning baseline \citep{rhoencryption} already shows that the single-epoch latency reach hundreds of hours for tasks in GLUE, we believe that it is sufficient to demonstrate our optimizations on a 2-layer model. 

\paragraph{Task justification.}
While the SOTA baseline of FHE fine-tuning~\citep{rhoencryption} evaluates six GLUE tasks, 
we limit our primary analysis to 3 tasks: \textbf{STS-B}, \textbf{MRPC}, and \textbf{RTE}, due to the extremely high computational overhead of the other tasks. 
According to Table 3 in~\cite{rhoencryption}, fine-tuning for \textbf{COLA}, \textbf{SST-2}, and \textbf{QNLI} on 8-GPU cluster requires approximately 82.5 hours (for 5 epochs), 130 hours (1 epoch), and 202 hours (1 epoch), respectively. 
Such durations are even more computationally impractical for our single-GPU environment. 

Furthermore, focusing on these three representative tasks is well justified, as prior studies consistently show that FHE introduces negligible numerical error and therefore has minimal impact on downstream performance. 
Frameworks such as \cite{zhang2024secure, cryptoeprint:2024/1881, park2024powerformer} have demonstrated that encrypted results remain closely aligned with plaintext counterparts. 

\textbf{Limitations.} Despite the proposed optimizations, while FHE enables non-interactive, privacy-preserving transformer fine-tuning, it remains computationally intensive and typically requires high-performance hardware (e.g., multiple GPUs) to mitigate latency, especially for large model and datasets.

\subsection{PhantomFHE vs. HEaaN in~\cite{rhoencryption}}
\label{app-lib-exp}
The implementation in~\cite{rhoencryption} is based on the proprietary (non-open-sourced) HEaaN library and evaluated on an NVIDIA GeForce RTX 4090 GPU, whereas our implementation is built on PhantomFHE and runs on an NVIDIA H200 GPU. 

To provide a more controlled comparison, in Table~\ref{tab:micro-lib}, we benchmark the primitive operations of the CKKS scheme in both libraries under same parameters.\footnote{The HEaaN results are taken from Table 5 of~\cite{rhoencryption}. $\mathrm{BTS}$, $\mathrm{ExtBTS}$, and $\mathrm{FuncBTS}$ denote bootstrapping, extended bootstrapping (only in HEaaN), and functional bootstrapping (only in our implementation), respectively.}

\begin{table}[h]
    \caption{Runtime (ms) of basic operations in CKKS with parameters: degree of polynomial ring = $2^{16}$, number of slots = $2^{15}$, multiplicative level before bootstrapping = $9$, security parameter = 128-bit. The result of HEaaN reported by \cite{rhoencryption}.}
    \label{tab:micro-lib}
    \centering
    \resizebox{\textwidth}{!}{
    \begin{tabular}{c|c|cccccc}
    \toprule
       Library & GPU  & $\add$ & $\plainmult$ & $\mult$ & $\ckksrot$ & $\mathrm{BTS}$/ $\mathrm{ExtBTS}$ &$\mathrm{FuncBTS}$ \\ \hline
       HEaaN & Nvidia GeForce RTX 4090 & 0.02 & 0.1 & 0.6 & 0.49 & 60/137 & - \\ \hline 
       PhantomFHE & NVIDIA H200 & 0.03 & 0.06 & 3.13 & 2.87 &476& 398 \\ \hline
       \multicolumn{2}{c|}{Ratio (PhantomFHE/HEaaN)} & $1.5\times$ & $0.6 \times$ & $5.2 \times$ & $5.8 \times$ & $3.5-7.9\times$ & - \\
       \bottomrule
       
    \end{tabular}
    }
\end{table}


Under same CKKS parameters, PhantomFHE on H200 achieves comparable or faster performance for simple operations (e.g., plaintext multiplication), but incurs significantly higher costs for expensive operations: (1) being about $5-6\times$ slower for ciphertext multiplication and rotations and (2) $3.5-7.9\times$ slower for bootstrapping, compared to HEaaN on RTX 4090.
The difference reflects the effects of hardware and library-level optimization differences, which are not the focus of this work. 
\newpage

\subsection{Breakdown of \textsc{Felid} encrypted fine-tuning}
\label{app:FT-time-exp}
Table~\ref{tab:breakdown-ft-time} lists the runtime of all fine-tuning operations over a batch of input in 1 layer, including forward pass, backward pass and optimizer. 

\begin{table}[htb]
    \caption{Breakdown fine-tuning time (seconds) over a batch of input with batch size = 16.}
    \label{tab:breakdown-ft-time}
    \centering
    \resizebox{\textwidth}{!}{
    \begin{tabular}{cccc}
    \toprule
      Block  & Operation &  Input size & Time (s) \\ \midrule
     \multirow{19}{*}{\makecell[c]{Forward:\\ Transformer\\ Layer}}  & $\times W_Q,W_K,W_V$ &  $(128\times 768) \times (768 \times 768) \times 3$ & 5.8 \\
       & $\times A*,\times B*$, $*=Q,K,V$ & $(128\times 768) \times (768 \times 2) \times (2\times 768)\times 3$ & 0.45 \\
       & LoRA addition & $((128\times 768)+(128 \times 768))\times 3$ & 0.001 \\
       &  Repack($\C \rightarrow \R$) & $(128 \times 64)\times 12 \times 2$ & 0.013 \\
       & Kernel $S$ & $(128 \times 64)\times 12 \times 2$ & 5.97 \\
       & Kernel $U$ (FuncBTS) & $(128 \times 128) \times 12$ & 38.31 \\
       & $\times V$ & $(128 \times 128) \times (128 \times 64) \times 12$ & 4.28 \\
       & Concatenation & $(128 \times 64) \times 12$ & 0.071 \\
       & $\times W_o$ & $(128\times 768) \times (768 \times 768)$ & 2.05 \\
       & Repack($\C \rightarrow \R$) & $128 \times 768$ & 0.005 \\
       & Layernorm & $128 \times 768$ & 1.95 \\
       & Bootstrapping & $128 \times 768$ & 11.45 \\
       & $\times W_{fc1}$ & $(128 \times 768)\times (768 \times 3072)$ & 7.6 \\
       & Repack($\C \rightarrow \R$) & $128 \times 3072$ & 0.023 \\
       & GELU (FuncBTS) & $128 \times 3072$ & 85.75 \\
       & $\times W_{fc2}$ & $(128 \times 3072)\times (3072 \times 768)$ & 7.2 \\
       & Repack($\C \rightarrow \R$) & $128 \times 768$ & 0.025 \\
       & Layernorm & $128 \times 768$ & 1.9 \\
       & Bootstrapping & $128 \times 768$ & 11.53 \\ \midrule 
       
     \multirow{6}{*}{\makecell[c]{Forward:\\ Classification\\ Head}}  & Extract the first row & $128 \times 768$ & 0.02 \\
        & $\times W_{c1}$  & $(1 \times 768) \times (768 \times 768)$ & 0.16 \\
        & Tanh (FuncBTS) & $1 \times 768$ & 0.67 \\
        & $\times W_{c2}$ & $(1 \times 768) \times (768 \times c)$ & 0.008 \\
        & Softmax & $1\times c$ & 1.86 \\
        & Bootstrapping & $1 \times c$ & 1.05 \\ \midrule

       \multirow{4}{*}{\makecell[c]{Backward:\\ Classification\\ Head}} & $\partial L / \partial W_{c2}$ & $1 \times c$ & 0.004 \\
        & $\partial L / \partial(\Vec{h}W_{c1}+\Vec{b}_{c1})$ & $1 \times 768$ \& $1 \times c$ & 0.007 \\
        & $\partial L / \partial W_{c1}$ & $(768 \times 1) \times (1 \times 768) $ & 0.52 \\
        & $\partial L / \partial X^{(2)}$ & $(1 \times 768)$ & 0.27 \\ \midrule 

       \multirow{14}{*}{\makecell[c]{Backward:\\ Transformer\\ Layer}} & $\partial L / \partial (X_{ffn}+X_{att})$ & $128 \times 768$ & 0.067 \\
        & $\partial L / \partial X_{att}$ & $ 128 \times 768 \& (128 \times 3072)\times 2$ & 26.7 \\
        & Bootstrapping & $128 \times 768$ & 11.45 \\
        & $\partial L / \partial(X_o+X)$ & $128 \times 768$ & 0.097 \\
        & $\partial L / \partial X_{cat}$ & $128 \times 768$ & 3.73 \\

        & $\partial L / \partial V$ & $(128 \times 128) \times (128 \times 64) \times 12$ & 9.82 \\
        & $\partial L / \partial U$ & $(128 \times 128) \times (128 \times 64)\times 12$ & 5.5 \\
        & $\partial L / \partial S$ & $(128 \times 128) \times 2 \times 12$ & 0.65 \\
        & $\partial L / \partial Q$ & $(128 \times 128) \times (128 \times 64) \times 12$ & 0.87 \\
        & $\partial L / \partial K$ & $(128 \times 128) \times (128 \times 64) \times 12$ & 11.8 \\
        & Bootstrapping & $(128 \times 768) \times 3$ & 33.43 \\
        & $\partial L / \partial A* $, $*=Q,K,V$ & $(128 \times 768)^{\intercal} \times (128 \times 768) \times (768 \times 2)^{\intercal} \times 3$ & 3.7 \\
        & $\partial L / \partial B* $, $*=Q,K,V$ & $(128 \times 2)^{\intercal} \times (128 \times 768) \times 3$ & 1.8 \\
        & $\partial L / \partial X $ & $ (128 \times 768) \times 4 $ & 35.7 \\ \midrule 

        Optimizer & AdamW-HE & $(768 \times c)$, $(768 \times 768)$, $(768 \times 2) \times 3$, $(2 \times 768) \times 3$ & 19.9 \\
        \bottomrule

    \end{tabular}
    }
\end{table}

\subsection{End-to-end results across multiple batch sizes}
We conducted additional experiments across batch sizes 4, 16, and 64 to showcase that the proposed flexible packing method is well-suited for transformer fine-tuning. 
\paragraph{Matrix operations and vector-wise function.}
The amortized time (Table \ref{tab:batch-matrix} decreases when batch size increases, which is consistent to our complexity overhead analysis in Appendix \ref{app:matrix} and Section \ref{sec:non-linear}. 
Note that for element-wise functions such as exp and GELU, we evaluate them by functional encryption, and the amortized time is independent of the batch size.
\begin{table}[htb]
    \centering
    
    \caption{Time costs (sec) of matrix operations and vector-wise function LayerNorm across multiple batch sizes.}
    \label{tab:batch-matrix}

    \resizebox{\textwidth}{!}{
    
    \begin{tabular}{l|cc|cc|cc}
    \toprule
\textbf{Operation} &	\multicolumn{2}{c|}{\textbf{Batch Size 4}}	 &	\multicolumn{2}{c|}{\textbf{Batch Size 16}}	 &	\multicolumn{2}{c}{\textbf{Batch Size 64}}	\\
 & \textbf{Total Time}&	\textbf{Amortized}	&\textbf{Total Time}&	\textbf{Amortized}	&\textbf{Total Time}	&\textbf{Amortized} \\
\midrule
CPMM	&0.26	&0.065&	0.97	&0.061	&3.11&	0.049\\
CCMM-LoRA	&0.018&	0.0045&	0.067	&0.0042&	0.169&	0.0026\\
CCMM-kernel&	0.88	&0.22	&3.42	&0.21&	10.2&	0.16\\
CCMM-V	&0.86	&0.22&	3.24&	0.20	&10.32&	0.16\\
LayerNorm	&1.02	&0.26&	1.90&	0.12&	6.02&	0.094\\
\bottomrule
    \end{tabular}
    
    }
\end{table}

\paragraph{2-layer BERT-style model.}
Across all tested batch sizes, our packing uses all available plaintext slots in the ciphertexts. The results in Table \ref{tab:batch-bert} show that the amortized latency decreases from 45.32s at batch size 4 to 36.67s at batch size 64. We further show that noise remains very close across all settings (Table \ref{tab:batch-noise}). 
\begin{table}[htb]
    \centering

    \caption{Time costs (sec) in 2-layer BERT-style model across multiple batch sizes. }
    \label{tab:batch-bert}
    \begin{tabular}{lcccc}
    \toprule
         & \textbf{Sub-operation} & \textbf{Batch Size 4}	&	\textbf{Batch Size 16}	&	\textbf{Batch Size 64}\\
         \hline
     \textbf{Forward pass} & Matrix & 17.10 & 67.79 & 216.90 \\
     & Non-linear & 	4.88	&9.56&	28.80\\
    &(Func) Boot	&73.80	&295.05	&1180.20\\
    \textbf{Backward pass}&	Matrix&	41.50	&164.04	&524.87\\
    &Non-linear	&1.02	&1.91	&6.02\\
    &(Func) Boot	&23.20&	92.45	&369.60\\
    \textbf{Optimizer}	&-	&19.76	&19.90	&20.24\\
    \textbf{Total}	&-	&181.26	&650.60&	2346.63\\
    \textbf{Amortized}	&-	&45.32	&40.66&	36.67\\
    \bottomrule
    \end{tabular}
\end{table}

\begin{table}[htb]
    \centering
    \caption{Mean and max of absolute noise across different batch sizes. }
    \label{tab:batch-noise}
    \begin{tabular}{lccc}
    \toprule
        & \textbf{Batch Size 4}	&	\textbf{Batch Size 16}	&	\textbf{Batch Size 64} \\
        \hline
        Mean of Absolute Noise& $2.32\times 10^{-8}$ & $2.54\times 10^{-8}$ & $2.64\times 10^{-8}$\\
        Max of Absolute Noise& $1.45\times 10^{-6}$ &$1.60\times 10^{-6}$ & $1.68\times 10^{-6}$\\
        \bottomrule
    \end{tabular}

\end{table}

\subsection{Additional cumulative component-level breakdown}
We in Section \ref{sec:micro} report separate improvements for matrix operations, functional bootstrapping, and bootstrapping counts. 
To makes the attribution clearer, we provide the following cumulative component-level breakdown in Table \ref{tab:breakdown-app}.
In summary:
\begin{itemize}
    \item Encrypted matrix evaluation contributes to largest absolute runtime reduction.
    \item Flexible interleaved packing and operation-specific algorithms substantially reduce expensive rotations and relinearizations (Table \ref{tab:comp-MM} in paper).
    \item Lower multiplicative depth and fewer required bootstrapping invocations reduce bootstrapping time.
    \item Functional bootstrapping further reduces the non-linear evaluation cost by folding element-wise function evaluation into the bootstrapping step.
\end{itemize}

\begin{table}[htb]
    \centering
    
    \caption{Cumulative component-level breakdown of each operation, in batch size 16. }
    \label{tab:breakdown-app}

    \resizebox{\textwidth}{!}{
    
    \begin{tabular}{lcccl}
    \toprule
\textbf{Operation}	&\cite{rhoencryption}	&\textsc{Felid} \textbf{(Ours)}&	\textbf{Reduction}	&\textbf{Main Cause}\\
\hline
Forward , Matrix	&1429	&67.79	&21.1×&	Packing + matrix algorithms\\
Forward , Non-linear&	291	&9.56	&30.4×&	Functional bootstrapping\\
Forward , (Func) Boot	&708	&295.05	&2.4×	&Reduced depth + fewer bootstrappings\\
Backward ,	Matrix	&2377&	164.04	&14.5×&	Packing + matrix algorithms\\
Backward , Non-linear&	47	&1.91&	24.6×	&Functional bootstrapping\\
Backward , (Func) Boot	&355	&92.45&	3.8×	&Reduced depth + fewer bootstrappings\\
\textbf{Optimizer}		&31	&19.90&	1.6×&	Compact packing (Appendix \ref{app:update})\\
\textbf{Total}	&	5238&	650.60&	8×	&Combined effect\\
\bottomrule
    \end{tabular}
    }
\end{table}

\subsection{Experiments across multiple hidden dimensions}
We include in Table \ref{tab:app-dim-matrix} and Table \ref{tab:app-dim-attention} that the amplification trend observed in Table 15, \cite{rhoencryption}. 
We additionally evaluate the amortized runtime with batch size 4 across multiple hidden dimensions. 
For CPMM and CCMM, we test them with $\R^{128\times d}\times \R^{d\times d}$. 
The results show that as the dimension $d$ increases, the performance gains from LoRA fine-tuning become larger.

We also follow the setting in Table 15 of \cite{rhoencryption} and evaluate the scaling behavior of Gaussian-kernel attention and softmax attention on a matrix in $\R^{n\times n}$. 
In \textsc{Felid}, the exponential function in Gaussian-kernel attention is evaluated by functional bootstrapping, which refreshes the ciphertext and evaluates the element-wise function in a single step. 
In contrast, softmax is a vector-wise function and cannot be directly evaluated by functional bootstrapping. 
Therefore, for a fair comparison, we compare Gaussian-kernel functional bootstrapping with the full “softmax evaluation + bootstrapping” flow, where both settings produce refreshed ciphertexts after function evaluation. The results also show that as the dimension $n$ increases, the performance gains from GK become larger.

\begin{minipage}[t]{0.5\linewidth}
\centering
    \captionof{table}{Operation time cost (s) in LoRA \\ fine-tuning across multiple hidden dimensions. }
    \label{tab:app-dim-matrix}
    \begin{tabular}{lccc}
    \toprule
    \textbf{Dim(d)} &	\textbf{256}	&\textbf{512}	&\textbf{768}\\
    \hline
    CPMM&	0.007	&0.027&	0.061\\
    CCMM&	0.067	&0.39	&1.13\\
    Factor	&9.57x&	14.4x	&18.5x\\
    \bottomrule
    \end{tabular}
\end{minipage}
\begin{minipage}[t]{0.5\linewidth}
    \centering
    \captionof{table}{Gaussian-kernel attention and softmax attention time cost (sec), across multiple dimensions. }
    \label{tab:app-dim-attention}

    \resizebox{\textwidth}{!}{
    
    \begin{tabular}{lccc}
    \toprule
    \textbf{Dim(n)} &	\textbf{128}	&\textbf{256}	&\textbf{512}\\
    \hline
    GK (Functional bootstrapping)&	0.3	&0.6	&2.39 \\
    Softmax+bootstrapping&	0.735	&1.726&	8.95 \\
    Factor&	2.45x&	2.88x	&3.74x \\
    \bottomrule
    \end{tabular}
    }
\end{minipage}

\subsection{CKKS approximation error during full training }
We performed an additional noise-propagation experiment to evaluate how CKKS approximation error accumulates through a full training step. we use the trained 2-layer Gaussian-kernel checkpoint and cyclically replicate its layers to construct deeper models.  we run forward pass, backward pass, and AdamW-HE update on the first batch of MRPC, RTE, and STS-B with batch size 16, and report the mean and maximum value of the absolute error (Table \ref{tab:CKKSerror}). These results show that \textsc{Felid} is robust to CKKS approximation error in the evaluated full-step training flow. The error does not blow up through forward propagation, backward propagation, or encrypted optimizer updates, and remains small compared with the scale of model activations and updates.
\begin{table}[htb]
    \centering
    \caption{The mean and maximum of the absolute error during full training. }
    \label{tab:CKKSerror}
    \resizebox{\textwidth}{!}{
    
    \begin{tabular}{lllll}
    \toprule
    \textbf{Number of layers}  &  \textbf{2} & \textbf{4} & \textbf{8} & \textbf{12} \\
    \hline
    \textbf{MRPC}&	\makecell{Mean: $2.54\times 10^{-8}$ \\Max: $1.9\times 10^{-6}$}&	\makecell{Mean: $3.14\times 10^{-8}$ \\Max: $2.9\times 10^{-6}$}	&\makecell{Mean: $5.47\times 10^{-8}$ \\Max: $9.8\times 10^{-6}$} &	\makecell{Mean: $3.03\times 10^{-7}$ \\Max: $8.6\times 10^{-5}$}\\
    \hline 
    \textbf{RTE}&	\makecell{Mean: $1.96\times 10^{-8}$ \\Max: $1.6\times 10^{-6}$}&	\makecell{Mean: $2.63\times 10^{-8}$ \\Max: $1.4\times 10^{-6}$}&	\makecell{Mean: $6.97\times 10^{-8}$ \\Max: $5.0\times 10^{-5}$}&	\makecell{Mean: $2.63\times 10^{-7}$ \\Max: $9.3\times 10^{-5}$}\\
    \hline
    \textbf{STS-B}&	\makecell{Mean: $5.51\times 10^{-7}$ \\Max: $5.1\times 10^{-5}$}	&\makecell{Mean: $7.18\times 10^{-7}$ \\Max: $7.4\times 10^{-5}$}	&\makecell{Mean: $7.48\times 10^{-7}$ \\Max: $7.5\times 10^{-5}$}	&\makecell{Mean: $1.14\times 10^{-6}$ \\Max: $1.6\times 10^{-4}$}\\
    \bottomrule
    \end{tabular}
    }
\end{table}


\end{document}